\pdfoutput=1
\documentclass[usenatbib]{mn2e}
\usepackage{apjfonts}
\usepackage{amssymb}
\usepackage{amsmath}
\usepackage{ctable}
\usepackage{fixltx2e} 

\newcommand{\be}{\begin{equation}}
\newcommand{\ee}{\end{equation}}

\newcommand{\msun}{M_{\sun}}

\newcommand\plotonesize[2]
 {\centering \leavevmode \includegraphics[width={#2\columnwidth}]{#1}}
\newcommand\plotone[1]
 {\centering \leavevmode \includegraphics[width={0.99\columnwidth}]{#1}}

\newcommand{\acknowledgments}{\begin{small}\section*{Acknowledgments}\end{small}}
\newcommand\altaffilmark[1]{$^{#1}$}
\newcommand\altaffiltext[1]{$^{#1}$}
\title[ISM Structure \&\ Growth Histories]{An Excursion-Set Model for the 
Structure of GMCs and the ISM\vspace{-0.5cm}}

\author[Hopkins]{
\parbox[t]{\textwidth}{ 
Philip F.~Hopkins\thanks{E-mail:phopkins@astro.berkeley.edu}\altaffilmark{1}
}
\vspace*{6pt} \\
\altaffiltext{1}{Department of Astronomy and Theoretical Astrophysics Center, University of California Berkeley, Berkeley, CA 94720}}

\date{Submitted to MNRAS, October 18, 2011}
\begin{document}
\maketitle
\label{firstpage}

\begin{abstract}
The ISM is governed by supersonic turbulence on a range of scales. We use this simple fact to develop a rigorous excursion-set model for the formation, structure, and time evolution of dense gas structures (e.g.\ GMCs, massive clumps and cores). Supersonic turbulence drives the density distribution in non-self gravitating regions to a lognormal with dispersion increasing with Mach number. We generalize this to include scales $\gtrsim h$ (the disk scale height), and use it to construct the statistical properties of the density field smoothed on a scale $R$. We then compare conditions for self-gravitating collapse including thermal, turbulent, and rotational (disk shear) support (reducing to the Jeans/Toomre criterion on small/large scales). We show that this becomes a well-defined barrier crossing problem. As such, an exact ``bound object mass function'' can be derived, from scales of the sonic length to well above the disk Jeans mass. This agrees remarkably well with observed GMC mass functions in the MW and other galaxies, with the only inputs being the total mass and size of the galaxies (to normalize the model). This explains the cutoff of the mass function and its power-law slope (close to, but slightly shallower than, $-2$). The model also predicts the linewidth-size and size-mass relations of clouds and the dependence of residuals from these relations on mean surface density/pressure, in excellent agreement with observations. We use this to predict the spatial correlation function/clustering of clouds and, by extension, star clusters; these also agree well with observations. We predict the size/mass function of ``bubbles'' or ``holes'' in the ISM, and show this can account for the observed HI hole distribution without requiring any local feedback/heating sources. We generalize the model to construct time-dependent ``merger/fragmentation trees'' which can be used to follow cloud evolution and construct semi-analytic models for the ISM, GMCs, and star-forming populations. We provide explicit recipes to construct these trees. We use a simple example to show that, if clouds are not destroyed in $\sim1-5$ crossing times, then all the ISM mass would be trapped in collapsing objects even if the large-scale turbulent cascade were maintained.
\end{abstract}


\section{Introduction}
\label{sec:intro}

The origins and nature of structure in the interstellar medium (ISM) and 
giant molecular clouds (GMCs) represents one of the most important unresolved 
topics in both the study of star formation and galaxy formation. 
In recent years, there have been several major advances in our understanding of the relevant processes. It is clear that a large fraction of the mass in the ISM is supersonically turbulent over a wide range of scales, from 
the sonic length ($\sim0.1\,$pc) through and above the disk scale height ($\sim$kpc). 
A generic consequence of this super-sonic turbulence -- so long as it can be maintained -- is 
that the density distribution converges towards a lognormal PDF, with a dispersion that scales weakly with mach number \citep[e.g.][]{vazquez-semadeni:1994.turb.density.pdf,
padoan:1997.density.pdf,scalo:1998.turb.density.pdf,
ostriker:1999.density.pdf}. 

Without continuous energy injection, this turbulence would dissipate in a single crossing time, 
and the processes that ``pump'' turbulence (generally assumed to be related to feedback from 
massive stars) remain poorly understood \citep[see e.g.][and references therein]{mac-low:2004.turb.sf.review,mckee:2007.sf.theory.review,hopkins:fb.ism.prop}. 
However, provided this turbulence can be maintained, it is able to explain the relatively 
small fraction of mass which collapses under the runaway effects of self-gravity and cooling 
\citep{vazquez-semadeni:2003.turb.reg.sfr,li:2004.turb.reg.sfr,li:2006.turb.reg.sfr,
padoan:2011.new.turb.collapse.sims}. 
In this picture, 
star formation occurs within dense cores, themselves typically embedded inside 
giant molecular clouds (GMCs), which represent regions where turbulent density fluctuations become sufficiently overdense so as to be marginally self-gravitating and collapse 
\citep{evans:1999.sf.gmc.review,gao:2004.hcn.sfr.relation,bussmann:2008.hcn32.sfr}.
Some other process such as stellar feedback 
is believed to be responsible for disrupting the clouds, after a few crossing times \citep[e.g.][]{evans:2009.sf.efficiencies.lifetimes}. 
The turbulent cascade has also been invoked to explain GMC scaling relations, such 
as the size-mass and linewidth-size relations \citep{larson:gmc.scalings,
scoville:gmc.properties}. 

However, despite this progress, there remains no rigorous analytic theory that can simultaneously predict these properties, as well as other key observables such as the GMC mass function, and the spatial distribution of gas over and under-densities in the ISM. 

The approximately Gaussian distribution of the logarithmic density field, though, suggests that considerable progress might be made by adapting the excursion set or ``extended Press-Schechter'' formalism. 
This has proven to be an extremely powerful tool in the study of cosmology and galaxy evolution. 
The seminal work by \citet{pressschechter} derived the form of the halo mass function via a simple (albeit somewhat ad hoc) calculation of the mass fraction expected to be above a given threshold for collapse, expected in a Gaussian overdensity distribution with the variance as a function of scale derived from the density power spectrum. \citet{bond:1991.eps} developed a rigorous analytic (and statistical Monte Carlo) formulation of this, defining the excursion set formalism for dark matter halos. Famously, this resolved the ``cloud in cloud'' problem, providing a means to calculate whether structures were embedded in larger collapsing regions. Since then, excursion set models of dark matter have been studied extensively: they have been generalized and used to predict -- in addition to the halo mass function -- the 
spatial distribution/correlation function of halos \citep{mowhite:bias}, 
the distribution of voids \citep{sheth:2004.void.eps}, 
the evolution and structure of HII regions in re-ionization \citep{haiman:2000.reion.bubble.eps,
furlanetto:2004.reion.bubble.eps}, 
and many higher-order properties used as cosmological probes. By incorporating the time-dependence of the field, they have been used to study the growth and merger histories of halos and to construct Monte Carlo ``merger trees'' \citep{bower:1991.eps.groups,laceycole:halo.merger.rates.93}. These trees formed the basis for the extensive field of semi-analytic models for galaxy formation, in which analytic physical prescriptions for galaxy evolution are ``painted onto'' the background halo evolution 
\citep[e.g.][]{somerville:merger.trees,cole:durham.sam.initial}. It is not an exaggeration to say that it has proven to be one of the most powerful theoretical tools in the study of large scale structure and galaxy formation.

There have been other, growing suggestions of similarities between the mathematical structure of the ISM and that invoked in excursion set theory. 
The mass function of GMCs, for example, has a faint-end slope quite similar to that of galaxy halos 
(both close to ${\rm d}n/{\rm d}M\propto M^{-2}$), suggestive of hierarchical collapse.
\citet{vazquez-semadeni:1994.turb.density.pdf} attempted to rigorously examine whether the structure of the ISM should be ``hierarchical,'' although they strictly define this as the probability of many independent fluctuations dominating the ``peaks'' in the density distribution (which does not technically need to be satisfied in excursion set theory). This is related to (but not equivalent to) the large body of work on the quasi-fractal structure of the ISM \citep[see e.g.][and references therein]{elmegreen:2002.fractal.cloud.mf}. On smaller scales, \citet{krumholz.schmidt} suggested that the fraction of a lognormal PDF above a ``collapse threshold'' at the sonic length could explain the fractional mass forming stars per free-fall time, inside of GMCs. \citet{padoan:1997.density.pdf,padoan:2002.density.pdf} suggested that the distribution of lognormal density fluctuations above a threshold overdensity could explain the shape of the stellar initial mass function (IMF).
\citet{scalo:1998.turb.density.pdf} explicitly discuss the analogy between this and cosmological 
density fluctuations, and \citet{hennebelle:2008.imf.presschechter} expanded upon the \citet{padoan:1997.density.pdf} argument using a derivation almost exactly analogous to the original \citet{pressschechter} derivation, and showed that it agreed well with the standard IMF.

But despite these suggestions, and the enormous successes of the excursion set model in cosmological applications, there has been no attempt to translate the excursion set formalism to the problem of the ISM and GMC evolution. At first glance, it is obvious why. The cosmological excursion 
set theory is applied to small fluctuations of the linear density field, in the linear regime, 
to dark matter (collisionless) systems, with Gaussian, nearly scale-free fluctuations seeded by inflation, 
and to Lagrangian ``halos'' which (modulo mergers) are conserved in time. 
The Gaussian distribution of ISM densities represent large fluctuations in the logarithmic 
density field, which are a product of a fully non-linear, turbulent, gaseous (collisional) medium, 
and evolve both rapidly and stochastically in time. 

However, in this paper, we will show that although the {\em physics} involved are very different, 
none of these differences fundamentally invalidates the underlying mathematical formalism 
of the excursion set theory. 

Here, we develop a rigorous excursion-set model 
for the formation, structure, and time evolution of structures in the ISM and within GMCs. 
We show that this is possible, and that it allows us to develop statistical predictions of ISM properties in  a manner analogous to the predictions for the halo mass function. 
In \S~\ref{sec:model} we describe the model. First (\S~\ref{sec:model:collapse}), 
we derive the conditions for self-gravitating collapse in a turbulent 
medium (the ``collapse threshold''), in a manner generalized to both small (sonic length) and large 
(above the disk scale height) scales. 
Next (\S~\ref{sec:model:densities}), we discuss the density field, 
and, assuming it has a lognormal character, construct the statistical properties of the field 
smoothed on a physical scale $R$, which allows us to define the excursion set ``barrier crossing'' 
problem. 
In \S~\ref{sec:model:barrier:mf}, we use this to derive an exact ``self-gravitating object'' mass function, 
over the entire range of masses (from the sonic length to disk mass), and show that 
it agrees remarkably well with observed GMC mass functions and depends only very weakly on the exact turbulent properties of the medium (including deviations from a lognormal PDF). 
In \S~\ref{sec:model:barrier:larsons}, we show that the model also predicts the 
linewidth-size and size-mass relations of GMCs, and their dependence on external galaxy properties. 
We also examine how this depends on the exact properties of the turbulent cascade. 
In \S~\ref{sec:model:barrier:clustering}, we extend the model to predict the spatial correlation function 
and clustering properties of clouds (and, by extension, young star clusters), 
and compare this to observations.
In \S~\ref{sec:model:barrier:bubbles}, we predict the size and mass distributions of 
underdense ``bubbles'' or ``holes'' in the ISM which result simply from the same 
normal turbulent motions. We show that this can explain most or all of the distribution of 
HI ``holes'' observed in nearby galaxies, without explicitly requiring any feedback mechanism 
to power the hole expansion.
In \S~\ref{sec:model:trees}, 
we generalize the model to construct time-dependent ``GMC merger/fragmentation trees'' 
which follow the time evolution, growth histories, fragmentation, and mergers of clouds. 
In \S~\ref{sec:model:trees:mechanism} we provide simple recipes to construct these trees, and discuss how they can be used to build semi-analytic models for GMC and ISM evolution and star formation, in direct analogy to semi-analytic models for galaxy formation. We use a very simple example of this to predict the rate at which the gas in the ISM collapses (absent feedback) into bound structures, show that this agrees well with the results of fully non-linear turbulent box simulations, and argue that feedback must destroy clouds on a short timescale (a few crossing times) to prevent runaway gas consumption. 
Finally, in \S~\ref{sec:discussion}, we summarize our results and conclusions and discuss a number of possibilities for future work, both to improve the accuracy of these models and to enable predictions for additional properties of the ISM.

\section{The Model}
\label{sec:model}

The fundamental assumption of our model is that non-rotational velocities 
are dominated by super-sonic turbulence (down to some sonic length), with some power spectrum 
$P(k)$ or $E(k)$\footnote{
There are different conventions in the turbulence and excursion-set literature 
for the normalization and $k$-dependence in the definition of $P(k)$. 
To simplify matters, we will refer to the velocity power spectrum 
by means of $E(k)$, which with the assumption of isotropic turbulence gives 
the differential energy per mode as ${\rm d}E=E(k)\,{\rm d}k$.
} which is maintained by any process (presumably stellar 
feedback) in approximate statistical steady-state.
As we discuss in \S~\ref{sec:discussion}, all other assumptions we make are convenient 
approximations to simplify our calculations, but it is possible to generalize the model.

The two key quantities we need to calculate the cloud mass function 
and other properties are the conditions for ``collapse'' of a cloud (i.e.\ conditions under which self-gravity 
can overcome turbulent forcing) and the power spectrum of density 
fluctuations. Below, we show how these can be calculated for a turbulent medium 
from the velocity power spectrum; however, 
in principle they can be completely arbitrary (for example, specified 
ad hoc from numerical simulations or observations). So long as they are known, 
the rest of our model proceeds identically.

\subsection{Collapse in a Turbulent Medium}
\label{sec:model:collapse}

First, for simplicity, consider gas in a 
galaxy whose average properties are evaluated on a scale 
$R$ where the velocity dispersion is highly supersonic ($R\gg\ell_{\rm sonic}$, where 
$\ell_{\rm sonic}$ is the sonic length), 
but where shear from the disk rotation and large-scale density gradients 
can be neglected ($R\ll h$, where $h$ is the disk scale-height). 
The turbulent dispersion on these scales is $\langle v_{t}(R)^{2}\rangle \sim k\,E(k)$. 
If the turbulence has a power-law cascade over this interval then 
$E(k)\propto k^{-p}$. 
If the region has some mean density $\rho$ (on the same scale $R$), 
then the potential from self-gravity is $|U| \approx \beta\,G M/R \approx \beta G\rho (4\pi/3)\,R^{2}$ 
while the kinetic energy in turbulence is $(1/2)\,\langle v_{t}(R)^{2}\rangle$; 
the region will be gravitationally bound and ``trapped'' when 
$\rho \gtrsim \rho_{c} = (3/8\pi\beta)\,\langle v_{t}(R)^{2}\rangle/G\,R^{2}$, 
where $\beta\sim1$ depends on the shape (internal structure) of the 
density perturbation. 
Formally, we also need to check whether the momentum ``input'' rate 
from the turbulent cascade (equal to the dissipation rate in steady-state) 
is less than the gravitational force, and whether the energy input rate 
is less than the rate at which a gravitationally collapsing object will 
dissipate. 
However, because for super-sonic turbulence, 
the timescale for energy or momentum dissipation 
on a scale $R$ just scales with the crossing time $t_{\rm cross}=R/v_{t}(R)$, 
we obtain the identical dimensional scaling for all of these criteria.\footnote{The 
energy injection rate in the turbulence is $\dot{u} = (1/2)\,v_{t}^{2}(R)/(\eta\,t_{\rm cross}) = 
(1/2)\,v_{t}^{2}/(\eta\,R/v_{t})$ 
where $\eta\sim1$ is constant. A virialized object where cooling 
is rapid (i.e.\ pressure forces can be neglected), where the virial 
motions are turbulent, will then just lose energy at a rate
$=(1/2)\,|U|/(\eta\,R/\sqrt{\beta\,G\,M/R})$ -- equating these gives 
an identical dimensional requirement on $\rho$ to the binding criterion, but with a slightly different 
coefficient. 
Equating the turbulent momentum input rate ${\rm d}(Mv_{t})/{\rm d}t = M\,v_{t}/(\eta\,t_{\rm cross})$ 
to the gravitational force $F_{\rm grav} \approx G\,M/R^{2}$ 
again gives the identical result. 
We should take the most stringent normalization from 
these as the relevant criterion, but this is entirely degenerate with the value of $\beta\sim1$. For a rigorous derivation of each of these criteria, see \citet{bonazzola:1987.turb.jeans.instab}.}

These are simply a restatement of the Jeans criterion, 
for wavenumber $k \sim 1/R$, 
but with the sound speed $c_{s}$ replaced by the turbulent velocity dispersion $v_{t}$. 
For an individual $k$-mode (sinusoidal density perturbation), 
the criteria becomes
\be 
\rho(R) \ge \frac{k^{2}\langle v_{t}^{2}(k) \rangle}{4\pi\,G} \propto k^{3-p} \propto R^{p-3}
\ee
where the latter equalities assume a power-law spectrum \citep{vazquez-semadeni:1995.turb.jeans.instab}. 
If the system is marginally stable with density $\rho_{0}$ on scale $R_{0}$, 
then this simply becomes
$\rho(R) \ge \rho_{0}\, (R/R_{0})^{p-3}$. 
If we are in the super-sonic regime, then we expect 
something like Burgers turbulence \citep{burgers1973turbulence}, with $p\approx 2$; 
but we will discuss this further below.

Now generalize this to a more broad range of radii. 
On small scales, we need to include the effects of 
thermal pressure: this amounts to a straightforward 
modification of the Jeans criterion with 
$v_{t}^{2}\rightarrow c_{s}^{2} + v_{t}^{2}$ 
\citep{chandrasekhar:1951.turb.jeans.condition,bonazzola:1987.turb.jeans.instab}.\footnote{It is likely that the power spectrum of velocities $v_{t}$ will change 
as we go to scales below the sonic length; 
however, since (by definition) $v_{t}<c_{s}$ in this regime, 
such corrections have essentially no effect on our results.
Moreover the change -- expected to be e.g.\ a transition from 
$p=2$ to $p=5/3$, is small for our purposes. 
}
On large scales, we need to include the effects of rotation 
stabilizing perturbations. 
If we focused only on very large ($R\gg h$) scales, 
where we can neglect the disk thickness, 
then we simply re-derive the 
\citep{toomre:spiral.structure.review}
dispersion relation and collapse conditions, 
with the gas ``dispersion'' $\sigma_{g}^{2}=v_{t}^{2}+c_{s}^{2}$.
More generally, 
\citet{begelman:direct.bh.collapse.w.turbulence} note that 
the dispersion relation for growth of density perturbations in a 
turbulent disk (with finite thickness $h$) can be written: 
\begin{align}
\label{eqn:dispersion.relation}
\omega^{2} &= \kappa^{2} + \sigma_{g}(k)^{2}\,k^{2} - \frac{2\pi\,G\,\Sigma\,|k|}{1+|k|\,h} \\ 
&= \kappa^{2} + \sigma_{g}(k)^{2}\,k^{2} - \frac{4\pi\,G\,\rho\,|k|\,h}{1+|k|\,h}
\end{align}
where $\Sigma\equiv 2\,h\,\rho$ is the disk surface density, 
$\rho$ the average density on scale $k$, and $h$ the disk scale-height, 
$v_{t}$ the turbulent velocity dispersion, and $\kappa$ the usual epicyclic 
frequency. This differs from the infinitely thin-disk 
dispersion relation by the term $(1+|k|\,h)^{-1}$, which accounts for 
the finite scale height for modes with scales $\lambda \lesssim h$ 
\citep{vandervoort:1970.dispersion.relation,elmegreen:1987.cloud.instabilities,
romeo:1992.two.component.dispersion}.\footnote{
Eqn.~\ref{eqn:dispersion.relation} 
is an exact solution for a disk with an 
exponential vertical profile. It is also always asymptotically 
exact at small and large $|k|$ and tends to be within $\sim10\%$ of the 
exact solution at all $|k|$ for the range of observed vertical profiles 
\citep[see][]{kim:2002.mhd.disk.instabilities}.}
Note that this relation nicely interpolates between the Jeans criterion 
which we derived above on small scales ($k\gg h^{-1}$), 
and the Toomre (thin-disk) dispersion relation on large 
scales ($k\ll h^{-1}$). 

If the average density is $\rho_{0}$, 
and corresponding average surface density $\Sigma_{0}$, 
then we can define the usual Toomre $Q$ at the scale $h$
\be
Q_{0}(h)\equiv \frac{\kappa\,\sigma_{g}(h)}{\pi\,G\,\Sigma_{0}}
=\frac{h\,\tilde{\kappa}\,\Omega^{2}}{\pi\,G\,\Sigma_{0}}
\ee
where the second equality follows from $\sigma_{g}(h)=h\,\Omega$, 
which is be true for any disk in vertical equilibrium, 
and we define $\tilde{\kappa} \equiv \kappa/\Omega$ ($=\sqrt{2}$ for a 
constant-$V_{c}$ disk).
If we define the convenient dimensionless form of $k$, $\tilde{k}\equiv |k|\,h$, 
we can write the criterion for instability 
($\omega^{2}<0$) as
\begin{align}
\label{eqn:rhocrit}
\frac{\rho}{\rho_{0}} &\ge \frac{\rho_{c}}{\rho_{0}} \equiv \frac{Q_{0}(h)}{2\,\tilde{\kappa}}\,(1+\tilde{k})
{\Bigl[} \frac{\sigma_{g}(k)^{2}}{\sigma_{g}(h)^{2}}\,\tilde{k}  + \tilde{\kappa}^{2}\,\tilde{k}^{-1}{\Bigr]} 
\end{align}
Note that the assumption of a finite $Q_{0}$ ensures that so long as there is any non-gaseous component of the potential, the gas alone is not self-gravitating on arbitrarily large scales (this is important below, to un-ambiguously define the largest self-gravitating scales of clouds). 
Again, on small scales $kh\gg1$, this reduces to 
the Jeans criterion $\rho_{c}= k^{2}\,\sigma_{g}(k)^{2}/(4\pi\,G) \propto R^{p-3}$, 
and on large scales $kh\ll1$ it becomes $\rho_{c} = \Sigma_{c}/2h = (k\,h)^{-1}\kappa^{2}/(4\pi G)\propto R$. 

\citet{kim:2002.mhd.disk.instabilities} 
note that is straightforward to further generalize this criterion to include the 
effects of magnetic fields by taking $\sigma_{g}^{2}=v_{t}^{2}+c_{s}^{2}+v_{\rm A}^{2}$, 
where $v_{\rm A}^{2}$ is the Alfv{\`e}n speed.  If we follow the usual convention 
in the literature and assume $\beta\equiv c_{s}^{2}/v_{\rm A}^{2}$ is constant, 
then changing the strength of magnetic fields is identically equivalent to 
changing the sound speed/mach number (which we explicitly consider below). 
Even if we allow $\beta$ to have an arbitrary power spectrum, the 
results are quite similar to this renormalization -- for any power spectrum 
where the magnetic energy density is peaked on large scales, it is nearly equivalent 
to renormalizing the turbulent velocities; for a power spectrum peaked 
on small scales, equivalent to renormalizing the sound speed. 
We therefore will not {\em explicitly} consider magnetic fields in what follows, 
but emphasize that they are straightforward to include if their power spectrum 
is known.

Formally, the turbulent velocity power spectrum $E(k)$ 
must eventually flatten/turn over on large scales $R\gtrsim h$, 
both by definition (since $h$ itself traces the maximal three-dimensional 
dispersions) and to avoid energy divergences. If it did not, 
we would recover $v_{t}\gtrsim V_{c}$ on large scales in gas-rich systems! 
Constancy of energy transfer and energy conservation 
require that the slope become at least as shallow as $E(k)\propto k^{-1}$. 
A good approximation to the behavior seen in simulations is obtained 
by generalizing the exact correction for $k$ near the lowest 
wavenumbers in the inertial scale 
in Kolmogorov turbulence \citep[][]{bowman:inertial.range.turbulent.spectra}, 
taking $E(k)\rightarrow E(k)\,(1+|k\,h|^{-2})^{(1-p)/2}$, which interpolates 
between these regimes.
This may not be exact. 
Fortunately however, even if we ignored this correction entirely, we can see immediately from 
Equation~\ref{eqn:rhocrit} that for any reasonable 
power spectrum ($p<3$), the dominant velocity/pressure term on scales 
$\gtrsim h$ is the disk shear ($\sim \kappa\,R$), not $v_{t}$. 
We therefore include this turnover, but stress that it is 
not necessary to our derivation and has only weak effects 
on our conclusions.

\subsection{The Density Distribution}
\label{sec:model:densities}

The other required ingredient for our model is 
an estimate of the density PDF/power spectrum. 
We emphasize that the our methodology is robust to 
the choice of an {\em arbitrary} PDF and/or power spectrum in 
$\rho$. We could, for example, simply extract a density power spectrum 
(or fit to it) from simulations or observations. 
This is, however, less predictive -- so in this paper, we chose to 
focus on the case of supersonic turbulence in which case it 
is possible to (at least approximately) construct the density 
PDF knowing only the velocity power spectrum information. 

As discussed in \S~\ref{sec:intro},
in idealized simulations of supersonic turbulence with a 
well-defined mean density $\rho_{0}$ 
and mach number $\mathcal{M}$ on a scale $k\sim1/R$, 
the distribution of densities 
tends towards a lognormal distribution 
\begin{align}
\label{eqn:lognormal}
{\rm d}p(\delta\,|\,k) &= \frac{1}{\sigma_{k}\sqrt{2\pi}}\,
\exp{{\Bigl(} - \frac{\delta^{2}}{2\,\sigma_{k}^{2}} {\Bigr)}}\,
{\rm d}{\delta} \\ 
\delta &\equiv \ln{{\Bigl(}\frac{\rho}{\rho_{0}} {\Bigr)}} - {\Bigl \langle}{\ln{{\Bigl(}\frac{\rho}{\rho_{0}} {\Bigr)}}} {\Bigr \rangle}
\end{align}
where because $\rho_{0}$ is the mean density,  
\be
{\Bigl \langle}{\ln{{\Bigl(}\frac{\rho}{\rho_{0}} {\Bigr)}}} {\Bigr \rangle} = -\frac{\sigma_{k}^{2}}{2}
\ee
This form of the PDF and our results are identical whether we define 
all quantities as volume-weighted or mass-weighted, so long as we are consistent throughout: 
here it is convenient to define all properties as {\em volume-weighted} 
(otherwise $\rho_{0}$ is scale-dependent).

The dispersion in these simulations is a function of the 
rms (one-dimensional) Mach number averaged on the same scale 
$\mathcal{M}(k)$, 
\be
\label{eqn:sigma.mach}
\sigma_{k} \approx {\Bigl(}\ln{{\Bigl [}1 + \frac{3}{4}\,{\mathcal{M}(k)^{2}}{\Bigr]}} {\Bigr)}^{1/2}
\ee
which is naturally expected for supersonic turbulence with efficient cooling 
(because the variance in $\ln{(\rho)}$ 
in ``events'' -- namely strong shocks -- scales as $\ln{(\mathcal{M}^{2})}$).\footnote{The exact coefficient in front of $\mathcal{M}^{2}$ in this scaling does depend on e.g.\ the form of turbulent forcing and other details \citep{federrath:2010.obs.vs.sim.turb.compare,price:2011.density.mach.vs.forcing}.  For our purposes, however, this is entirely degenerate with the normalization of the velocity/scale height of the disk and enters very weakly (sub-logarithmically). It is potentially more important, however, on small scales near the sonic length.}

If the turbulence obeys locality -- i.e.\ if the density distribution averaged on some small 
scale $R_{1}$ depends only on the local gas properties on that scale as opposed to e.g.\ the 
structure on much larger scales $R_{2}\gg R_{1}$ -- then 
the distribution of densities $\delta({\bf x},\,R)$ averaged over any spatial scale $R$ with 
some window function $W({\bf x},\,R)$
is also a lognormal in $\delta$, with 
variance
\be
\label{eqn:variance.R}
\sigma^{2}(R) = \int {\rm d}\ln{(k)}\,\sigma_{k}^{2}(\mathcal{M}[k])\,|W(k,\,R)|^{2}
\ee
where $W(k,\,R)$ is the Fourier transform of $W(x,\,R)$. 
This is easy to see if we recursively divide an initially large volume (e.g.\ the 
entire disk) into sub-regions with different mean $\rho_{0}$ and $\mathcal{M}$ 
on scale $R$; each of these sub-regions is a ``box'' that should obey 
the density distributions above, and so on. 
Because it greatly simplifies the algebra, we will generally follow the standard practice in 
the excursion set literature and choose $W(k,\,R)$ to be a 
Fourier-space top-hat: $W(k\,|\,R_{w}) = 1$ if $k\le R_{w}^{-1}$ and 
$W(k\,|\,R_{w}) = 0$ if $k> R_{w}^{-1}$.
This choice is arbitrary, but so long as it is treated consistently, 
our subsequent results are essentially identical (we will show, for example, 
that using a Gaussian window function makes a small difference 
in all predicted quantities).\footnote{As has been discussed 
extensively in the EPS literature, 
this does introduce some ambiguity in 
the definition of ``mass'' in the mass function, since the real-space window volume is 
not well-defined. In practice, if we adopt a fixed definition of volume 
$ = (4\pi/3)\,R_{w}^{3}$, the corresponding 
systematic differences are relatively small ($<10\%$) between 
different window function 
crossing distributions \citep[see][]{zentner:eps.methodology.review}.}

It should immediately be clear, however, that if we simply 
extrapolated $\mathcal{M}^{2}=v_{t}^{2}/c_{s}^{2}\propto R^{(p-1)/2}$, 
the dispersion would be divergent! Physically, this would imply 
ever larger fluctuations in $\log{\rho}$ on arbitrarily large scales; but this cannot 
be true once the scale $R$ approaches that of the entire disk. As 
$kh\rightarrow0$, the fact that the disk has finite mass means that $\sigma_{k}\rightarrow0$. 
The resolution of this apparent dilemma is evident in Equation~\ref{eqn:dispersion.relation}: 
what matters in $\mathcal{M}$ in the dispersion is the effective ``pressure'' from 
$c_{s}^{2}$; on sufficiently large scales $kh\ll1$, the differential rotation $\kappa/k$ plays an identical 
physical role. 
We can therefore generalize Equation~\ref{eqn:sigma.mach} as
\begin{align}
\label{eqn:sigma.mach.general}
\sigma_{k} &\approx {\Bigl(}\ln{{\Bigl [}1 + \frac{3}{4}\,
\frac{v_{t}^{2}(k)}{c_{s}^{2} + \kappa^{2}\,k^{-2}} 
{\Bigr]}} {\Bigr)}^{1/2}\\
&={\Bigl(}\ln{{\Bigl [}1 + \frac{3}{4}\,
\frac{\mathcal{M}^{2}(k)}{1 + \mathcal{M}_{h}^{2}\tilde{\kappa}^{2}/|kh|^{2}} 
{\Bigr]}} {\Bigr)}^{1/2}
\end{align}
where $\mathcal{M}_{h}\equiv \sigma_{g}(h)/c_{s}$. 
This ensures the correct physical behavior, $\sigma_{k}\rightarrow0$ 
as $k\rightarrow0$ for all plausible turbulent $E(k)$. 

At some level, our assumptions must break down. 
And although it is well-established that the density PDF at the 
resolution limit in numerical simulations  
(in a ``box-averaged'' sense) approaches the behavior of 
Eqns.~\ref{eqn:lognormal}-\ref{eqn:sigma.mach}, it is less clear whether we can assume this 
on a $k$-by-$k$ basis and so derive Eqns.~\ref{eqn:variance.R}-\ref{eqn:sigma.mach.general}. 
The lognormal character of the density distribution holding on various smoothing 
scales as we assume is, however, supported in the investigations of \citet{lemaster:2009.density.pdf.turb.review} 
and \citet{passot:1998.density.pdf,scalo:1998.turb.density.pdf}.
And any distribution which is lognormal in either real space or $k$ space must be lognormal in both. 
Moreover, the robustness of this assumption is supported by the conservation of lognormality 
in resolution studies, since all simulations essentially measure the PDF smoothed over a window function corresponding to their resolution limits. To the extent that there is some violation of 
these assumptions in e.g.\ the higher-order-structure functions \citep[although they are 
largely consistent with locality when $\mathcal{M}_{h}$ is large; see][]{boldyrev:2002.sf.cloud.turb,padoan:2004.turb.structure.functions,schmidt:2008.turb.structure.fns}, 
this is really a question of the degree to which the density PDF globally departs from 
a lognormal, which we discuss below.

What is somewhat less clear is how Eq.~\ref{eqn:sigma.mach} generalizes on a scale-by-scale basis. 
Analytically, the same arguments that prove the density distribution of isothermal turbulence should 
converge to a lognormal with real-space variance $\sigma^{2}=\ln{(1+(3/4)\langle\mathcal{M}^{2}\rangle)}$  trivially generalize to a $k$-by-$k$ basis 
\citep[Eq.~\ref{eqn:sigma.mach}; see][]{passot:1998.density.pdf,nordlund:1999.density.pdf.supersonic}. 
If locality also holds, Eq.~\ref{eqn:variance.R} must follow.
This is the origin of the expectation for analytic models of the density power spectrum. 
Note that, as defined, $\sigma_{k}^{2}$ is equivalent to the logarithmic density 
power spectrum, $\sigma_{k}^{2}=k\,E_{\ln{\rho}}(k)$. When 
$\mathcal{M}$ is not large, $\sigma_{k}^{2}$ in Eq.~\ref{eqn:sigma.mach} scales 
$\propto \mathcal{M}(k)^{2}\propto v_{t}^{2} \sim k\,E(k)$, 
so $E_{\ln{\rho}}(k)\propto E(k)$. This is just the well-known expectation that 
in the weakly compressible regime, the log density power spectrum should have the 
same shape as the velocity power spectrum. \citet{kowal:2007.log.density.turb.spectra} 
and \citet{schmidt:2009.isothermal.turb} show that this is a good approximation for 
the $\ln{(\rho)}$ field in simulations of supersonic turbulence. At higher $\mathcal{M}$, this should 
flatten logarithmically, and this is seen in 
numerical simulations in \citet{kowal:2007.log.density.turb.spectra}, in excellent agreement 
with Eq.~\ref{eqn:sigma.mach}.
These behaviors, and the approximate normality of $\ln{(\rho)}$, 
appear to hold even in simulations which include explicitly non-local effects such as 
magnetic fields, self-gravity (excluding the collapsing regions), radiation pressure, 
photoionization, and non-isothermal gas with realistic heating/cooling 
\citep[see e.g.][]{ostriker:1999.density.pdf,
klessen:2000.pdf.supersonic.turb,
lemaster:2009.density.pdf.turb.review,hopkins:fb.ism.prop}.

Even if our analytic derivation is not exact, we can 
think of the resulting $\sigma^{2}(R)$ and implied log density power 
spectrum ($E_{\ln{\rho}}(k)\sim k^{-1}\sigma^{2}_{k}$) as a convenient approximation 
for the power spectrum measured in hydrodynamic simulations and observations. 
At sufficiently large $k$, where $\mathcal{M}$ is small, 
$E_{\ln{\rho}}(k)\propto k^{-1}\mathcal{M}^{2}\propto k^{-p}$; 
a steep falloff with $k$ for typical $p\approx2$; 
at smaller $k$ (but still smaller scales than the disk scale-height) $\mathcal{M}$ is large and 
this flattens to $E_{\ln{\rho}}(k)\propto k^{-1}\ln{\mathcal{M}^{2}} \propto k^{-1}$ with 
a small logarithmic correction. 
This is exactly the behavior directly measured in numerical simulations 
\citep{kowal:2007.log.density.turb.spectra,schmidt:2009.isothermal.turb}. 
Qualitatively similar behavior is seen in the linear density spectrum, but it 
is important to distinguish the two, since it is well known that 
large fluctuations at higher $\mathcal{M}$ 
will further flatten the linear spectrum 
\citep[see][]{scalo:1998.turb.density.pdf,
vazquez-semadeni:2001.nh.pdf.gmc,
kim:2005.density.pwrspec.turb,kritsuk:2007.isothermal.turb.stats,
bournaud:2010.grav.turbulence.lmc}.
It is also consistent with observations of the projected surface density power 
spectrum in local galaxies 
and star-forming regions \citep{stanimirovic:1999.smc.hi.pwrspectrum,
padoan:2006.perseus.turb.spectrum,
block:2010.lmc.vel.powerspectrum}.
If we integrate to get $\sigma(R)$ we obtain $\sigma\rightarrow$constant 
as $R\rightarrow0$, with an absolute value of 
$\sigma(R)\approx1.25 - 1.9$\,dex for a range $p=5/3-2$ and $\mathcal{M}_{h}=10-50$. 
This range is quite similar to the range measured in $\sigma(R)$ 
on the smallest resolved scales in a wide range of simulations that have a sufficiently 
large dynamic range in scales to probe the typical Mach numbers in GMCs and 
disk scale heights \citep[see][]{vazquez-semadeni:1994.turb.density.pdf,
nordlund:1999.density.pdf.supersonic,
ostriker:2001.gmc.column.dist,mac-low:2004.turb.sf.review,
slyz:2005.kpc.box.sf.feedback,hopkins:fb.ism.prop}. 
It also agrees well with measured 
values of the dispersion in the real ISM \citep{wong:2008.gmc.column.dist,
goodman:2009.gmc.column.dist,federrath:2010.obs.vs.sim.turb.compare}.

\begin{figure}
    \centering
    \plotone{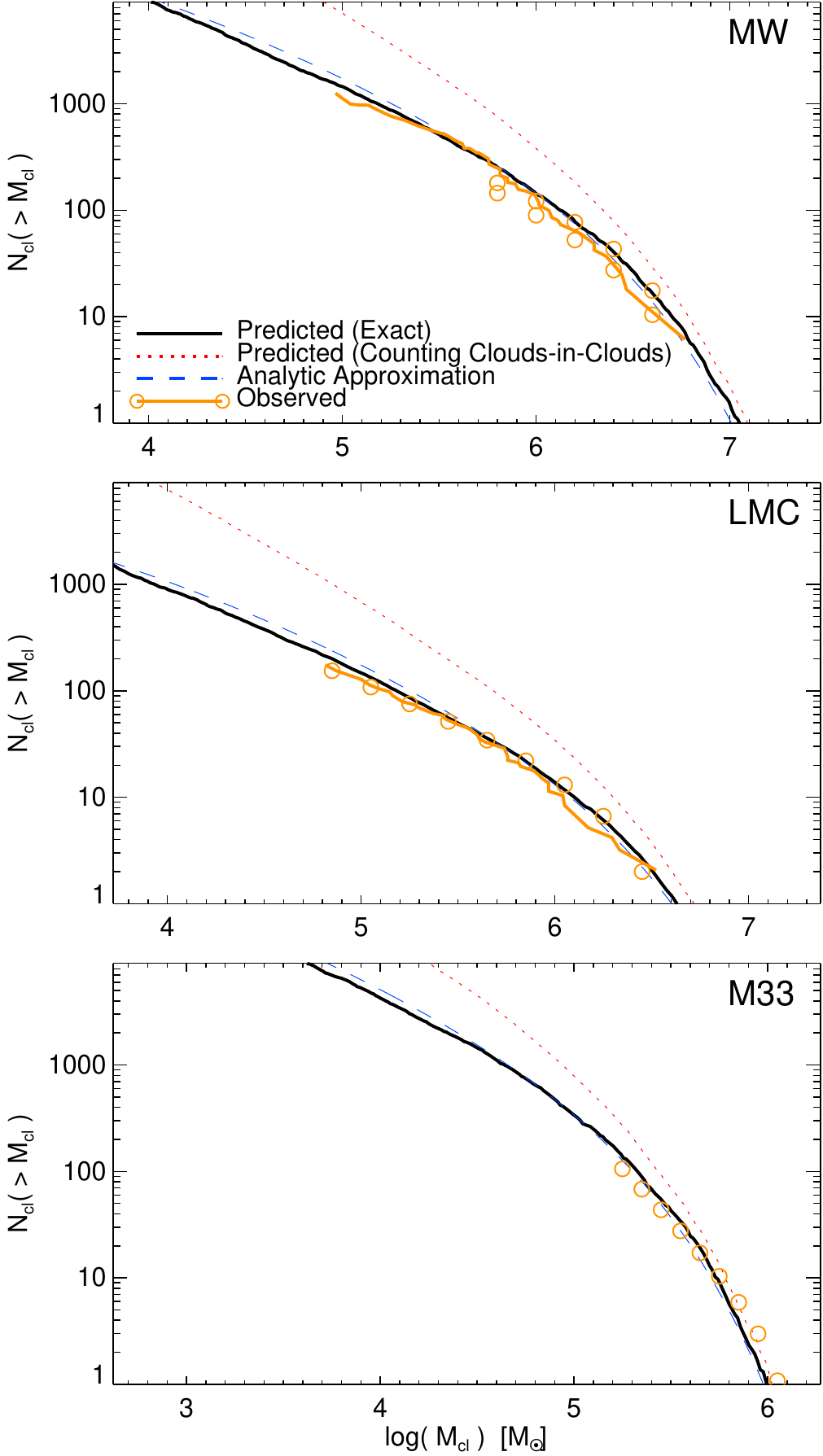}
    \caption{Predicted \&\ observed (orange) 
    GMC mass functions. The generally predicted mass 
    function is dimensionless; we normalize it to the observed 
    surface density $\Sigma_{\rm gas}$, gas density (or scale height) $n_{0}$, and total 
    gas mass $M_{\rm gas}$. Together with the assumption that $Q\sim1$, 
    this completely specifies the model. 
    For each case, we show the exact (Monte Carlo) mass function (solid black), and the 
    mass function if we ignore the ``cloud in cloud'' problem by counting 
    bound mass on all scales (dotted red); and the analytic fit to the mass function 
    in Eq.~\ref{eqn:mf.approx} (dashed blue).
    {\em Top:} Milky Way. The observed MFs are taken from 
    \citet{williams:1997.gmc.prop} (solid line) and 
    \citet{rosolowsky:gmc.mass.spectrum} (orange points in each panel), 
    \&\ model normalized to $(\Sigma_{\rm gas},\,n_{0},\,M_{\rm gas})=
    (10\,\msun\,{\rm pc^{-2}},\,1\,{\rm cm^{-3}},\,3\times10^{9}\,\msun)$.    
    {\em Middle:} LMC. Observed MF from \citet{fukui:2008.lmc.gmc.catalogue} 
    (line), normalized to $(8\,\msun\,{\rm pc^{-2}},\,0.8\,{\rm cm^{-3}},\,3\times10^{8}\,\msun)$ 
    \citep[see][]{wong:2009.lmc.co.detection}.
    {\em Bottom:} M33. Normalized to 
    $(5\,\msun\,{\rm pc^{-2}},\,1.5\,{\rm cm^{-3}},\,1\times10^{9}\,\msun)$ 
    \citep[see][]{engargiola:2003.m33.gmc.catalogue}.
    \label{fig:mf.pred}}
\end{figure}

\begin{figure}
    \centering
    \plotone{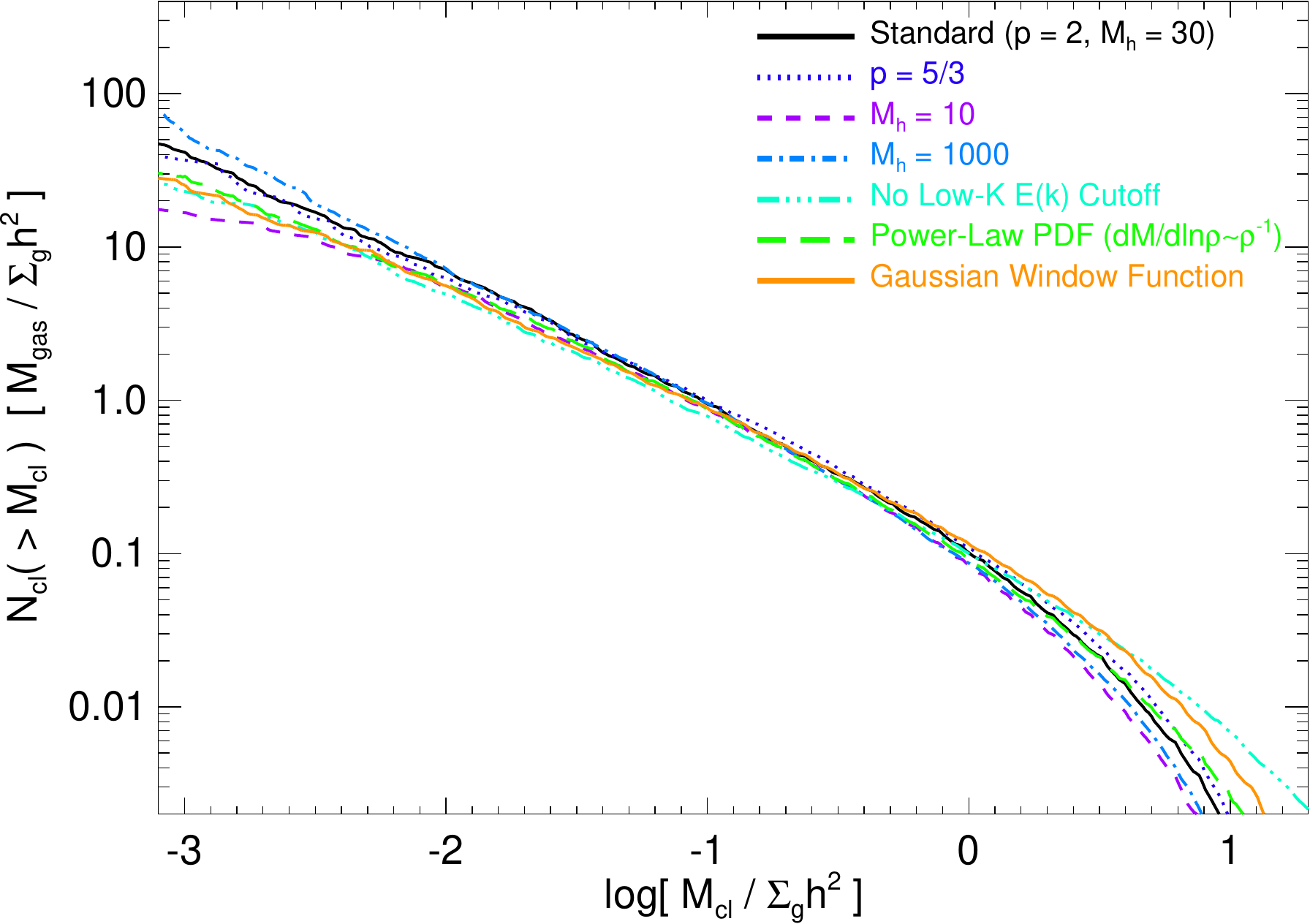}
    \caption{Variation in the predicted GMC 
    mass function with model assumptions. The MFs are plotted in dimensionless units. 
    We compare the standard model (from Fig.~\ref{fig:mf.pred}), which assumes 
    a turbulent spectral index $p=2$, and Mach number at scale $\sim h$ of $\mathcal{M}_{h}=30$. 
    Assuming $p=5/3$ instead slightly flattens the slope at intermediate masses. 
    Changing $\mathcal{M}_{h}=10,\,1000$ increases/decreases the sonic 
    length, below which the MF flattens, but near the MF break, the assumption of $Q\sim1$ 
    means that $\mathcal{M}_{h}$ factors out. 
    Removing the assumed cutoff in the turbulent power spectrum at scales $\gg h$ 
    makes the cutoff in the MF shallower at large masses. 
    Using a Gaussian window function to smooth the density field (instead of the usual $k$-space 
    tophat) makes the MF slightly more shallow, because for the same window volume (same 
    mass definition), the radii which contribute fluctuations are shifted.
    In all these models, the density PDF is assumed to be lognormal; if we instead assume it 
    is a pure power-law distribution (Eq.~\ref{eqn:pwrlaw.rho.pdf}), but assume the same 
    variance in $\ln{\rho}$, the result is nearly identical. 
    In all cases, the variations in the MF are very small -- the marginal stability assumption, and 
    weak (logarithmic) running of density variance with scale mean the MF shape is largely 
    independent of even substantial model assumptions. 
    \label{fig:mf.var}}
\end{figure}

\section{The Mass Function}
\label{sec:model:barrier:mf}

The question of the mass collapsed on different scales is now a well-posed barrier crossing 
problem. The quantity $\delta(R)$ -- the logarithm of the density smoothed on the 
scale $R$ -- is distributed as a Gaussian random 
field with variance $\sigma^{2}(R)$ and zero mean, with a well-defined barrier 
\be
\delta_{c}(R)\equiv 
\delta(\rho_{c},\,R) = \ln{{\Bigl(}\frac{\rho_{c}}{\rho_{0}} {\Bigr)}} - {\Bigl \langle}{\ln{{\Bigl(}\frac{\rho}{\rho_{0}} {\Bigr)}}} {\Bigr \rangle}
\ee 
which, upon crossing, leads to collapse. 
The mass of a collapsed object is simply the integral of the density over 
the effective volume of a window of effective radius $R_{w}$ 
in real space. If the medium were infinite and homogenous, 
this would just be $M(R_{w}) \equiv (4\pi/3)\,\rho_{c}(R_{w})\,R_{w}^{3}$; 
however, we need to account for the 
finite vertical thickness of the disk. For the same vertical exponential profile 
that gives rise to the dispersion relation in Equation~\ref{eqn:dispersion.relation}, 
the total mass inside $R_{w}$ is 
\be
\label{eqn:mass.size}
M(\rho\,|\,R_{w}) \equiv 4\,\pi\,\rho(R_{w})\,h^{3}\,
{\Bigl[}\frac{R_{w}^{2}}{2\,h^{2}} + {\Bigl(}1+\frac{R_{w}}{h}{\Bigr)}\,\exp{{\Bigl(}-\frac{R_{w}}{h}{\Bigr)}}-1 {\Bigr]}
\ee
where $\rho(R_{w})$ is the midplane density (chosen for consistency with the dispersion relation).
This formula simply interpolates between 
$M = (4\pi/3)\,\rho\,R^{3}$ for $R\lesssim h$ and 
$M = \pi\,(2\rho\,h)\,R^{2}=\pi\,\Sigma\,R^{2}$ for $R\gtrsim h$, as it should.

The fraction of the total mass which is in collapsed objects, averaged over a 
given smoothing scale $R_{w}$, is then just
\be
\label{eqn:fcoll.simple}
F_{\rm coll}(R_{w}) = \frac{1}{M_{\rm tot}}\int_{\delta_{c}}^{\infty}\, M(\rho\,|\,R_{w})\,p(\delta\,|\,R_{w})\,{\rm d}\,\delta
\ee
where $\rho(\delta) = \rho_{0}\,\exp{(\delta-\sigma[R_{w}]^{2}/2)}$.
Naively, we would equate this to the mass function of such 
objects with the relation $M_{\rm tot}\,{\rm d}\,F_{\rm coll}/{\rm d}\,M = M\,{\rm d}N(M)/{\rm d}M$. 
Indeed, up to a normalization factor, that is exactly the original approach of 
\citet{pressschechter}. However, this neglects the ``cloud in cloud'' problem: 
namely, it does not resolve whether or not a collapsed region on a scale $R_{1}$ is 
independent, or is simply a random sub-region of a larger object collapsed on a 
scale $R_{0}>R_{1}$. For the case of a constant $\delta_{c}$, accounting for this 
amounts to a simple re-normalization; but there is no simple closed-form 
analytic solution for the complicated $\delta_{c}$ here, and we will show 
that accounting for this behavior is critical.

\subsection{Exact Solution}
\label{sec:model:barrier:mf:exact}

To derive the {\em exact} mass function solution we turn to the standard 
Monte Carlo excursion-set approach. 
Consider the density field at some arbitrary location ${\bf x}$, 
smoothed over some window corresponding to the radius $R$ (and mass $M$) 
$\delta({\bf x}\,|\,R_{w})$. 
This is the convolution $\delta{({\bf x}\,|\,R_{w})} \equiv \int{{\rm d}^{3}x^{\prime}\,
W(|{\bf x}^{\prime}-{\bf x}|,\,R_{w})\,\delta{({\bf x}^{\prime})}}$; 
so if we Fourier transform, we obtain 
$\delta{({\bf k}\,|\,R_{w})} \equiv W({\bf k}\,|\,R_{w})\,\delta({\bf k})$. 
In other words, the amplitude $\delta{({\bf x}\,|\,R_{w})}$ is simply the 
integral of the contribution from all Fourier modes $\delta({\bf k})$, 
weighted by the Fourier-space window function. 

In this sense, we can think of the (statistical) evaluation of the density 
field as the results of a ``random walk'' through Fourier space. 
\citet{bond:1991.eps} show that this integration becomes particularly simple 
for the case of a Gaussian field with a Fourier-space tophat window, 
in which case the probability of a 
transition from 
$\delta_{1}$ to $\delta_{2}\equiv \delta_{1}+\Delta \delta$ 
as we step from a scale 
$k_{1}$ to $k_{2}$ is given by 
\begin{align}
p(\delta_{1}+\Delta\,\delta)\,{\rm d}\,\Delta\,\delta &= 
\frac{1}{\sqrt{2\pi\,\Delta\,S}}\,\exp{{\Bigl(}-\frac{(\Delta\,\delta)^{2}}{2\,\Delta\,S}{\Bigr)}}
{\rm d}(\Delta\,\delta) \\ 
\Delta\,S &\equiv S_{2}-S_{1} \equiv \sigma^{2}(R_{2})-\sigma^{2}(R_{1})
\end{align}
where we define the variance
\be
S(R) \equiv \sigma^{2}(R)
\ee
i.e.\ the increment $\Delta\,\delta$ is a Gaussian random variable 
with standard deviation $\sqrt{\Delta\,S}$.

If we begin on some sufficiently large initial scale $k\rightarrow0$ 
($R\rightarrow\infty$), then the overdensity $\delta$ and density variations $\sigma(R)$ 
must go to zero. We then have the well-defined initial conditions for the walk, 
$\delta(R_{\rm max}\rightarrow\infty)=0$, 
$S(R_{\rm max}\rightarrow\infty)=0$. Starting at some 
arbitrarily large $R_{\rm max}$, and moving to progressively smaller scales 
with increments\footnote{
The walks defined in this way will always converge as $\Delta R\rightarrow0$. 
In practice, the value of $\Delta R$ should be sufficiently small to ensure 
multiple barrier crossings are not missed -- i.e.\ 
so that the probability of crossing the barrier in a given step is 
small, $\Delta S \ll \delta_{c}(R)$.} 
in $R$ or $S$ ($\Delta\,R_{i}$ or $\Delta\,S_{i}$) 
we can then compute the {\em trajectory} 
$\delta(R)$ or $\delta(S)$, 
\be
\delta(R_{i}) \equiv \sum_{j}^{R_{j}>R_{i}} \Delta\,\delta_{j}\ .
\ee
At each scale $R_{i}$, we then evaluate whether or not the barrier has 
been crossed, 
\be
\delta(R_{i}) \ge \delta_{c}(R_{i}) \ .
\ee
If this is satisfied, we then associate 
that trajectory with a collapse on the scale $R_{i}$ 
and mass $M(\rho_{c}[R_{i}]\,|\,R_{i})\equiv M(R_{i})$. 

Recall, we are sampling the field $\delta({\bf x}\,|\,R_{w})$, 
so the fraction of trajectories that cross the barrier in 
some interval $\Delta R_{i}$ or (equivalently) $\Delta M(R_{i})$ 
represents the probability of an Eulerian {\em volume} element 
being collapsed on that scale. 
This corresponds to a differential {\em mass} 
$df_{\rm mass} = \rho(\delta\,|\,R_{w})\,df_{\rm vol} = \rho_{c}[R_{i}]\,df_{\rm Vol}$. 
Since the total mass associated with the mass function is 
$M_{\rm tot}\,{\rm d}N(M)/{\rm d}M$, we have the predicted mass function or 
``first-crossing distribution'': 
\be
\frac{{\rm d}n}{{\rm d}M} = 
\frac{\rho_{c}(M)}{M}\,\frac{{\rm d}f}{{\rm d}M}
\ee
where ${\rm d}f/{{\rm d}M}$ is the differential fraction of trajectories 
that cross $\delta_{c}$ between $M$ and $M+{\rm d}M$.

This formalism has several advantages. It provides an 
exact solution that also allows us to rigorously calculate the 
normalization and shape of the mass function. 
It also allows us to explicitly resolve the ``cloud-in-cloud'' problem, 
i.e.\ to address the situation where a trajectory crosses the 
barrier $\delta_{c}$ multiple times.
Figure~\ref{fig:mf.pred} plots the resulting 
mass function (for a few choices of 
parameters, which just determine the 
normalization of the mass function and will be discussed below). 
We also compare the mass function if we were to 
ignore the ``cloud in cloud'' problem -- i.e.\ where we treat 
{\em every} crossing above $\rho_{c}$ on a smoothing scale 
$R$ as a separate cloud. 
At the highest masses, the difference is 
small -- this is because the variance is small and 
$\delta_{c}$ is large, so the probability of being inside a ``yet larger'' 
cloud vanishes. However, at lower masses, 
the difference rapidly becomes quite large (order-of-magnitude) -- much larger than 
the factor $=2$ of the Press-Schechter mass function. 
This owes to the complicated behavior 
of $\delta_{c}$, which increases again on small scales. 
Failure to properly account for the cloud-in-cloud problem and moving barrier
will clearly lead to large inaccuracies.

\subsection{Key Behaviors}
\label{sec:model:barrier:mf:behavior}

If the barrier $\delta_{c}$ were constant, the mass function of collapsed objects would 
then simply follow the Press-Schechter formula; 
\be
\label{eqn:ps.mf}
\frac{{\rm d}n_{\rm PS}}{{\rm d}M} = 
\frac{\rho_{0}}{M^{2}}\,\sqrt{\frac{2}{\pi}}\,
\frac{\delta_{c}}{\sigma}\,
{\Bigl |}\frac{{\rm d}\ln{\sigma}}{{\rm d}\ln{M}} {\Bigr|}\,
\exp{{\Bigl(}-\frac{\nu^{2}}{2}{\Bigr)}}
\ee
where $\nu\equiv \delta_{c}/\sigma(M)$ is the collapse threshold in units of the 
standard deviation ($\sigma(M)$) of the smoothed density field on the 
scale $R$ corresponding to $M(R)$. 

However, the barrier here is {\em not} constant (it depends on $R$). 
A reasonable approximation to the first-crossing distribution, however, 
is given by 
\begin{align}
\label{eqn:mf.approx}
\frac{{\rm d}n}{{\rm d}M} &\approx \frac{\rho_{c}}{M^{2}}\,\sqrt{\frac{2}{\pi}}\,
\frac{\tilde{B}}{\sigma}\,
{\Bigl |}\frac{{\rm d}\ln{\sigma}}{{\rm d}\ln{M}} {\Bigr|}\,
\exp{{\Bigl(}-\frac{\nu^{2}}{2}{\Bigr)}} \\ 
\tilde{B} &\equiv 
   \left\{ \begin{array}{ll}
      \ln{(\rho_{c,\,{\rm min}}/\rho_{0})} & M < M(\rho_{c,\,{\rm min}}) \\
      \ln{(\rho_{c}/\rho_{0})} &  M \ge M(\rho_{c,\,{\rm min}})\ 
\end{array}
    \right.
\end{align}
where $\rho_{c,\,{\rm min}}\equiv {\rm MIN}(\rho_{c}[M])$ is 
the critical density at the most-unstable scale.
This is motivated by the exact solution for the first-crossing 
distribution for a linear barrier with 
$\delta_{c}=\delta_{1} + \sigma^{2}/2$, 
but with $\tilde{B}=\delta_{1}$ held constant below $M(\rho_{c})$.\footnote{The 
fitting function from \citet{sheth:2002.linear.barrier}:
\begin{align}
\frac{{\rm d}f}{{\rm d}M}\,{\rm d}M=
f(S)\,{\rm d}S &=|T(S)|\exp{[-\delta_{c}(S)^{2}/2S]}\,\frac{{\rm d}\ln{S}}{\sqrt{2\pi S}} \\ 
T(S) &= \sum_{n=0}^{5}\,\frac{(-S)^{n}}{n!}\,\frac{\partial^{n}\delta_{c}(S)}{\partial S^{n}}
\end{align}
gives a similar answer, but it is less straightforward to interpret.
An approximate solution for the case {\em neglecting} the cloud-in-cloud 
problem is given by 
\be
\label{eqn:mf.neglect.cic}
\frac{{\rm d}n}{{\rm d}M} \approx \frac{\rho_{c}}{M^{2}}\,
\frac{3}{\sqrt{2\pi}\,\sigma}\,
{\Bigl [}
{\Bigl |}\frac{{\rm d}\ln{\rho_{c}}}{{\rm d}\ln{M}}{\Bigr |}
+\nu\,{\Bigl |}\frac{{\rm d}{\sigma}}{{\rm d}\ln{M}}{\Bigr |}
{\Bigr ]}
\exp{{\Bigl(}-\frac{\nu^{2}}{2}{\Bigr)}}
\ee
which can be derived (up to a normalization) 
from differentiating Equation~\ref{eqn:fcoll.simple}.
}
Because the deviation from a constant barrier is only 
logarithmic, these formulae do not differ too severely, 
and we can gain considerable insight from their functional forms. 

Consider the behavior of both $\delta_{c}$ and $\nu$, which define 
three primary regimes. 
On scales above the sonic length but below $\sim h$, most of the dynamic 
range for GMCs, 
$\mathcal{M}^{2}\propto v_{t}^{2} \propto R^{p-1}$ (for power-law 
turbulent cascades) is large, so
$\sigma$ is a very weak function of $R$ (most of the contribution 
comes from the largest scales, since $p-1>0$)
while $\rho_{c}$ decreases with $R$ $\propto R^{p-3}$ 
so $\delta_{c} = \ln{\rho_{c}/\rho_{0}} \rightarrow -(3-p)\,\ln{R}$. 
Therefore $\nu \propto \delta_{c} \propto -(3-p)\,\ln{R}\propto -((3-p)/p)\,\ln{M}$ is a 
(logarithmically) decreasing function of mass.
So we expect an approximately power-law mass function 
${\rm d}n/{\rm d}M\propto M^{\alpha}$ with slope 
$\alpha\sim-2$. 
This implies similar mass per logarithmic interval in mass 
and simply follows from gravity -- which is self-similar -- 
being the dominant force (since the turbulence is super-sonic). 
To the extent that the slope deviates from $-2$, it is because the barrier 
$\nu$ gets larger towards lower $M$. 
From the above equation, 
$M^{-2}\,\exp{(-\nu^{2}/2)} \propto M^{-2}\,\exp{[-(3-p)^{2}\,[\ln{(M/M_{0})}]^{2}\,/2p^{2}\sigma^{2}]} 
\propto M^{\alpha}$ with 
\begin{align}
\label{eqn:faintslope}
\alpha &\approx -2 + \ln{(M_{0}/M)}\,(3-p)^{2}/2\,p^{2}\,\sigma^{2}\\
&\approx -2 + 0.1\,\log{(M_{0}/M)}
\end{align}
(where $M_{0}$ is approximately 
the location of the mass function ``break''; formally $(4\pi/3)\rho_{0}\,h^{3}\approx10^{6}\,\msun$ for MW-like systems). In other words, we expect a slope 
$\alpha$ which is shallower than $-2$ by a small logarithmic correction, $\alpha\sim1.7-1.9$, 
as observed.

At very small scales we approach the sonic length, 
$\mathcal{M}\rightarrow 1$; 
the growth in $\sigma(R)$ becomes vanishingly small 
($\sim \sqrt{3}\mathcal{M}/4 \propto R^{(p-1)/2}$)
while $\rho_{c}$ continues to increase 
logarithmically as before. The mass function must 
therefore flatten or turn over, with a rapidly decreasing mass in  
clouds below the sonic length (although the absolute {number} may still rise weakly).

At large scales above $\sim h$, 
$\sigma(R)$ decreases rapidly with increasing $R$ -- 
the contribution from large scales goes as $\sim \sqrt{\ln{(1+(3/4)\,v_{t}^{2}/\kappa\,^{2}\,R^{2})}} 
\propto R^{-4+p}$ as $R\rightarrow\infty$, 
while now $\rho_{c}$ also increases $\propto R$ (so $\delta_{c}\propto \ln{R}$), 
so the mass function 
is exponentially cut off as $\propto \exp{(-c\,M^{1-p/4})}$.
We caution that at the largest size/mass scales, global gradients in galaxy properties -- which are currently neglected in our derivation of the collapse criterion -- may become significant. However, the number of clouds in this limit is small.

\subsection{Comparison with Observations}
\label{sec:model:mf.obs}

Figure~\ref{fig:mf.pred} plots the predicted mass function: 
we show the exact solution, both excluding and including 
``clouds in clouds,'' and the approximations 
in Equation~\ref{eqn:mf.approx} \&\ \ref{eqn:mf.neglect.cic}. 
For our ``standard'' model, we will assume the disk 
is marginally stable ($Q_{0}(h)=1$), and that the 
turbulence, being supersonic and rapidly cooling, 
should have $p\approx 2$ (see the discussion in \S~\ref{sec:intro}). 
Motivated by observations, we normalize the turbulent 
spectrum by assuming a Mach number on large scales 
$\mathcal{M}_{h}\approx30$ (though we will show this exact choice 
has very weak effects, provided $\mathcal{M}_{h}\gg 1$). 
With these choices, the model is completely fixed in dimensionless 
terms. To predict an {\em absolute} number and 
mass scale of the mass function, we require some normalization 
for the galaxy properties: some measure of the 
local gas properties (mean density, velocity dispersion, surface density, etc, 
to set the mass and 
spatial scales) and total galaxy mass or size (to know the gas 
mass available). 
Because of our assumption of marginal stability, many of 
these properties are implicitly related -- we need only specify 
e.g.\ a total disk mass, gas fraction, and spatial size. 
Or equivalently, a mean density, velocity dispersion, 
and total mass.

Taking typical observed values for the total gas mass, 
mean density, and velocity dispersion in the Milky Way, 
we plot the resulting predicted GMC mass function and 
compare to that observed. 
Because we are considering the total gas mass 
of the inner MW, we need to compare with a GMC mass 
function corrected to the same effective volume -- we therefore 
compare with the values in \citet{williams:1997.gmc.prop} (who attempt 
to construct a ``galaxy-wide'' GMC mass function for the same 
total volume). 
We then repeat the experiment with the average properties 
observed in the LMC and M33, and compare with the 
mass function compilations in \citet{rosolowsky:gmc.mass.spectrum,
fukui:2008.lmc.gmc.catalogue}, corrected to the appropriate survey area.

In each case, the predicted mass functions agree 
remarkably well with the observations. 
We emphasize that although the observed densities 
and masses enter into  
the normalization of the mass function, 
the {\em shape}, which agrees extremely well, is entirely an 
a priori prediction. Moreover, the assumed densities 
do not entirely determine the normalization -- because 
the barrier and variance are finite at all radii, 
the models here specifically predict that not all mass 
is in bound units. In fact, only $\sim20-30\%$ of the 
total mass is predicted to be in such units -- 
for the MW, the total bound GMC mass is predicted to be 
$\approx 10^{9}\,\msun$, in good 
agreement with that observed \citep{williams:1997.gmc.prop}. 
Likewise, the details of our stability and collapse conditions determine where, 
relative to the Jeans mass, the ``break'' in the mass function occurs.\footnote{The predicted high-mass cutoff in the GMC MF is steep, but there is some suggestion that the GMC MF terminates or truncates more sharply at the maximum cloud mass in some systems \citep[e.g.\ the MW; see][]{williams:1997.gmc.prop}. As noted above, including the corrections from global gradients in galactic properties in our collapse condition may steepen the predicted cutoff. However, since the distinctions appear over a narrow range in mass (factor $<2$) where the expected number of clouds in is in the Poisson regime (and consistent with zero within $2\,\sigma$), it is difficult to discriminate between different models.}

We should caution that it is not entirely obvious that our predicted mass function is the same as that observed. The mass function here is well-defined because we restrict to self-gravitating objects and resolve the cloud-in-cloud problem, knowing the three-dimensional field behavior (and assuming spherical collapse). In the observations,  the methods used to distinguish substructures and the choice of how to average densities (in spherical or arbitrarily shaped apertures) can make non-trivial differences to the mass function \citep{pineda:2009.clumpfind.issues}. This may be considerably improved by the use of more sophisticated observational techniques that attempt to statistically identify only self-gravitating structures \citep[see e.g.][]{rosolowsky:2008.dendrograms}; preliminary comparison of these methods in hydrodynamic simulations and observations suggests that most of the identified GMCs are indeed self-gravitating structures so the key characteristics of the GMC MFs in our comparison should be robust, although details of individual clouds may change significantly \citep{goodman:2009.dendrogram.sims}.

\subsection{Effects of Varying Assumptions}
\label{sec:model:barrier:mf:vary.assumptions}

Of course, it is important to check how 
sensitive the predicted mass functions are to 
the assumptions in our model. 
Figure~\ref{fig:mf.var} 
shows the results of varying these assumptions. 
We plot the mass function in dimensionless units 
($\rho_{0}=h=1$, with the absolute mass being 
an arbitrary normalization). 

If we assume Kolmogorov turbulence ($p=5/3$ instead of $p=2$), 
the predicted mass function is nearly identical 
at intermediate and high masses, but flattens more 
rapidly at low masses, because the velocity 
drops more slowly at small scales so $\rho_{c}\propto R^{p-3}$ rises 
more steeply. The difference agrees well with the scaling in Equation~\ref{eqn:faintslope}, 
which predicts a faint end slope  
$\alpha\approx -2+0.3\,\log{(M_{0}/M)}$ for $p=5/3$ 
instead of $\alpha\approx -2+0.1\,\log{(M_{0}/M)}$ 
for $p=2$.

If we vary the Mach number on large scales $\mathcal{M}_{h}$ (or, equivalently, 
the assumed sound speed or magnetic field strength), 
the differences are very small at 
almost all masses, because the assumption that the disk as a whole 
is marginally stable effectively scales out the absolute value of $\mathcal{M}_{h}$.
What $\mathcal{M}_{h}$ does determine is 
the (dimensionless) scale of the sonic length ($R_{\rm sonic}
\sim h\,\mathcal{M}_{h}^{-2/(p-1)}$), 
below which the mass function will flatten. With 
lower $\mathcal{M}_{h}=10$, this happens at higher 
masses -- but still quite low in absolute terms ($R\sim0.01\,h$, or 
$\gtrsim3$\,dex below the maximum GMC masses). 

As noted above, the exact manner in which the velocity power spectrum $E(k)$ 
should flatten at large scales $kh\lesssim1$ is uncertain. 
We therefore re-calculate the mass function ignoring such flattening 
entirely -- i.e.\ assuming $E(k)\propto k^{-p}$ for all $k$. This 
makes the very high-mass end of the mass function slightly more shallow, 
but has a negligible effect at all other masses. 
Since the only difference will be in the regime where the number of 
clouds is $\sim1$ (so subject to large Poisson fluctuations) it is difficult to constrain 
this from observations.

Re-calculating our results with a different window function makes 
little difference. We test this with a Gaussian window function (convenient 
as it remains a Gaussian in real and Fourier space). As discussed in 
\citet{zentner:eps.methodology.review}, this makes the calculation more 
complex because we can no longer treat the Fourier-space trajectory 
as having uncorrelated steps; following \citet{bond:1991.eps} the 
first-crossing distribution is computed by  numerically 
integrating a Langevin equation. However, we hold our mass 
definition fixed; with this choice, for fixed $R_{w}$, 
the exact choice of window shape about $k_{w}\sim1/R_{w}$ 
introduces only small ($\sim10\%$) corrections 
(we refer to the discussion therein 
and \citealt{maggiore:2010.path.integral.non.tophat.eps.filter} for more detailed discussion of 
the effects of different window functions).

What if the density distribution is not a lognormal? 
It has been suggested, for example, that 
for systems which have significant gas pressure 
and whose equations of state are non-isothermal, or 
which have large magnetic fields, the density distribution 
may more closely resemble a power-law \citep[see e.g.][]{passot:1998.density.pdf,
scalo:1998.turb.density.pdf,
ballesteros-paredes:2011.dens.pdf.vs.selfgrav}. 
This is certainly still treatable with the excursion set formalism: 
there has been considerable discussion in the literature 
regarding the halo mass function and bias with non-Gaussian 
primordial fluctuations \citep[see][and references
therein]{matarrese:2000.nongaussian.numdens,
afshordi:2008.nongaussian.collapsestatistics,
maggiore:2010.nongaussian.eps}. However most of these 
rigorous approaches assume the non-Gaussianity is 
small and can be treated in perturbation theory. 
For large deviations from Gaussianity it is not trivial to construct a 
fully self-consistent theory. For example, if $P(\rho\,|\,R_{w})$ 
were locally power-law at each ``step'' in $k$-space 
in a random walk, the resulting $P(\rho)$ evaluated 
on each scale would no longer be a power-law; some violation 
of locality would be required so the distribution could 
``self-correct.'' In any case, if we simply {\em assume} some 
pre-specified $P(\rho\,|\,R_{w})$ at all scales, 
it is still straightforward to evaluate the first-crossing distribution. 
The distribution $f(S)$ in $({\rm d}f/{\rm d}M)\,{\rm d}M = f(S)\,{\rm d}S$ 
is given by the solution to the integro-differential equation: 
\begin{align}
f(S) &= -P^{\prime}(\delta_{c}\,|\,S)\,\frac{{\rm d}\delta_{c}(S)}{{\rm d}S} - 
\int_{-\infty}^{\delta_{c}(S)}\,\frac{\partial P^{\prime}(\delta\,|\,S)}{\partial S}\,{\rm d}\delta\\
P^{\prime}(\delta\,|\,S) &\equiv 
P(\delta\,|\,S) - \int_{0}^{S}\,{\rm d}S^{\prime}\,f(S^{\prime})\,P(\delta-\delta_{c}[S^{\prime}]\,|\,
S-S^{\prime})
\end{align}
where ${\rm d}p(\delta)=P(\delta)\,{\rm d}\delta$.
This is essentially just the collapsed mass given by $P(\delta>\delta_{c}\,|\,S)$, 
corrected by the probability that the collapse occurred on a larger scale (smaller $S$), 
and can be solved numerically for any specified $P$.

Consider the following form for the density PDF: 
\be
\label{eqn:pwrlaw.rho.pdf}
\frac{{\rm d}p(\rho)}{{\rm d}\ln{\rho}} \propto \exp{{(} - \gamma\,|\ln{[\rho/\bar{\rho}]}|{)}} = 
   \left\{ \begin{array}{ll}
      (\rho/\bar{\rho})^{\gamma}  & \rho < \bar{\rho} \\
      (\rho/\bar{\rho})^{-\gamma} & \rho \ge \bar{\rho}\ 
\end{array}
    \right.
\ee
where $\bar{\rho}=(1-\gamma^{-2})\,\rho_{0}$. 
The exact functional 
form is arbitrary, of course, but convenient because it is a pure power-law 
symmetric in $\ln{\rho}$, and has a well-defined variance: 
$\langle (\ln{\rho})^{2} \rangle = 2\gamma^{-2} + (\ln{[1-\gamma^{-2}]})^{2}$
We can therefore map this one-to-one to our assumed density 
power spectrum by assuming $\gamma=\gamma(R)$, with  
$2\gamma^{-2} + (\ln{[1-\gamma^{-2}]})^{2} = \sigma^{2}(R) = S$. Note 
that this gives $\gamma\sim1$ over much of the dynamic range of interest, 
quite similar to the best-fit distributions in the references above.
At low and high masses, the predicted mass function is slightly more 
shallow than our standard model. At high masses this is because of the 
more extended power-law tail to high $\delta$; at low masses this 
is both an effect of 
more first-crossings at larger scales 
and a result of some of the mass being moved from the ``core'' of the distribution 
to those tails. 
However, the differences are quite small. This is because a lognormal (unlike 
a pure normal distribution) is very similar to a single power law over a wide dynamic 
range. Moreover, the collapsed mass fraction is not extremely small, so it is not 
sampling some extreme tail of the distribution. 
So, for the same variance $S$, deviations from lognormal behavior have 
only small effects.

\subsection{The Core Mass Function}
\label{sec:model:mf.cores}

In this paper, we choose to focus on the mass function of GMCs and other large-scale structures in the ISM. Part of the reason for this is that we can focus on the first-crossing distribution (the largest scales on which structures are self-gravitating) and so have a well-defined mass function. Although there are certain similarities, this is not the same as the mass function of self-gravitating dense cores within GMCs, as calculated using qualitatively similar arguments in e.g.\ \citet{padoan:2002.density.pdf} and \citet{hennebelle:2008.imf.presschechter}. 

In principle, our model can be extended iteratively to smaller scales to investigate the mass function of cores and make a direct comparison with these previous predictions as well as observations, and in a companion paper \citep{hopkins:excursion.imf} we attempt to do so. This is not trivial, however. The difficulty is that, because cores are substructures, the mass function definition (the resolution to the ``cloud in cloud'' problem) is somewhat ambiguous: we cannot simply isolate first-crossing. Even in simulations where the full three-dimensional properties are known, it is not trivial to find a unique mass function of such substructure in a turbulent medium \citep[see e.g.][]{ballesteros-paredes:2006.imf.turb.sims,anathpindika:2011.shell.frag}. The approach of \citet{hennebelle:2008.imf.presschechter} is to treat this ambiguity as an effective normalization term (and to truncate the problem at larger scales -- treating the properties of the ``parent'' GMC as assumed/given and restricting to much smaller spatial scales); as such their derivation is similar to the original \citet{pressschechter} derivation as discussed in \S~\ref{sec:intro}. That in \citealt{padoan:2002.density.pdf} more simply makes some general scaling arguments. But as we show in Fig.~\ref{fig:mf.pred}, this is not necessarily a good approximation. We therefore require some more detailed criteria to inform our definition of cores, for example some estimate of the scales on which fragmentation below the core scale will not occur (defining the ``last-crossing,'' as opposed to ``first-crossing'' distribution). This is a topic of considerable interest, but is outside the scope of our comparisons here.

\begin{figure}
    \centering
    \plotone{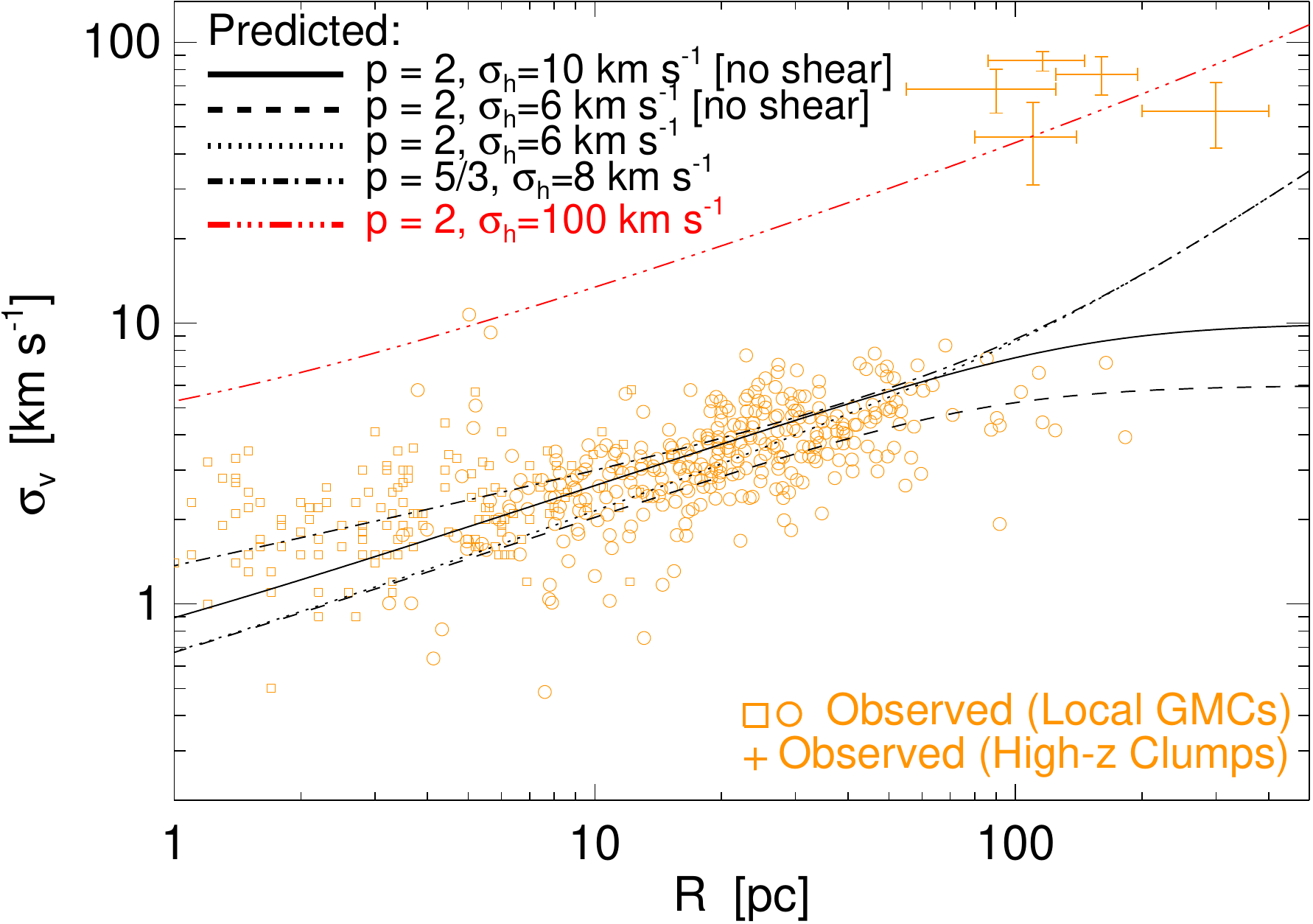}
    \caption{Predicted GMC linewidth-size relation. 
    Different lines correspond to different model assumptions: specifically 
    we vary the turbulent spectral index ($p$), the absolute normalization 
    of the system (amounting to the velocity dispersion $\sigma_{h}^{2}$ at scale $=h$), 
    and whether or not we include disk shear in the ``velocity'' $\sigma_{v}$. 
    Note that in the models here, $\sigma_{h}$ is not freely varied, it is 
    {\em predicted} from the global parameters of the system via our marginal 
    stability assumption. The velocity $\sigma_{v}$ is the one-dimensional 
    linewidth (using $\sigma^{2}=c_{s}^{2}+v_{t}^{2}$) for each cloud at the time 
    of collapse, $R$ is the three-dimensional collapse radius.
    On scales below $\sim h$, the Monte Carlo results are approximately a 
    power-law with slope $\sigma_{v}\propto R^{0.5}$ (Eq.~\ref{eqn:linewidthsize}).
    We compare observations of clouds in the MW and local galaxies, 
    compiled in \citet[][circles]{bolatto:2008.gmc.properties} and 
    \citet[][squares]{heyer:2009.gmc.trends.w.surface.density}, appropriately corrected to the 
    same quantities. The agreement is good -- even for $p=5/3$, for which 
    large-scale effects make the relation slightly steeper than the naive 
    expectation $\sigma_{v}\propto R^{(p-1)/2}$; moreover the marginal 
    stability assumption predicts the normalization accurately. We also compare individual high-redshift 
    molecular ``clumps'' in extremely gas-rich, rapidly star-forming 
    lensed galaxies in \citet[][crosses with error bars]{swinbank:clumps}, 
    which form in much more dense disks (much larger $\Sigma_{\rm disk}$); these lie well above the extrapolation of the relation for MW-like properties. However, if we compare the predictions for a model with the observed $\sigma_{h}\approx100\,{\rm km\,s^{-1}}$ of their host disks, the agreement is good. Clouds in the MW center, which has intermediate $\Sigma_{\rm disk}$ between these extremes, lie correspondingly between these curves \citep[see][]{oka:2001.mw.center.gmcs}.
    \label{fig:linewidth.size}}
\end{figure}

\begin{figure}
    \centering
    \plotone{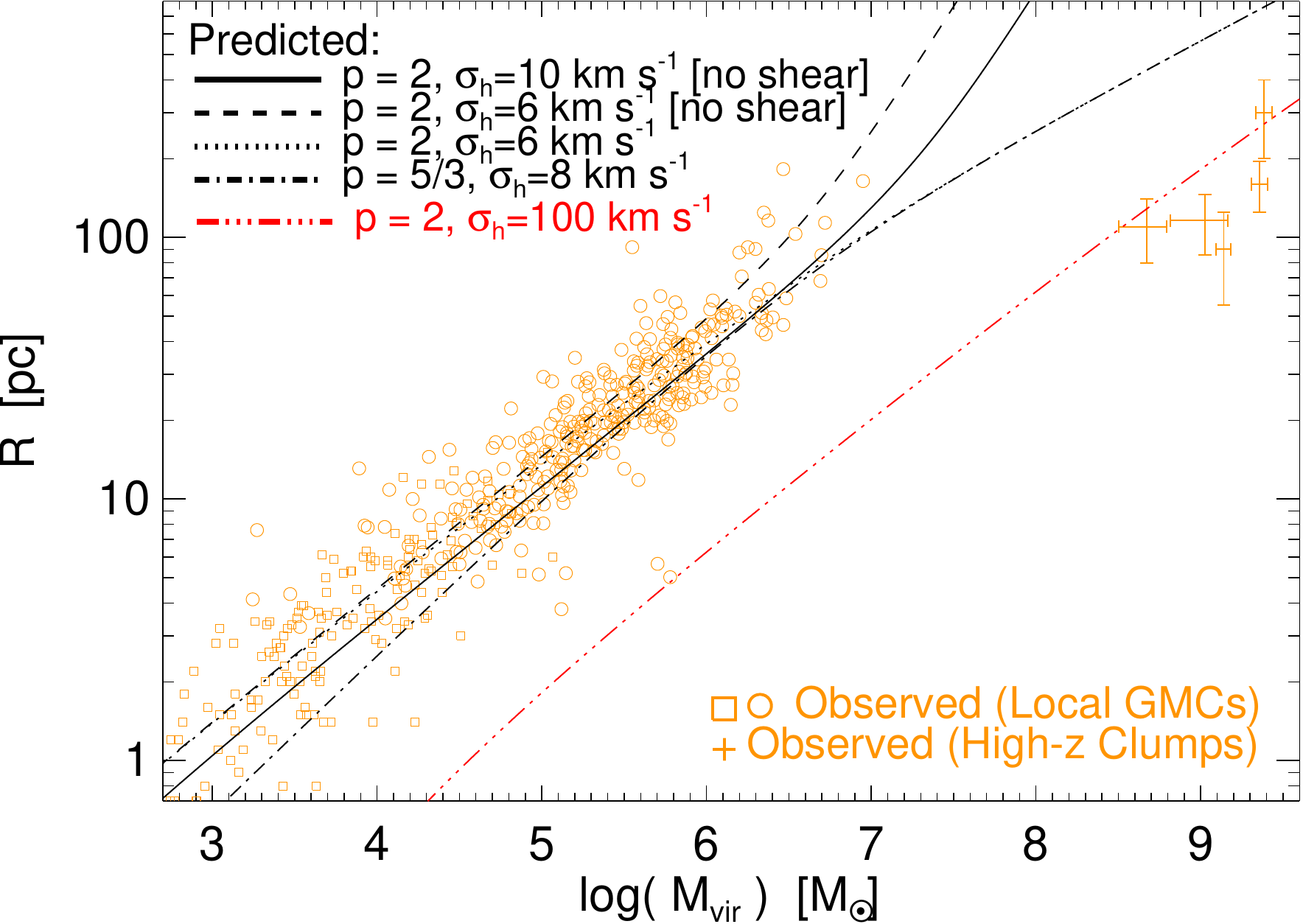}
    \caption{Size-mass relation of clouds, in the same style as 
    Figure~\ref{fig:linewidth.size}. 
    The observations from \citet{bolatto:2008.gmc.properties} use the virial mass estimator  $M_{\rm vir}\equiv 5\,\sigma_{v}^{2}\,R/G$; those from \citet{heyer:2009.gmc.trends.w.surface.density} and \citet{swinbank:clumps} are derived from the CO luminosity.
    The agreement with observations is good, 
    and the scaling is an approximate power-law 
    with slope $R\propto M^{0.5}$
    (approximately constant $\Sigma_{\rm cloud}\sim100\,\msun\,{\rm pc^{-2}}$; 
    Eq.~\ref{eqn:sizemass}). Again, the high-redshift clumps lie off the ``typical'' local galaxy relations; however a model of a more dense disk with $\sigma_{h}\approx100$ agrees well. Similarly, MW-center clumps lie between the extremes shown \citep{oka:2001.mw.center.gmcs}. 
    \label{fig:size.mass}}
\end{figure}

\begin{figure}
    \centering
    \plotone{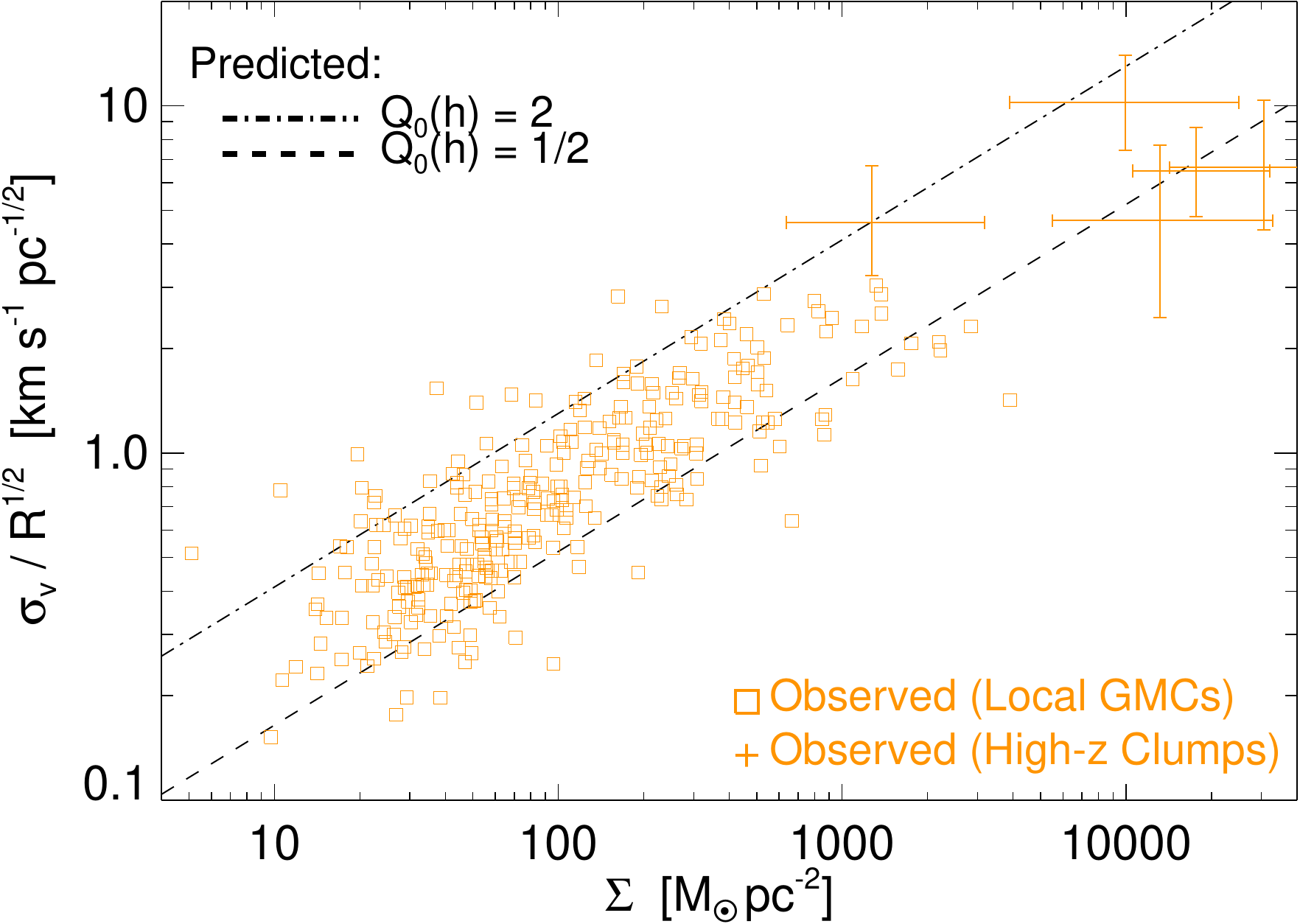}
    \caption{Residuals from the linewidth-size relation for {\em bound} clouds as a function of 
    disk/region surface density $\Sigma$, in the style of Figure~\ref{fig:linewidth.size}.
    Because we define 
    clouds as self-gravitating, the different predicted lines (from different 
    turbulent spectra) in Figs.~\ref{fig:linewidth.size}-\ref{fig:size.mass} 
    collapse to a single line in this plot. So we instead plot the predicted lines for an 
    assumed global stability parameter $Q_{0}\approx0.5-2.0$. Unbound clouds/overdensities 
    will have higher $\sigma_{v}$, but are not the collapsed objects followed here.
    \label{fig:size.mass.offsets}}
\end{figure}

\section{Size-Mass \&\ Linewidth-Size Relations}
\label{sec:model:barrier:larsons}

We can also use our model to predict 
the scaling laws obeyed by GMCs ``at collapse.''

The linewidth-size relation follows trivially 
from our assumed turbulent power spectrum. 
The exact $\sigma_{v}(R)$
relation is plotted in Figure~\ref{fig:linewidth.size} for power-law turbulent 
slopes of $p=5/3$ and $p=2$, with the normalization set by 
requiring a marginally stable disk with MW-like surface density. 
We can define the line width either as just the turbulent width, or 
the turbulent width plus the contribution 
from disk shear $\sigma_{v}^{2}(R)=v_{t}^{2} + \kappa^{2}\,R^{2}$; the distinction 
is unimportant for typical observed scales, but shear 
is predicted to contribute significantly to the velocities of the 
largest GMCs when $R\gtrsim h$.
We compare with observations compiled from the 
MW and other local group galaxies from \citet{bolatto:2008.gmc.properties} and \citet{heyer:2009.gmc.trends.w.surface.density}.\footnote{Because \citet{heyer:2009.gmc.trends.w.surface.density} caution that more detailed studies in nearby clouds \citep[e.g.][]{goldsmith:2008.taurus.gmc.mapping} suggest their LTE masses may be low by a factor $\sim2-3$ at intermediate column densities, we plot the results for the ``high density'' cuts in cloud area defined therein (the ``A2'' sample) within which the LTE approximation should be valid.}

In the regime above the sonic length and 
below the scale height, this is just a simple 
power law with 
$\sigma_{v}(R) \propto R^{(p-1)/2}$, 
i.e.\ $\approx 0.5$ for $p=2$ or $\approx 0.33$ for $p=5/3$. 
This is essentially an assumption of our model (although 
it follows from basic turbulent conditions); 
more interesting is that the normalization can 
be predicted from the assumption of marginal stability ($Q\approx1$), 
giving 
\be
\label{eqn:linewidthsize}
\sigma_{v}(R) \approx 0.4\,{\rm km\,s^{-1}}\,
{\Bigl (} \frac{\langle \Sigma_{\rm disk} \rangle}{10\,\msun\,{\rm pc}^{-2}} {\Bigr)}^{0.5}
{\Bigl (} \frac{R}{{\rm pc}} {\Bigr)}^{0.5}
\ee
This agrees well with the observed relation. 
In the full solution, because of the change in dimensionality above the 
scale $h$, the relationship flattens if we 
consider only turbulent velocities; it becomes steeper, however, 
with the inclusion of the shear term. 

This model also specifically predicts a residual dependence in the normalization of the 
linewidth-size relation that scales as $\langle \Sigma_{\rm disk} \rangle^{1/2}$, 
where $\langle \Sigma_{\rm disk} \rangle$ is the 
large-scale mean disk surface density. We stress that this is {\em not} necessarily the same as a dependence 
on the local cloud $\Sigma_{\rm cloud}$ (over a wide dynamic range, in fact, $\Sigma_{\rm disk}$, hence $\Sigma_{\rm cloud}$, is similar). This is also, by definition, for bound objects, not for un-bound overdensities on small scales. The predicted dependence is shown indirectly in Figure~\ref{fig:linewidth.size}, and directly in Figure~\ref{fig:size.mass.offsets}, 
where we compare with the observations compiled in 
\citet{heyer:2009.gmc.trends.w.surface.density} in local 
galaxies and in \citet{swinbank:clumps} for massive star-forming molecular complexes 
in lensed, high-redshift galaxies. These sample extremely different 
environments, and are indeed offset in the linewidth-size relation. 
However, the magnitude of their offsets is in good agreement with 
that predicted here.\footnote{If the predicted clouds perfectly followed $M\propto R^{2}$ and $\sigma\propto R^{1/2}$, they would collapse to a single point in this Figure. They do not, because of the changes below the sonic length and above $\sim h$. However, because the clouds are defined as self-gravitating, the models collapse to a line (with most of the clouds concentrated near the ``typical'' point for intermediate scales.} 
The galaxies in \citet{swinbank:clumps} have an average surface density of $\sim10^{3}\,\msun\,{\rm pc^{-2}}$, and a correspondingly very large measured $\sigma_{h}\approx100\,{\rm km\,s^{-1}}$ (as expected for $Q_{0}(h)\approx1$); normalizing the predicted linewidth-size relation for these properties, we expect an order of magnitude larger $\sigma_{v}$ at fixed size. 
Clouds observed in the MW center \citep{oka:2001.mw.center.gmcs}, which has a higher mean surface density than the local neighborhood but generally lower than estimated for the high-redshift systems, lie neatly between the predicted curves for the local and high-redshift cases (a mean offset of $\sim3-5$ relative to the local clouds, corresponding to a factor of $\sim10-30$ higher $\Sigma_{\rm disk}$, about what is expected for the observed exponential profile). Similar offsets  are known in other local galaxies with high surface densities, such as mergers and starburst galaxies \citep{wilson:2003.supergiant.clumps,rosolowsky:2005.gmcs.m64}.

As discussed in \citet{hopkins:fb.ism.prop}, 
a dependence of exactly this sort is seen in high-resolution hydrodynamic 
simulations as well. 
In the observations, this normalization dependence has 
sometimes been interpreted as a consequence of magnetic support or confining external pressure
\citep[see the discussion in][]{blitz:h2.pressure.corr,bolatto:2008.gmc.properties,
heyer:2009.gmc.trends.w.surface.density}, 
but in this context magnetic fields and pressure confinement are not explicitly present -- 
such a scaling is a much more broad consequence of the simple 
Jeans requirements 
for collapse in {\em any} marginally stable environment. 

The size-mass relation follows from 
the critical density $\rho_{c}$ derived in 
\S~\ref{sec:model:collapse}, by 
simply inverting Eqn~\ref{eqn:mass.size}. 
We plot the exact prediction in Figure~\ref{fig:size.mass}.
In the regime above the sonic length but 
below the disk scale-height, recall that 
a power-law turbulent cascade gives the 
simple condition 
$\rho_{c} = k^{2}\,v_{t}(k)^{2}/(4\pi G)
\propto R^{p-3}$, so 
$R\propto M^{1/p}$, i.e.\ 
$R\propto M^{1/2}$ for $p\approx 2$, 
very similar to the observed 
power-law scaling. 
The normalization also follows -- for MW-like global conditions
\be
\label{eqn:sizemass}
R_{\rm cloud}(R\gg{\ell}_{\rm sonic},\,R\ll h)
\approx 1.4\,\sigma_{0.4}^{-1}\,{\rm pc}\,
{\Bigl(}\frac{M_{\rm cloud}}{300\,M_{\rm sun}}{\Bigr)}^{0.5}
\ee
where $\sigma_{0.4}$ is the normalization of the 
turbulent velocities
$v_{t}= \sigma_{0.4}\times{0.4\,{\rm km\,s^{-1}}}\,(R/{\rm pc})^{1/2}$. 
This corresponds to an approximately constant cloud surface density in 
agreement with Larson's laws: 
in projection $\Sigma_{\rm cloud}\approx100\,\msun\,{\rm pc^{-2}}$ at the 
time of collapse.
Note that re-calculating this for $p=5/3$ only changes the slope 
from $0.5$ to $0.6$, well within the observational uncertainty.
This will also alter behavior at the highest masses, 
but this is not 
significant until well above the mass function break. 
There does however appear to be tentative evidence for such a 
transition in the observations shown in Figure~\ref{fig:size.mass}. 
As expected from the behavior of the linewidth-size relation, clouds in high density environments -- which will have a higher $\sigma_{0.4}$ in Eqn.~\ref{eqn:sizemass} above -- are offset to lower $R$ at fixed $M_{\rm cloud}$; we show the same model prediction for the high-redshift systems in good agreement with the observations. Once again, MW center clouds and other local systems in environments with higher densities are similarly offset. 

As discussed in \S~\ref{sec:model:mf.cores}, fully extending the models here to the scales of dense cores is beyond the scope of this paper. However, we expect these cores, if self-gravitating, to obey the scaling in Figure~\ref{fig:size.mass.offsets}. This means that if they form inside of high-density GMCs, we can (approximately) think of the ``parent'' GMC surface density as similar to the background $\langle \Sigma_{\rm disk}\rangle$ term in Eqn.~\ref{eqn:linewidthsize}, and might expect them to have higher dispersions at fixed sizes. This has been seen in suggested from observations \citep{ballesteros-paredes:2011.core.collapse.sizemass}, as part of a quasi-hierarchical gravitational collapse, similar to the predictions here. Of course, some regions can have much higher $\sigma_{v}$ at fixed $R$ and be simply not self-gravitating; these will not lie on the relation in Figure~\ref{fig:size.mass.offsets} (they will be offset to higher $\sigma_{v}/R^{1/2}$). This may, in turn, give rise to a dependence of the linewidth-size relation on the tracers and extinction threshold adopted, as observed \citep{goodman:1998.lws.dependence.on.tracers,lombardi:2010.larsonlaws.extinction.lognormal}

\begin{figure}
    \centering
    \plotone{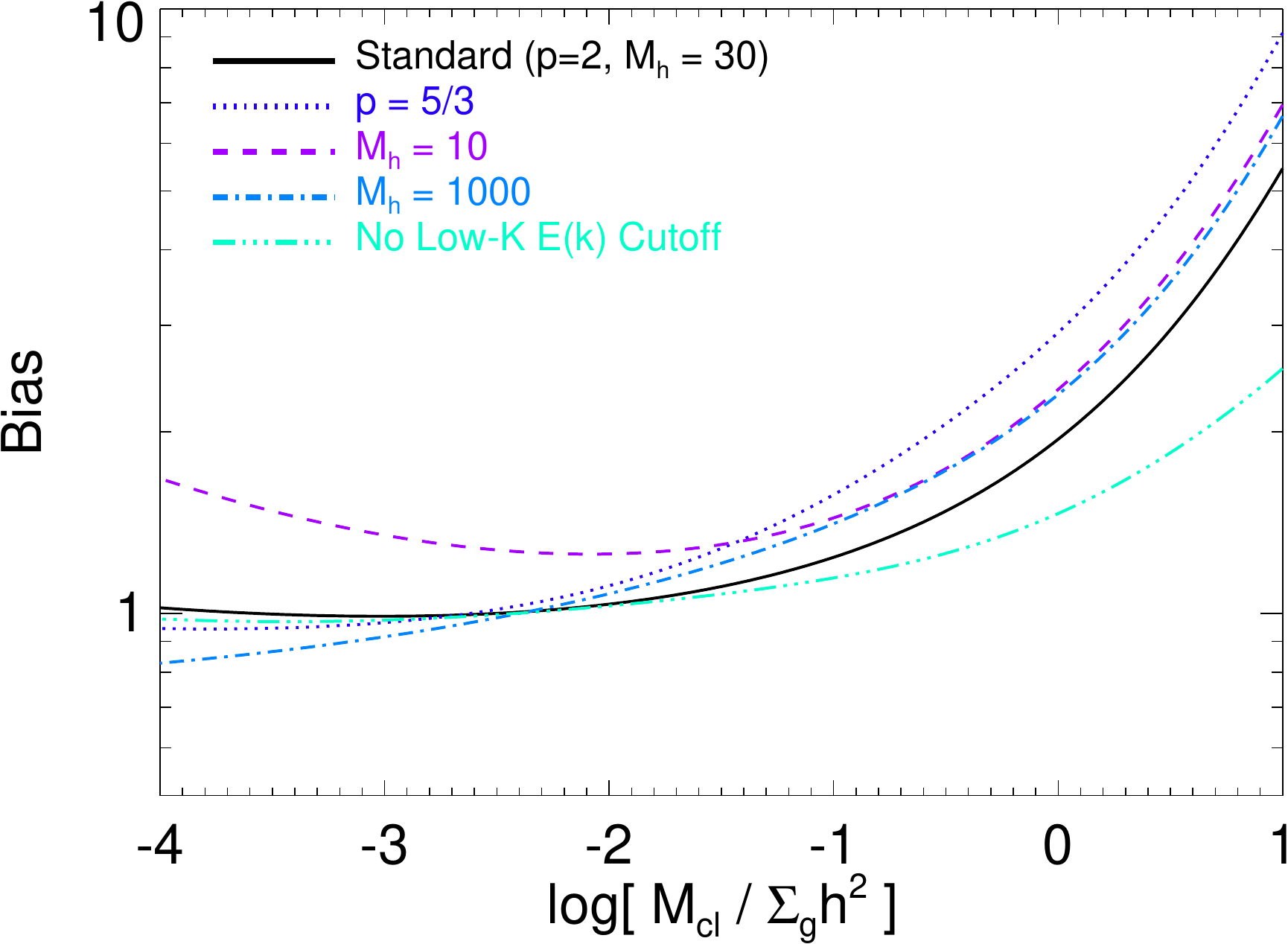}
    \caption{Predicted linear bias $b$ -- i.e.\ the amplitude of spatial clustering -- 
    as a function of GMC mass (allowing for 
    clouds inside of bound over densities).
    We plot this for our standard model and model variations in 
    Figure~\ref{fig:mf.var}, in dimensionless units. 
    Low-mass GMCs are weakly biased or anti-biased -- they simply trace the 
    dense gas. The highest-mass GMCs are strongly clustered -- they preferentially 
    trace global overdensities (e.g.\ spiral arms, galaxy nuclei, etc.).    
    \label{fig:bias}}
\end{figure}


\begin{figure}
    \centering
    \plotonesize{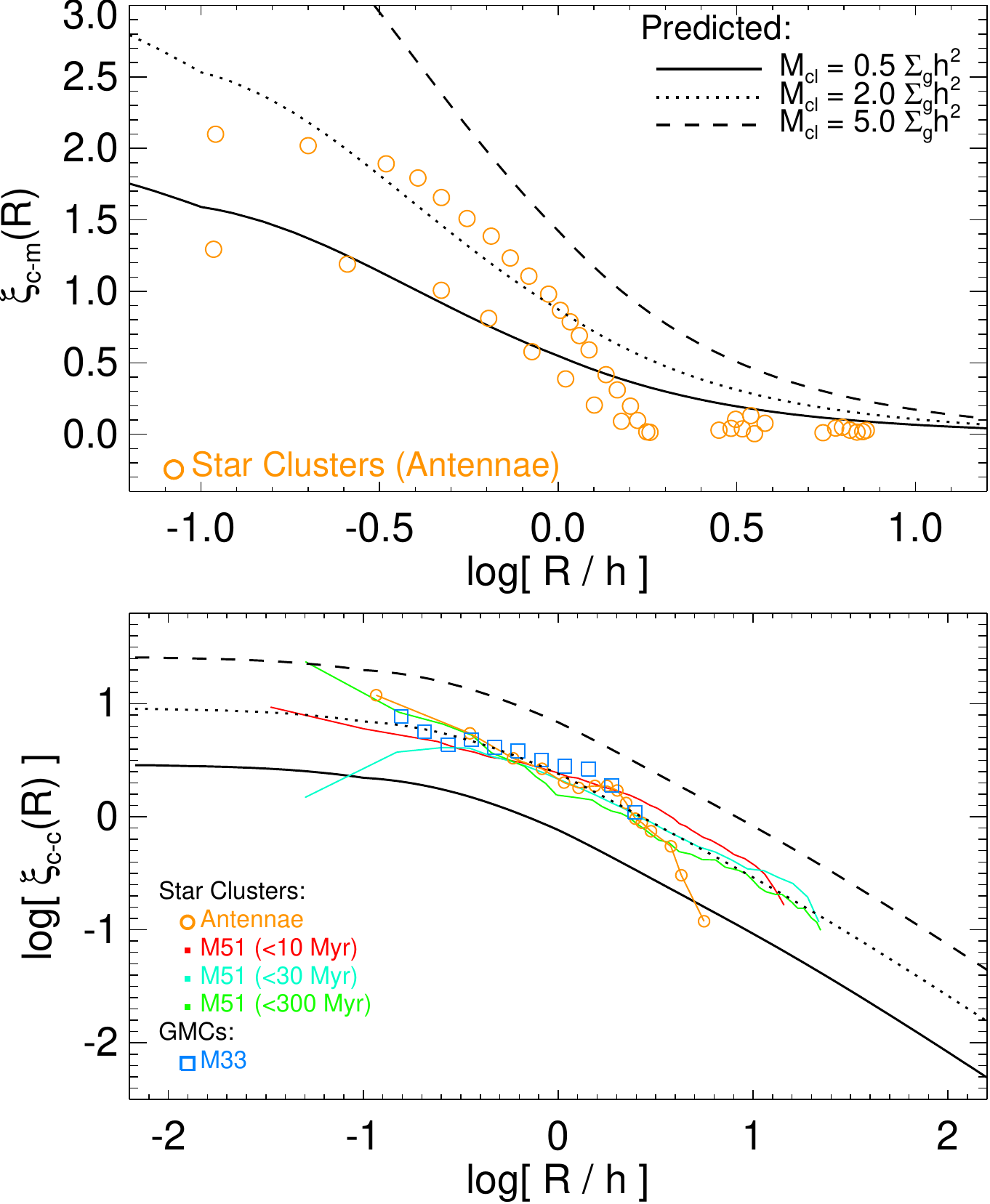}{1.02}
    \caption{Comparison of the predicted correlation function $\xi(R)$
    of bound gas objects/GMCs (lines) with the observed correlation functions 
    of young star clusters. We show the predicted values for three different 
    masses (essentially a normalization difference in the correlation function, 
    corresponding to the bias $b$ in Fig.~\ref{fig:bias}). For lower masses, 
    $b$ changes weakly, and for higher masses the MF is exponentially 
    suppressed, so this covers the interesting range. We plot radii in units of 
    the scale height $h$, in which the correlation function is dimensionless.
    {\em Top:} $\xi_{\rm cm}$, the cross-correlation between bound objects and gas mass 
    (defined in Eq.~\ref{eqn:xi.cm.defn}). 
    We compare the observed cross-correlation between young star clusters 
    (which should trace the locations -- regardless of how efficiently they form -- of 
    their ``parent'' GMCs) and CO gas maps measured in the Antennae by 
    \citet{zhang:2001.antennae.starcluster.clustering}. We compare the 
    two youngest $\sim100$ cluster samples (sampling two different regions and mass ranges), 
    with ages $\lesssim5\,$Myr and $\sim3-16$\,Myr. We do not know the masses 
    of the progenitor GMCs, but they are likely to be in this range, since these 
    are the most massive young star clusters in the galaxy (the more massive sample 
    has the higher $|\xi_{\rm cm}|$). 
    Despite this being a disturbed system, the agreement is reasonable.
    {\em Bottom:} $\xi_{\rm mm}$, the auto-correlation function of bound objects. 
    We again compare this measured for the young star clusters in the 
    Antennae (orange circles). We also compare the young $\sim1000$ star cluster 
    autocorrelation function in M51 (which is not disturbed), measured by 
    \citet{scheepmaker:2009.m51.cluster.clustering} 
    for age intervals $2.5-10$, $10-30$, and $30-300$\,Myr (red, cyan, 
    green, respectively). Especially in the youngest samples, 
    the agreement is good. We compare the same, measured directly for GMCs 
    with mass $\sim 2\,\Sigma_{g}\,h^{2}$ in M33 from \citet{engargiola:2003.m33.gmc.catalogue}; 
    again the agreement is good.
    \label{fig:bias.obs}}
\end{figure}

\section{Spatial Clustering of GMCs}
\label{sec:model:barrier:clustering}

In analogy to dark matter halos, we can use the excursion set 
formalism to also determine the spatial clustering and correlation function 
strength of these bound sub-units. 
Following \citet{mowhite:bias}, the 
excess abundance of collapsing objects (relative to the mean abundance) 
in a sphere of radius $R_{0}$ with mean density $\delta_{0}$ 
is 
\be 
\delta_{\rm coll}(R_{1},\,\delta_{c,\,1}\,|\,R_{0},\,\delta_{0}) \equiv 
\frac{\mathcal{N}(1|0)}{n(M_{1})\,V_{0}} - 1 
\ee
where $n(M_{1})$ is the average abundance of objects of mass $M_{1}$ 
(from the mass function) and 
$\mathcal{N}(1|0)$ is the number of collapsing objects in a region of radius $R_{0}$ 
(variance $S_{0}$) 
with fixed overdensity $\delta_{0}$.

\subsection{Linear Bias}
\label{sec:model:barrier:clustering:bias}

If $\delta_{c}$ were constant, $\mathcal{N}(1|0)$ can be determined 
analytically and is simply
\be
\mathcal{N}(1|0) = \frac{\rho_{c,\,1}\,V_{0}}{M_{1}}\,
\frac{\delta_{c,\,1}-\delta_{0}}{\sqrt{2\pi}\,(S_{1}-S_{0})^{3/2}}\,
\exp{{\Bigl[}-\frac{(\delta_{c,\,1}-\delta_{0})^{2}}{2\,(S_{1}-S_{0})} {\Bigr]}}\,
\frac{{\rm d}\,S_{1}}{{\rm d}M_{1}}
\ee
\citep{bond:1991.eps}.
In the regime where $R_{0}\gg R_{1}$, so $\Delta_{0}\ll \Delta_{1}$, 
this simplifies to 
\be
\label{eqn:bias}
\delta_{\rm coll} \approx {\Bigl (} \frac{\nu_{1}^{2}-1}{\delta_{c,\,1}}{\Bigr)}\,\delta_{0} = b(M_{1})\,\delta_{0}
\ee
where $b(M_{1})$ is defined as the linear bias of objects of mass $M_{1}$.\footnote{The 
expression for bias here is different from that for dark matter halos by a linear 
offset of unity. That offset arises in the dark 
matter case because of the expansion of the Universe and 
subsequent mapping from ``initial'' (Lagrangian) coordinates to ``observed'' 
(Eulerian) coordinates. It does not appear here because the terms are all 
evaluated instantaneously (the expression here is equivalent to the 
``initial time'' expression for $b$ in halos).} 

The barrier $\delta_{c}$ here is not constant. However, 
for arbitrary $\delta_{c}(M)$, we can calculate $\mathcal{N}(1|0)$ exactly 
by repeating our Monte Carlo excursion from 
\S~\ref{sec:model:barrier:mf:exact}, but instead of beginning with 
initial conditions $S=0$, $\delta=0$ for each walk, 
we begin at scale $S=S_{0}$ with density $\delta=\delta_{0}$. 
The bias $b(M_{1})$ is then just the ratio of 
$\delta_{\rm coll}/\delta_{0}$ for small $\delta_{0}$. 

Figure~\ref{fig:bias} plots the bias as a function 
of cloud mass.
A couple of key properties are clear. 
At high masses above the exponential cutoff in the mass function, 
the bias increases rapidly. This is qualitatively similar to 
what is seen for dark matter halos: because such systems are exponentially 
rare, they will tend to be strongly biased towards the few regions 
with substantial large-scale over densities. Physically, this corresponds 
to gas overdensities in the disk on scales {\em larger} than the scale-height 
$h$, i.e.\ a preferential concentration of the 
most massive GMCs in {\em global} instabilities such as spiral 
arms, bars, and $\sim$kpc-scale massive star-forming complexes, 
rather than their being randomly distributed across 
the gas. 
At intermediate masses below the mass function break, where most of 
the cloud mass lies, the bias is weak (order unity), so most of the 
mass in clouds simply traces most of the gas mass in general. 
We stress that this does {\em not} necessarily mean 
clouds are randomly distributed over the disk as a whole; 
it means they are unbiased relative to the gas mass distribution. 
But at low masses, the bias again rises (weakly). 
This is related to the anti-hierarchical nature of cloud formation: 
the bias here is driven by clouds which form via fragmentation 
from other clouds.

We can approximate these exact results using our previous approximate fitting functions for the mass function 
(Equation~\ref{eqn:mf.approx} \&\ \ref{eqn:mf.neglect.cic}) modified (as with the case of 
a linear barrier) 
so $S\rightarrow S_{1}-S_{0}$ and $\nu^{2}\rightarrow(\delta_{c}-\delta_{0})^{2}/(S_{1}-S_{0})$.
Neglecting the clouds-in-clouds problem (i.e.\ including those clouds), 
we obtain the approximate  
\be
\label{eqn:bias.approx.1}
b_{\rm cic}(M_{1}) \approx \frac{\nu_{1}^{2}-1+\nu_{1}\,{\rm d}\ln\rho_{1}/{\rm d}\sigma_{1}}
{\delta_{c,\,1}\,(1+\nu_{1}^{-1}{\rm d}\ln\rho_{1}/{\rm d}\sigma_{1})}
\ee
Which in practice is a small ($\sim10-20\%$) correction to Equation~\ref{eqn:bias}. 
If we exclude clouds-in-clouds, 
\be
\label{eqn:bias.approx.2}
b(M_{1}) \approx \frac{1}{\delta_{c,\,1}}\,{\Bigl [}\nu_{1}^{2} - \frac{\delta_{c,\,1}}{\tilde{B}} {\Bigr]}
\ee
(where $\tilde{B}$ is defined in Equation~\ref{eqn:mf.approx}). 
This is identical to Eqn.~\ref{eqn:bias} at high masses, but it allows for {\em negative} bias at 
low masses, if $B_{\rm min}\equiv\ln{(\rho_{c,\,{\rm min}}/\rho_{0})}<2$ 
and $\delta_{c} < \sigma^{2}\,(B_{\rm min}^{-1} - 1/2)$. Physically, 
the fact that Equation~\ref{eqn:bias.approx.1} is always positive 
means that the number of bound regions of mass $M_{1}$ inside a 
large-scale overdensity always increases with $\delta_{0}$. 
However, for some values of $M_{1}$ and $\delta_{0}$, increasing $\delta_{0}$ 
more rapidly increases the probability that these regions are themselves inside a 
larger collapsed region.
For a more detailed discussion of the leading-order corrections 
when considering a moving as opposed to constant barrier $\delta_{c}$, 
we refer to \citet{shethtormen}.

\subsection{The Correlation Function: Theory}
\label{sec:model:barrier:clustering:corrfn}

Recall, the physical over-density is $\rho/\rho_{0}=\exp{(\delta-\sigma(R)^{2}/2)}$. 
The correlation function $\xi_{\rm cm}$ between collapsed objects of mass $M_{1}$ 
and background mass, 
as a function of radius $R_{0}$, 
is defined by  
\begin{align}
\label{eqn:xi.cm.defn}
1+\xi_{\rm cm}(R_{0},\,M_{1}) &\equiv 
\frac{\langle \mathcal{N}(1\,|\,R_{0})\,|\,\rho\rangle}{n(M_{1})\,V_{0}\,\rho_{0}} \\ 
&= \langle (1+\delta_{\rm coll})\,|\,\exp{(\delta_{0}-S[R_{0}]/2)} \rangle_{R_{0}}  \\  
\label{eqn:xi.exact}&= \int\,\frac{\mathcal{N}(1|0)}{n(M_{1})\,V_{0}}\,e^{(\delta_{0}-S_{0}/2)}\,
q(\delta_{0}\,|\,S_{0})\,{\rm d}\delta_{0}
\end{align}
where the integral is over all $\delta_{0}<\delta_{c}(R_{0})$, 
and $q(\delta_{0}\,|\,S_{0})$ is a weighting factor defined 
in \citet{bond:1991.eps} as the probability that the overdensity at a random 
point, smoothed on a scale $R_{0}$, is $\delta_{0}$ and does not exceed 
$\delta_{c}(R_{0})$ on any larger smoothing scale.
\footnote{
Note that the equations here are modified from those used in 
the cosmological case because we use $\delta$ to represent the logarithmic 
(not linear) density field. However for small $\delta_{0}$ they are identical, 
which is why we recover similar scalings for the bias and 
correlation functions.}

Equation~\ref{eqn:xi.exact} can be evaluated numerically 
with the Monte Carlo solution for $\mathcal{N}(1|0)$ and 
$q(\delta_{0}\,|\,S_{0})$.
But at large $R_{0} \gg R_{1}$ (provided 
$S_{0}\rightarrow0$ as $R\rightarrow\infty$), 
it simplifies to just
\begin{align}
\label{eqn:expansion}
1+\xi_{\rm cm}(R_{0},\,M_{1}) &\approx 1 + b(M_{1})\,\sigma^{2}(R_{0})\ \ \ \ \ (R_{0}\gg R_{1})\\
&= 1 + b(M_{1})\,\xi_{\rm mm}(R_{0})
\end{align}
This can be shown for any first-crossing distribution 
by first taking $q\rightarrow p(\delta_{0}\,|\,S_{0})$ since the 
probability of collapse on larger scales is negligible, and then noting 
$\exp{(\delta_{0}-S_{0}/2)}\,p(\delta_{0}\,|\,S_{0})
= 1/\sqrt{2\pi\,S_{0}}\,\exp{(-(\delta_{0}-S_{0})^{2}/2\,S_{0})}$, 
which becomes a delta function centered at $\delta_{0}=S_{0}$ as $S_{0}\rightarrow0$.


The auto-correlation function of the mass 
$\xi_{\rm mm}$ is given by $1+\xi_{\rm mm}\equiv \langle \rho^{2} \rangle/\rho_{0}^{2}
= \exp{(S_{0})}$, so 
$\xi_{\rm mm}=\exp{(S_{0})}-1 \approx S_{0} = \sigma^{2}(R_{0})$ at 
large $R_{0}$ is just the 
variance in the mass field.
So the collapsed object-mass correlation function on large scales is then just the 
bias times the mass autocorrelation function.
It is straightforward to verify that 
the auto-correlation function of collapsed objects 
is just given by 
\be
\xi_{\rm cc} \approx b^{2}(M_{1})\,\xi_{\rm mm}\ \ \ \ (R_{0}\gg R_{1})
\ee

The correlation functions discussed above are 
the {\em three-dimensional} correlation functions. However, 
with rare exceptions, it is in general much 
easier to determine the projected correlation function $\xi_{2d}(R_{p})$, 
defined so the probability of finding another object in a 2-d annulus 
$d^{2}{\bf r}$ around a given object is $\langle {\rm d}N/{\rm d}A\rangle\,(1+\xi_{2d})\,d^{2}{\bf r}$. 
This is straightforward to calculate
\be
w_{p}\equiv \xi_{2d}(R_{p}) = 
\frac{\int_{-\infty}^{\infty}\,n_{0}(z)\,\xi_{3d}(\sqrt{R_{p}^{2}+z^{2}})\,{\rm d}z}
{\int_{-\infty}^{\infty}n_{0}(z)\,{\rm d}z}
\ee
where $z$ is the line-of-sight direction and $n_{0}(z)=n(M)$ is 
the average abundance. For the typical case of an approximately face-on disk 
with the exponential vertical profile we have adopted, 
$n_{0}(z)\propto \exp{(-z/h)}$; however, accounting for this, we 
should also slightly modify our calculation of $\xi_{3d}$, integrating over 
$\rho_{0}$ at all central positions with $\mathcal{N}(1\,|\,0\,|\,{\bf x})$ 
a function of $\rho_{0}({\bf x})$ (since our derivation 
up to this point implicitly assumed a homogenous background). In either case, at large radii this is just
$w_{p}\propto (R/h)\,\xi_{\rm 3d}$.


\subsection{Observed GMC \&\ Star Cluster Correlation Functions}
\label{sec:model:barrier:clustering:obs}

In Figure~\ref{fig:bias.obs}, we compare the predicted (two-dimensional) 
correlation functions to observations. Unfortunately, at present there 
are no published observations of the GMC-GMC correlation function. 
However, various groups have measured the correlation functions of 
young, massive star clusters in nearby systems. 
Statistically, the positions of such star clusters should trace those of 
their ``parent'' GMCs (with greater fidelity as we consider younger star clusters). 
And although clusters will disperse or be destroyed with time, the 
correlation function should not be affected so long as this ``infant mortality'' 
is not strongly position-dependent (though that is uncertain, if it depends on 
e.g.\ tidal fields). This also has the advantage that star clusters can be 
much longer-lived than GMCs, so allow better statistics. 
The major uncertainty is that, without knowing the (uncertain) star formation 
efficiency, the exact mass of the progenitor GMCs is undetermined. 
However, since the observed systems sample the brightest clusters, we can 
safely assume that their progenitors were the most massive 
GMCs (and since the mass function cuts off exponentially, should reflect 
masses $\sim1-10$ times the ``break'' in the mass function). 

\citet{scheepmaker:2009.m51.cluster.clustering} measure the star cluster-star cluster auto-correlation 
function (which we should compare to the GMC autocorrelation $\xi_{\rm cc}$) 
in M51 for the brightest $\sim1000$ star clusters, 
in three age intervals ($2.5-10$, $10-30$, and $30-300$\,Myr). 
The cluster masses range from $10^{3.5-5}\,M_{\sun}$, which for a few percent 
star formation efficiency indeed corresponds to the most massive 
GMCs. The mass scale only affects the bias (normalization) -- it 
is more important to compare the shape of $\xi_{\rm cc}$ -- this is invariant in 
units of $R/h$. With a large number of clusters and a nearly face-on projection, 
this is the most robust probe over large dynamic range. 
\citet{zhang:2001.antennae.starcluster.clustering} measure in the Antennae the star cluster-star cluster 
autocorrelation function and the star cluster-gas cross correlation function 
(tracing the gas in CO maps, 
which -- since the system is quite dense -- account for 
most of the gas mass). Here the geometry is obviously 
much more complex so the results should be interpreted with additional caution, 
but the authors do attempt to account for the global structure of the system, 
and separately measure the correlation functions in different regions. 
We specifically consider their youngest cluster samples (R and B1), 
with the brightest $\sim100$ and $\sim1000$ objects 
at ages $\lesssim5\,$Myr and $\sim3-16\,$Myr, respectively 
(masses $\sim10^{4-6}\,M_{\sun}$). 
Finally, we attempt to follow the procedure in \citet{scheepmaker:2009.m51.cluster.clustering} to construct the auto-correlation function for GMCs in M33, using the catalogue in \citet{engargiola:2003.m33.gmc.catalogue}, which is both face-on and has a well-defined survey area and completeness limit. Since we cannot properly account for survey edge effects or the global density profile, we simply truncate the correlation function at half the radius inside of which $75\%$ of the identified GMCs are found. Here, we can determine the mean mass in the distribution, which is approximately $\sim2\,\Sigma_{h}\,h^{2}$ estimated using the parameters from Figure~\ref{fig:mf.pred} -- this is almost exactly the value in the model which gives the best-fit predicted normalization of $\xi(R)$.
Given the uncertainties in both observations and the 
cluster-GMC mapping, the agreement is striking.

\begin{figure}
    \centering
    \plotone{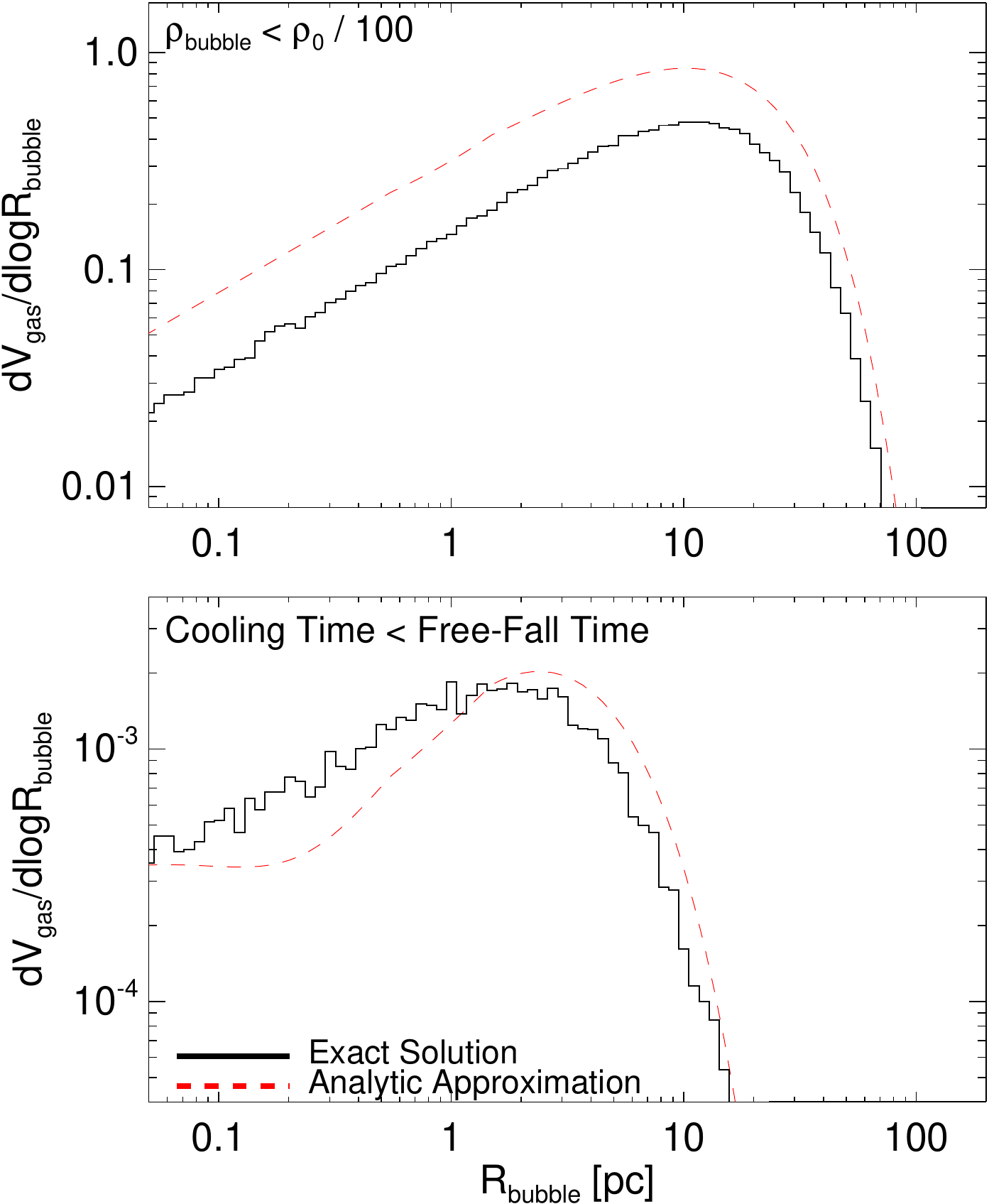}
    \caption{Predicted differential volume fraction 
    in underdense ``bubbles'' as a function 
    of bubble radius $R$. For illustrative purposes we assume 
    $h=200\,$pc but the scale $R_{\rm bubble}$ scales $\propto h$.
    {\em Top:} Bubbles defined as a proportional under-density $\rho \le \rho_{0}/100$. 
    {\em Bottom:} Bubbles defined as regions where the post-shock 
    cooling time at velocities $\sim v_{t}(R)$ exceeds the free-fall 
    time $\sim 1/\sqrt{G\rho}$. Because determining the cooling time requires 
    absolute units, we normalize the model by 
    assuming $\Sigma=10\,{\msun\,{\rm pc^{-2}}}$.
    The exact solution (black solid lines) is compared to the 
    approximate analytic solution (red dashed) from Eq.~\ref{eqn:bubbledist}.
    A broad distribution of underdense regions should be present simply 
    from turbulent velocity divergences, which can have sizes $\sim h$ 
    and contain a large fraction of the disk mass. However, only a small 
    fraction will ``self-heat'' to temperatures where they cannot cool -- 
    ``hot'' bubbles require energy input from some source (e.g.\ stellar 
    feedback).
    \label{fig:bubbles}}
\end{figure}

\begin{figure}
    \centering
    \plotone{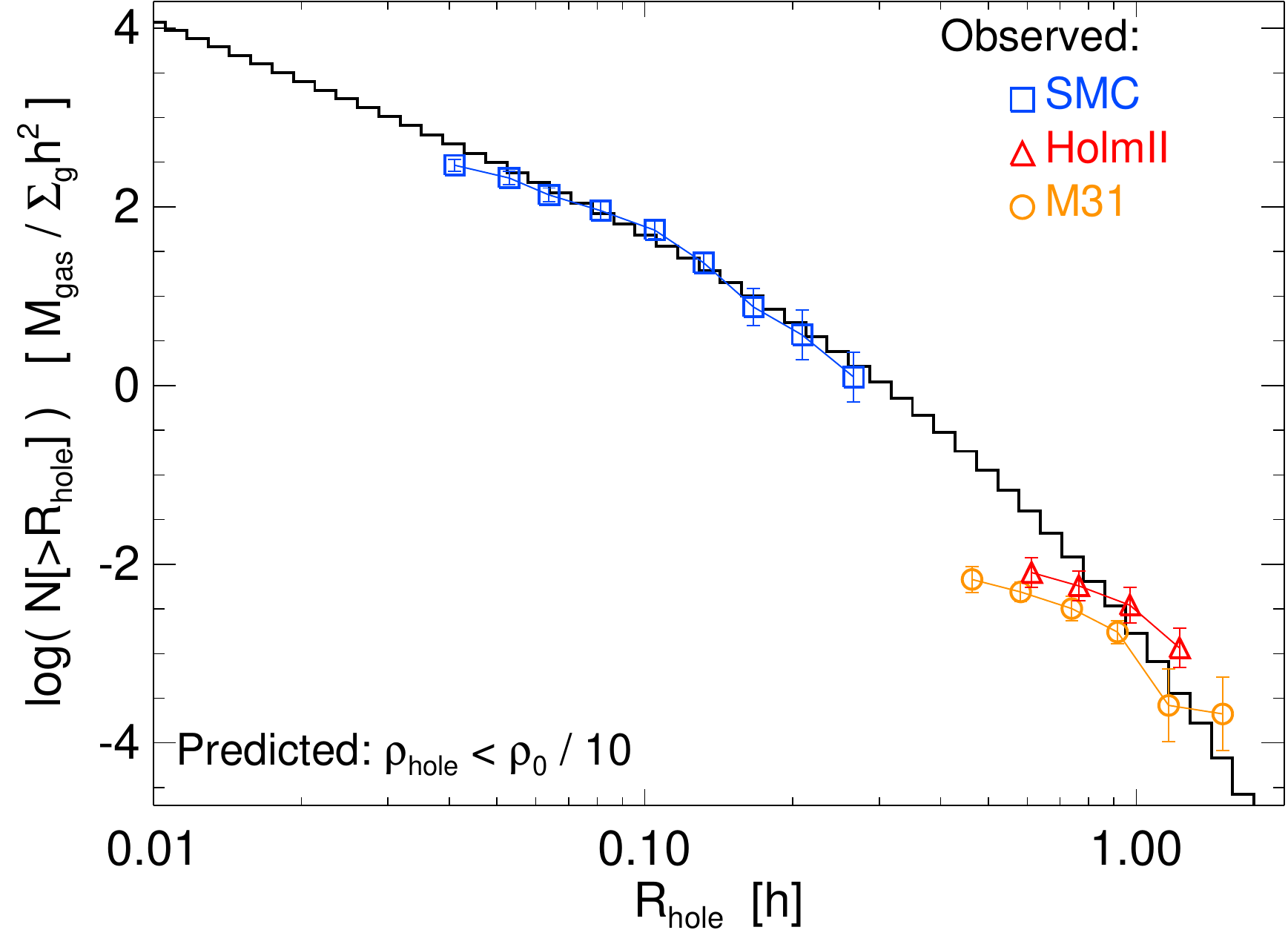}
    \caption{Comparison of predicted and observed hole/bubble radii. 
    We plot the predicted cumulative number of bubbles as a function of 
    bubble size for our standard model, in dimensionless units 
    (bubble size in units of $h$). 
    Here, we assume a simple order-of-magnitude proportional bubble under-density 
    $\rho \le \rho_{0}/10$. For typical galaxy properties, this also corresponds to 
    densities at which the diffuse galactic UV background will fully ionize the bubble.
    This allows us to plot all 
    observed systems on the same Figure. 
    We compare the observed HI hole radius functions from 
    radius functions from the SMC \citep{staveley-smith:1997.smc.HI.holes}, 
    Holmberg II \citep{puche:1992.holmbergII.hI.holes}, 
    and M31 \citep{brinks:1986.m31.HI.holes}, 
    and normalize them with the observed $M_{\rm gas}$, $\rho_{0}$, $h$ 
    from the same sources. 
    The agreement is good -- most, if not all, of the HI ``holes'' are a natural 
    consequence of turbulent density fluctuations and require no input energy 
    source to ``clear them out.''
    \label{fig:bubbles.obs}}
\end{figure}

\section{The Distribution of Underdense Bubbles}
\label{sec:model:barrier:bubbles}

Just as we used the excursion set formalism to predict the mass function 
of clouds by identifying objects {\em above} a critical over-density $\delta_{c}$, 
we can also use it to predict the abundance of under-dense regions (``bubbles'') by 
identifying regions {\em below} a critical under-density $\delta_{b}$. 
We will follow \citet{sheth:2004.void.eps}, who apply this formalism to the dark 
matter halo context to study the distribution of voids. 

Generally, the procedure is the same, but considering the mass/radii 
below $\delta_{b}$ instead of above $\delta_{c}$. However, some additional 
complications arise. 
First, unlike the case of collapsing objects where the counting of 
``clouds in clouds'' was potentially valid, 
here we should clearly count ``voids in voids'' as simply part of 
the larger, parent void/bubble. So we again need to specify to the first crossing 
distribution (the distribution of the largest radii on which trajectories cross 
$\delta_{b}$). 
Second, 
we must also ensure that the void/bubble region is not itself 
contained inside of a collapsing region (i.e.\ that 
$\delta<\delta_{c}$ on all scales above the 
$\delta_{b}$ crossing), since that would 
``overwhelm'' or ``squeeze'' the bubble.\footnote{The details of 
the criterion for this can be subtle and more complex than simply 
being in a collapsing region, since smaller overdensities can 
also ``squeeze'' voids. This is discussed in detail in \citet{sheth:2004.void.eps}. However, 
because we do not need to map here between initial and 
final overdensities, many of these ambiguities are avoided.}
Third, and most critical for our purposes, a ``void'' or ``bubble'' is 
not obviously well-defined in this context. Because there is no linear expansion 
here, we cannot derive the equivalent of the shell crossing criterion 
used for dark matter halo voids, and there is no obvious 
threshold which is physically as robust 
as the self-gravity criterion for collapse. 
We will return to this question and 
consider different plausible, but ultimately somewhat 
arbitrary choices of under-density criterion. 

If the ``bubble'' barrier $\delta_{b}$ and the collapse barrier 
(which must be avoided on scales above the bubble) $\delta_{c}$ 
were constant, 
then \citet{sheth:2004.void.eps} show that 
the first crossing distribution can be analytically re-derived subject to these 
boundary conditions, to give the fraction of trajectories in bubbles per logarithmic 
interval ${\rm d}\ln{\nu_{b}}$ 
\begin{align}
\label{eqn:bubbledist}
\nu_{b}\,f_{b}(\nu_{b}) &= 
\sum_{n=1}^{\infty}\, \frac{2n\pi\mathcal{D}^{2}}{\nu_{b}^{2}}\,
\sin{(n\pi\mathcal{D})}\,\exp{{\Bigl(}-\frac{n^{2}\pi^{2}\mathcal{D}^{2}}{2\nu_{b}^{2}} {\Bigr)}}\\
\mathcal{D} &\equiv \frac{|\delta_{b}|}{\delta_{c} + |\delta_{b}|} \ , \ \ \ \ 
\nu_{b} \equiv \frac{|\delta_{b}|}{S(R)^{1/2}}
\end{align}
Recalling that we are sampling the 
Eulerian space, we can then trivially translate this to the 
number density of bubbles per unit radius or mass, e.g.\
\be
\frac{{\rm d}n}{{\rm d}\ln{R}} = 
\frac{1}{V_{b}}\,
\frac{{\rm d}f}{{\rm d}\ln{R}} = \nu_{b}\,f_{b}(\nu_{b}) \,
\frac{1}{V_{b}}\,
\frac{{\rm d}\ln{\nu_{b}}}{{\rm d}\ln{R}}
\ee
where $V_{b}$ is the effective volume of the bubble.

Again, we stress that the barrier is not constant, so we do not know that this 
will be an accurate approximation. More rigorously, it is straightforward to 
derive the same first-crossing distribution using the Monte Carlo approach 
in \S~\ref{sec:model:barrier:mf:exact}. We follow the identical procedure, but simply record 
the first crossing of $\delta_{b}(R)$ for those trajectories that cross $\delta(R)<\delta_{b}(R)$ 
and have not crossed $\delta_{c}(R)$ at any larger scale. 

The results of this exact calculation, and the analytic approximation 
from Equation~\ref{eqn:bubbledist}, are shown in Figure~\ref{fig:mf.var}, for two 
different choices of $\delta_{b}$. 
First, we consider a simple under-density criterion: here 
$\rho_{b} \le \rho_{0}/100$. 
There is a very broad distribution of bubbles which satisfy this criterion: it 
includes several tens of percent of the total mass. 
The characteristic spatial ``bubble scale'' is at a factor of $\sim0.1\,h$, 
which (for the definitions used here) corresponds very closely to the scale 
at which the local contributions to density fluctuations ($\Delta S$) are 
maximized. A large population of such fluctuations must arise
for a density distribution similar to Equation~\ref{eqn:lognormal}: because 
the distribution is lognormal, the median density is $\ln{(\rho_{\rm med}/\rho_{0})}=-\sigma^{2}/2$; 
i.e.\ for $\sigma\sim1.3\,$dex fluctuations, $\rho_{\rm med}\approx 0.01\,\rho_{0}$ 
so of order half the volume should be in underdense regions. 
For any fixed (fractional) density threshold $\rho_{b}/\rho_{0}$, 
the behavior is qualitatively similar, but shifts systematically to smaller 
scales $R$ and smaller normalization (the total mass in such 
regions scaling as 
$\sim \exp{(-\nu_{b}^{2}/2}$) as $\rho_{b}/\rho_{0}$ decreases. 

There is nothing physically ``special'' about such regions -- they are simply the 
low-$\rho$ portion of the density PDF. A more meaningful threshold 
might be to define ``bubbles'' as regions where the cooling time becomes 
longer than the free-fall time.
The isothermal temperature $c_{s}$ is however quite low, so 
this will not be satisfied unless the temperature suddenly increases; for this, 
consider the shocks occurring in the turbulent medium 
at $v_{s}\sim v_{t}(R)$. Knowing $E(k)$, we can estimate the 
distribution of post-shock temperatures and densities for a 
random Lagrangian parcel, and compare the resulting cooling 
time to the free-fall time $t_{\rm ff}\approx0.54/{\sqrt{G\,\rho}}$.
Since we are interested in the regime where the cooling time 
will be long, we can simplify the problem by assuming a strong 
adiabatic shock and that thermal 
Brehmsstrahlung dominates the cooling. 
In this regime, $t_{\rm cool} \equiv n\,k_{B}\,T/\Lambda\,n^{2} \le t_{\rm ff}$ at densities 
\be
n_{b}(R)\lesssim 10^{-4}\,{\rm cm^{-3}}\,{\Bigl(}\frac{v_{t}(R)}{10\,{\rm km\,s^{-1}}} {\Bigr)}^{2}
\ee
If we normalize our model to MW-like 
conditions by assuming $\sigma_{g}(h)\approx10\,{\rm km\,s^{-1}}$ 
and $n_{0}=\rho_{0}/\mu m_{p}\approx 1\,{\rm cm^{-3}}$, then this 
defines $\delta_{b}$. The resulting 
distribution of bubbles is shown in Figure~\ref{fig:bubbles}. 
Qualitatively, the shape of the distribution is similar -- it truncates more 
rapidly at low $R$ because the decrease of turbulent 
velocities $\langle v_{t}(R) \rangle $ 
with decreasing $R$ means that the barrier becomes more difficult to cross.
The normalization is also significantly lower, corresponding to the 
lower absolute densities ($\rho_{b}/\rho_{0}\sim10^{-4}$) needed 
near scales $\sim h$ to reach this ``hot gas'' threshold. 

In both cases, the analytic 
approximation of Equation~\ref{eqn:bubbledist} works well 
for the largest voids (albeit with a factor $\sim1-1.5$ normalization offset), 
but is systematically offset for low-mass voids. This is directly a consequence of the 
moving barriers $\delta_{c}$ and $\delta_{b}$. 

In Figure~\ref{fig:bubbles.obs}, we compare the predicted size function
of bubbles (in dimensionless units) to observations of HI ``holes.'' We 
compile the observed HI hole size distributions in 
the SMC \citep{staveley-smith:1997.smc.HI.holes}, 
Holmberg II \citep{puche:1992.holmbergII.hI.holes}, 
and M31 \citep{brinks:1986.m31.HI.holes}, 
and scale the normalization 
of each according to the observed global galaxy properties measured at the 
radii enclosing half the ``holes.'' 
Observations of the LMC \citep{kim:1999.lmc.hi.holes}, IC2574 \citep{walter:1999.ic2574.hI.holes}, 
and M33 \citep{deul:1990.m33.HI.holes} give similar results.

There is no well-defined criterion for selection of ``holes'' and the density contrasts 
involved are typically modest, so we simply 
compare with the prediction for a constant density contrast $\rho_{b} \le \rho_{0}/10$. 
This is approximately consistent with the direct estimates of the interior bubble 
densities/density contrast, and also (for the global properties of these systems) 
corresponds to densities where even the largest (few hundred pc) holes would 
become fully ionized either from the diffuse 
galactic background or a single O/B star inside the ``hole.'' 
The agreement is good -- if anything, the model predicts more small 
``holes,'' but this may be a question of observational selection/completeness (or 
a deficit of sources to ionize them).
The characteristic hole size is predicted to scale with $h$ (the characteristic radius), 
giving larger holes in thicker galaxies -- a well-observed 
phenomenon \citep[see][and references therein]{oey:1997.HI.hole.models,
walter:1999.ic2574.hI.holes}. 

\section{Construction of GMC ``Merger Trees'' from This Formalism}
\label{sec:model:trees}

\subsection{General Considerations}
\label{sec:model:trees:notes}

One of the most powerful applications of the excursion set approach 
in galaxy formation is the construction of the extended Press-Schecter 
``merger trees,'' conditional mass functions, and formation histories 
for dark matter halo populations, which form the foundation of semi-analytic 
models. 
This provides a means to statistically link populations in time 
and self-consistently model their evolution, with whatever additional 
physics are desired. 
We now show that the same ``merger tree'' approach can be 
applied here, to derive the time evolution of the systems we have 
thus far considered static.

Before we describe the mechanics of constructing these 
trees, there are a couple of important physical distinctions that 
will necessitate a somewhat different treatment from the 
typical methodology in the dark matter halo EPS formalism. 

First, unlike with dark matter halos, there is no reason to believe that 
bound clouds are ``conserved'' (modulo their mergers into more 
massive clouds). In fact we expect from observations that they only 
live a short time, then are disrupted \citep{zuckerman:1974.gmc.constraints,
williams:1997.gmc.prop,evans:1999.sf.gmc.review,evans:2009.sf.efficiencies.lifetimes}.
So it makes no sense to begin from a present population of 
clouds and work backwards in time to construct the tree (as 
is typically done for halo merger trees). Instead, we need to 
forward-model the time evolution, to allow whatever model physics 
the user desires to determine whether or not such clouds 
survive.

Second, we cannot assume that all the mass is in collapsing 
objects. We must therefore track un-collapsed elements as well, 
allowing for their possible collapse at later times. 

Third, density fluctuations in a turbulent medium clearly do {\em not} 
evolve according to simple linear growth, in the manner of 
cosmological density perturbations. How, then, can we link a 
fluctuation at any one time to that at another time? 
To do this, we will assume that the turbulence is globally steady-state: 
i.e.\ that -- excepting the behavior of collapsing regions -- the 
turbulent velocity cascade is (statistically) maintained and, as a result, so is the global density 
PDF. We stress that we are not attempting to model {\em how} 
the turbulence is maintained.
In this regime, the density PDF for independent 
modes on different scales obeys a generalized Fokker-Planck equation, 
with a diffusion term giving the effectively ``random walk'' behavior 
of each Lagrangian density parcel (from small-scale encounters/shocks/accelerations) 
and a drift term corresponding to damping/relaxation (from viscosity, pressure forces, 
mixing, and the energetic cost associated with large velocity deviations). 
Under these conditions, if we know the stationary behavior of the PDF for 
some variable 
$x$ is a Gaussian distribution with standard deviation $\sigma_{x}$ 
and zero mean, then 
the probability distribution to find the system with value $x$ at 
time $t$ given an initial distribution with (delta-function) value $x_{0}$ at 
time $t_{0}=t-\Delta\,t$ is 
\begin{align}
\label{eqn:timeevol}
p(x,\,t)\,{\rm d}x &= \frac{1}{\sqrt{2\pi\,\tilde{\sigma}_{x}^{2}}}\,
\exp{{\Bigl(}-\frac{(x-\tilde{x}_{0})^{2}}{2\,\tilde{\sigma}_{x}^{2}} {\Bigr)}} \\ 
\tilde{\sigma}_{x} &\equiv \sigma_{x}\,\sqrt{1-\exp{(-2\,[t-t_{0}]/\tau)}}
\approx \sigma_{x}\,\sqrt{2\,\Delta t/\tau}\\ 
\tilde{x}_{0} &\equiv x_{0}\,\exp{(-[t-t_{0}/\tau])}
\approx x_{0}\,(1-\Delta t/\tau)
\end{align}
where the latter equalities are the series expansion for 
$\Delta\,t/\tau \ll 1$.\footnote{In addition to being convenient later, 
these series expansions have the properties that for small timesteps, they 
represent the {\em only} form that the evolution of $p(x,\,t)$ can 
take if we require that it conserve $\sigma_{x}$ in 
ensemble average and conserve the growth in 
variance between $x$ and $x_{0}$ independent of 
the choice of integration stepsize.}

The timescale $\tau$ here is the timescale on which the 
variance of $x(t)$ with respect to $x_{0}$ grows, normalized 
by the steady-state variance $\sigma_{x}$, i.e.\ the 
timescale of ``mixing'' in the distribution. More 
formally the amplitude of the correlation between 
values in time declines with exponential timescale $\tau$. 
In supersonic turbulence, this is simply the crossing 
time
\be
\tau = \eta\, t_{\rm cross} = \eta\,R/\langle v_{t}^{2}(R) \rangle^{1/2}
\ee
Where $\eta\approx1$ is a constant which can be calibrated from 
numerical simulations (\citealt{pan:2010.turbulent.mixing.times} find $\eta\approx0.90-1.05$ over the range 
$\mathcal{M}\sim1.2-10$).

\subsection{The Mechanism of Tree Construction}
\label{sec:model:trees:mechanism}

With these points resolved, it is straightforward to 
generalize our approach to construct a time-dependent 
``fragmentation tree.'' 
We outline the methodology below. 

(0) Define the variance $S\equiv\sigma^{2}(R)$ 
and collapse threshold $\delta_{c}(R)$ either directly or from 
the turbulent power spectrum $E(k)$.

(1) Begin by constructing the initial conditions. 
Consider a Monte Carlo ensemble 
of ``trajectories,'' as in \S~\ref{sec:model:barrier:mf:exact}. 
Each trajectory $\delta(R)$ is defined by 
the values $\Delta \delta_{j}$ on each 
scale $R_{j}\rightarrow R_{j}-\Delta R$. 
We are free to choose whatever values of $\Delta \delta_{j}$ 
define an appropriate initial condition. For example, 
we can assume that the medium has a density PDF 
corresponding to ``fully developed'' turbulence and 
generate the $\Delta \delta_{j}$ exactly in \S~\ref{sec:model:barrier:mf:exact}. 
Or we can begin with a perfectly smooth medium, 
setting all $\Delta \delta_{j}=0$, and treat all structures 
self-consistently as they develop. 
Critically, save the full trajectories $\Delta \delta_{j}$ 
(full $\delta(R)$) for each member of the Monte Carlo population, 
{\em including} those for which the region is ``uncollapsed'' 
($\delta(R)$ never crosses $\delta_{c}(R)$). 

(2) Evolve the system forward by one time step $\Delta t$. 
For a Fourier-space tophat window, we can  
evolve the system by perturbing 
each $\Delta \delta_{j}$ independently according to Equation~\ref{eqn:timeevol} 
(obtaining a new, 
perturbed trajectory $\delta(R,\,t+\Delta t)$). 
The probability distribution for the 
perturbed $\Delta \delta_{j}(R,\,t+\Delta t)$ 
is given by Equation~\ref{eqn:timeevol} with 
the appropriate substitutions: 
\begin{align}
\frac{{\rm d}p(\Delta\delta_{j}[t+\Delta t])}{{\rm d}(\Delta\delta_{j}[t+\Delta t])} &=\\ 
\nonumber
\frac{1}{\sqrt{2\pi\psi\,\Delta S}}\,
&\exp{{\Bigl(}-\frac{(\Delta\delta_{j}[t+\Delta t]-\Delta \delta_{j}[t]\sqrt{1-\psi})^{2}}
{2\psi\,\Delta S} {\Bigr)}} \\ 
\psi &\equiv 1 - \exp{(-2\,\Delta t/\tau)}\\ 
\tau &\equiv {R}/{\langle v_{t}(R)^{2} \rangle^{1/2}} 
\end{align}
This is equivalent to taking 
\begin{align}
\Delta \delta_{j}(t+\Delta t) &=  \Delta \delta_{j}(t)\,\exp{(-\Delta t/\tau)} \\ 
& + \mathcal{R} \, \sqrt{\Delta S\,(1-\exp{(-2\,\Delta t/\tau)})}\\ 
& \approx 
\Delta \delta_{j}(t)\,(1-\Delta t/\tau) + \mathcal{R}\,\sqrt{2\,\Delta S\,\Delta t/\tau}
\end{align}
where $\mathcal{R}$ is a Gaussian random number with unity variance.
This is done for all $\Delta \delta_{j}$ in the trajectory, 
giving a new trajectory 
\be
\delta(R,\,t+\Delta t) \equiv \sum_{j}^{R_{j}>R}\,\Delta \delta_{j}(R_{j},\,t+\Delta t)
\ee
which can now be evaluated.

(3) After each timestep, evaluate all trajectories $\delta(R)$ in the Monte Carlo ensemble. 
If the trajectory did {\em not} cross $\delta_{c}(R)$ at any $R$ in the previous 
time steps -- i.e.\ it represented an uncollapsed region -- then 
either it will remain uncollapsed ($\delta(R)<\delta_{c}(R)$ at all $R$) in the 
new time step, or it will now cross the barrier at some $R_{\rm c}$. The 
largest such $R_{\rm c}$ corresponds to the collapse scale, 
defining a new self-gravitating object with mass $M \equiv 4\pi/3\,\rho_{c}(R_{\rm c})\,R_{\rm c}^{3}$. 
Physically, this event corresponds to the random density fluctuations from e.g.\ shocks 
and other processes pushing this previously ``diffuse'' 
gas parcel to densities at which it becomes self-gravitating, and collapses.
The trajectory should still be saved, but the mass is now in a self-gravitating object, 
and the first-crossing scale on which it became self-gravitating should be recorded. 

If the trajectory already crossed the barrier at some $R_{\rm c}$, then there are two 
possibilities. If the trajectory no longer crosses the barrier (or crosses at some 
smaller radius $R < R_{\rm c}$), it has no effect (continue to save the trajectory, but do not 
modify the properties of the object). This corresponds to a decline in the ``background'' 
density field -- however, because the object is self-gravitating, this cannot simply ``random 
walk'' the collapsed region back into being uncollapsed. By definition, gravity will 
prevent such expansion modulo some stronger forces applied in the model (discussed 
below). However, 
if the trajectory now crosses the barrier $\delta_{c}(R)$ at 
some {\em larger} $R_{\rm c,\,new} > R_{\rm c}$, 
this corresponds to a mass growth event for the collapsed object. The 
mass of the cloud should be updated to 
$M_{\rm c,\,new} \rightarrow (4\pi/3)\,\rho_{c}(R_{\rm c,\,new})\,R_{\rm c,\,new}^{3} > M_{\rm c}$, 
and the first-crossing/collapse radius updated to 
$R_{\rm c} \rightarrow R_{\rm c,\,new}$. 
Unlike the case with dark matter halos (where all mass is locked into halos, so 
every growth event is a halo-halo merger), the fact that there is un-collapsed mass 
means that some of these events correspond to cloud-cloud mergers, 
while others correspond to previously ``diffuse'' gas reaching a self-gravitating threshold. 
If this distinction is needed, the method in 
\citet{somerville:merger.trees} can easily be generalized to decompose the mass growth 
$\Delta\,M=M_{\rm c,\,new}-M_{\rm c}$ into a ``progenitor cloud'' and ``diffuse'' mass function.\footnote{To 
first approximation, this has the same behavior as the halo case: namely that the ``progenitor'' 
mass function has a similar shape to the ``collapsed object'' mass function, here 
with a similar ``diffuse'' mass fraction.}

(4) Apply whatever cloud-specific physics are desired, in the timestep 
$\Delta t$, for the population of identified bound objects. 
This is where the essence of any semi-analytic model 
enters. One could assume clouds 
continue to collapse under gravity, 
that they form stars (either instantaneously, or with some 
efficiency in time, or 
with some association with clump-clump 
mergers), that they form molecules (based on 
e.g.\ their local column densities and SFRs), 
that they disrupt on some timescale or as some function of 
star formation/feedback properties, that they accrete ``diffuse'' 
material (e.g.\ Bondi-Hoyle accretion, which as a non-local 
effect is {\em not} included in the ``growth events'' in step (3)). 
There are obviously a huge range of model physics than can 
be included.

One particularly interesting application of this model to bound clouds 
is to consider recursively applying the same analysis {\em within} 
each cloud, to determine the bound sub-units into which it will fragment. 
For a given bound cloud, if the model defines some average 
density and turbulent power spectrum (for example, 
assuming they maintain their properties at collapse, 
virialize and contract by dissipation, etc), then the procedure to 
determine the mass function and other properties of these ``sub-clumps'' 
is exactly identical to the procedure for the ``parent'' clumps, but 
with the revised or re-normalized density/mass/velocity properties 
of the ``parent'' clump. 

(5) Repeat steps (2)-(4), to continue to evolve the system in time 
as desired.

We also note that despite our stated assumption of steady-state 
turbulence, it is perfectly possible to make the global parameters 
of the model (e.g.\ densities, masses, assumed structural 
properties, turbulent power spectrum) arbitrary functions of time and/or 
consequences of the explicit ``cloud physics'' put into the model. 
For example, allowing for continuous accretion and/or gas exhaustion 
to systematically change the normalization of the density with 
time, or allowing turbulent velocities to damp in the absence of 
some feedback from clouds to ``pump'' them. 
Likewise, it is possible to repeat or rescale these experiments 
in different ``intervals'' corresponding to the average properties 
at different radii in a galaxy disk (corresponding 
to e.g.\ an exponential profile) so that together, the Monte-Carlo 
ensemble can represent the properties of the entire galaxy disk. 
The only implicit assumption in the above is that these properties 
evolve slowly, relative to the local mixing/equilibration time 
for the turbulence (a crossing time).

\begin{figure}
    \centering
    \plotone{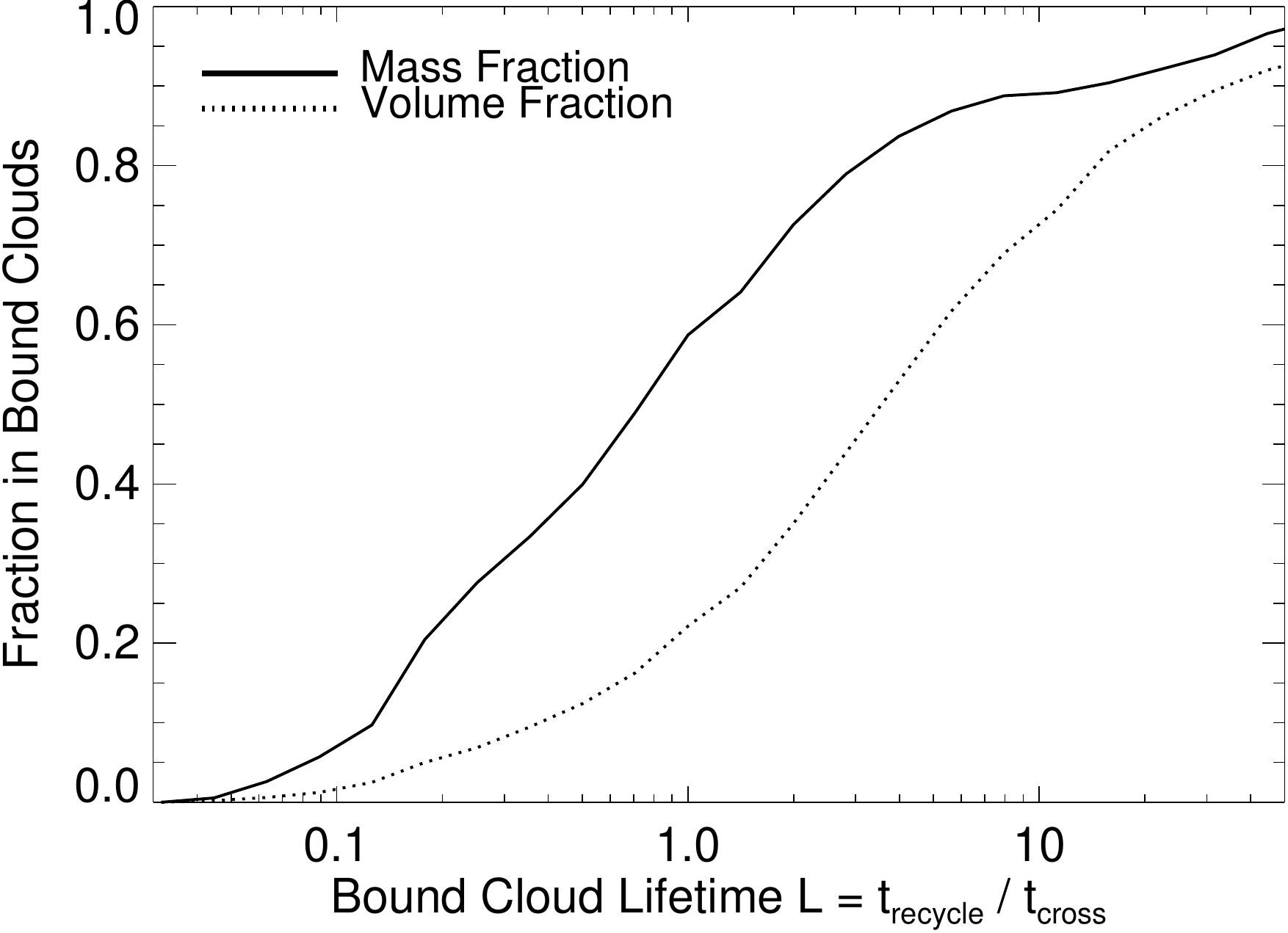}
    \caption{Fraction of the total ISM gas mass and volume in bound clouds, 
    as a function of the cloud lifetime (in units of the cloud crossing time). 
    We follow a full population of clouds through a time-dependent ``merger/fragmentation 
    tree'' constructed as described in \S~\ref{sec:model:trees:mechanism}. 
    When a bound region collapses, we allow it to remain collapsed for a 
    time $t_{\rm lifetime} = L\,t_{\rm dyn} = L/(R_{\rm c}/v_{t}[R_{\rm c}])$ 
    ($t_{\rm dyn}$ is the crossing time at the moment of becoming bound); 
    when this time expires the mass is returned to the ``diffuse'' (non-bound) 
    ISM. A lifetime $\sim 1-5\,t_{\rm dyn}$ gives a fraction in bound 
    units consistent with the observed ISM; larger values lock all mass into 
    bound units (and will over-predict the GMC MF), smaller values 
    the opposite.
    \label{fig:lifetimes}}
\end{figure}

\begin{figure}
    \centering
    \plotone{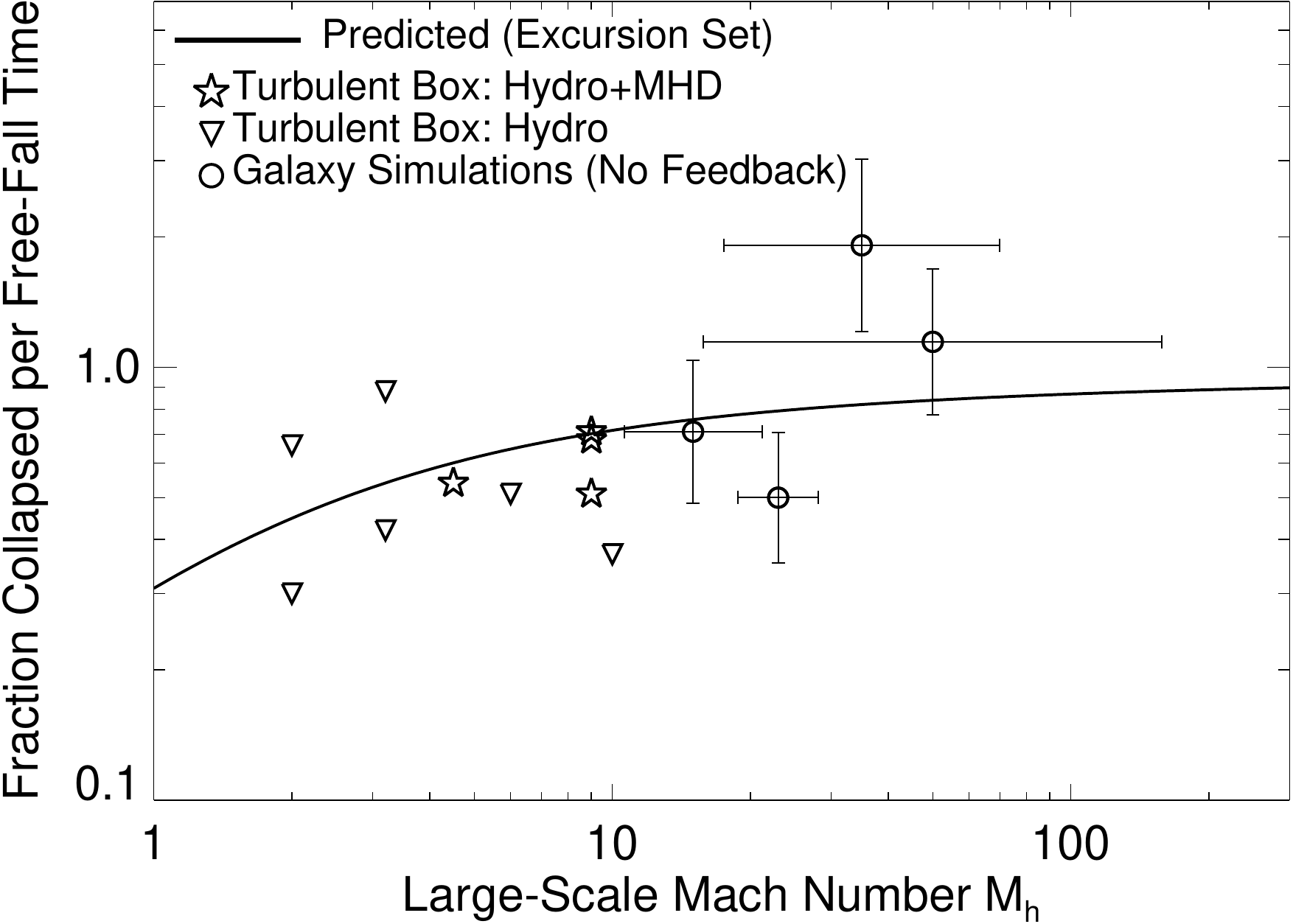}
    \caption{Mass fraction which collapses into bound sub-units per free-fall time 
    (defined as $t_{\rm ff}=0.54/\sqrt{G\,\rho_{0}}\approx h/\sigma(h) = \Omega^{-1}$, 
    if $Q=1$), 
    for systems which are marginally stable on large scales, 
    as a function of Mach number (normalized at large scales, $\mathcal{M}_{h}=\mathcal{M}(h)$).
    Solid line is the analytic prediction from running a suite of models as in Figure~\ref{fig:lifetimes}, 
    in which clouds remain bound (are removed from the ``diffuse'' medium).
    Given the same global stability, systems with higher $\mathcal{M}$
    have larger dispersions and more rapidly cross 
    the collapse barrier, but the scaling is weak.
    Without some additional physics to disrupt bound objects, 
    mass collapses at $\sim M_{\rm gas}/t_{\rm dyn}$ 
    independent of the maintenance of the large-scale cascade.
    We compare with the results of high-resolution 
    simulations of turbulence. 
    First, 
    idealized forced turbulent box simulations with collapse into 
    sink particles from \citet{vazquez-semadeni:2003.turb.reg.sfr} (hydro) 
    and \citet{padoan:2011.new.turb.collapse.sims} (hydro and MHD). 
    For each, we select the runs which most closely match the ``marginally stable'' 
    assumption ($\alpha_{\rm vir}\approx1$ for the box).
    Second, 
    full disk galaxy simulations with no stellar feedback from \citet{hopkins:rad.pressure.sf.fb} 
    with ``collapse rates'' defined as 
    the modeled star formation rate (which occurs in 
    bound gas at densities $n>1000\,{\rm cm^{-3}}$, on $\sim1\,$pc scales). 
    As shown therein, these all maintain $Q\approx1$ via gravitational instabilities; 
    $\mathcal{M}_{h}$ is taken as the mass-weighted average $\mathcal{M}$ for 
    the disk gas averaged over a spatial scale $=h$. 
    Error bars show the scatter in both quantities.
    The excursion-set model successfully predicts the results of fully nonlinear 
    hydrodynamic simulations, within the scatter between different simulations/realizations. 
    \label{fig:collapse.vs.mach}}
\end{figure}

\subsection{Example: The Rate of Collapse into 
Bound Units and Constraints on Cloud Lifetimes}
\label{sec:model:trees:examples}

It is not our intention here to present a full semi-analytic model for 
clouds. However, we briefly illustrate how such a model might be 
used with a highly simplified 
implementation. 

We follow steps (0)-(5) above, with the standard (dimensionless) 
parameters and $p=2$ power spectrum 
adopted throughout this paper. 
The specification of the power spectrum and assumption of 
marginal stability completely specify the model, except for the 
physics applied to bound objects, step (4). 

For these bound objects, we apply 
a toy ``zero physics'' model, with the only goal being to see 
the effects of different ``cloud lifetimes'' on the distribution 
of the ISM. When an object has collapsed, we allow it to 
remain collapsed for a ``lifetime'' $L\,t_{\rm cross}$ 
where $t_{\rm cross}$ is the crossing time of the cloud at the 
time of collapse $R_{\rm c}/v_{t}(R_{\rm c})\approx 1/\sqrt{G\rho_{c}}$.
When this time has elapsed, we ``destroy'' the cloud and recycle 
its material, in practice by ``resetting'' the associated path 
(setting all $\Delta \delta_{j} = 0$ for the designated path). 
The trajectory then re-grows with time according to 
Equation~\ref{eqn:timeevol}, essentially randomizing the density in 
a crossing time.

For any choice of $L$, the mass function and
mass fraction in bound objects will eventually 
converge to a steady state value (in practice, this requires only a 
couple of disk crossing times). 
Figure~\ref{fig:lifetimes} shows the resulting steady-state mass 
fraction in collapsed and bound objects, as a function of 
$L$ from $L\ll 1$ to $L\gg 1$. 
When $L\ll 1$, 
the mass fraction in collapsed objects is negligible, 
and declines $\propto L$ at lower $L$ (as expected for 
systems with a constant formation rate). 
When $L\gg1$, the mass in collapsed objects quickly saturates 
near unity (with an exponentially suppressed residual 
non-collapsed mass). In this regime, because clouds live 
much longer then the typical cloud-cloud merger time, 
the mass function also shifts to higher and higher masses 
(roughly shifting the break/maximum GMC mass $M_{\rm break}\propto L$). 

Only choices with $L\sim 1-5$ yield reasonable total 
collapsed masses in steady state (of order tens of percent, but 
with order tens of percent of the mass {\em also} in 
the inter-clump medium), and agreement with the observed 
GMC mass function. This is easy to understand: because 
the density distribution evolves on a crossing time, 
the rate of addition of mass to the GMC population is 
$\sim \exp{(-\nu(h_{\rm max})^{2}/2)}\,M_{\rm gas}/t_{\rm dyn}(h_{\rm max})$ 
where $h_{\rm max}\sim h$ represents the most unstable wavelength, 
where $\nu(h_{\rm max})$ is order unity. 
So the lifetime for an appropriate steady-state 
should be $L\sim\exp{(\nu_{\rm max}^{2}/2)}\sim$\,a few. 

This relates directly to idealized hydrodynamic simulations of 
turbulent boxes with self-gravity. 
These experiments have found that {\em even} when 
a forcing term is included to maintain the turbulent cascade at 
all times (for a box which is globally stable against collapse), 
a large fraction (tens of percent) of the mass in the box 
will reach densities where it becomes self-gravitating 
(presumably turning into stars, if there 
is no feedback) in a free-fall time 
\citep[see e.g.\ the discussion in][]{padoan:2011.new.turb.collapse.sims}. 
Here, we have calculated the exact same quantity analytically 
(on a galaxy-wide scale).

We can estimate the rate of collapse, in the absence of feedback, 
by simply assuming clouds are arbitrarily long-lived and then 
calculating the time for some fraction 
of the mass to be bound into clouds.
If we perform this exercise as a function of the 
dimensionless Mach number $\mathcal{M}_{h}$ (for a $p=2$ spectrum), 
we obtain
\be
\frac{t_{\rm consumption}}{t_{\rm dyn}(h)} 
\approx 1.5-0.34\sqrt{\ln{(1+3\mathcal{M}_{h}/8)}}
\ee
Where we define $t_{\rm consumption}$ as the time to $1/2$ of gas consumed 
and $t_{\rm dyn}(h) \equiv 
h/\langle v_{t}^{2}(h) \rangle^{1/2} \approx \Omega^{-1}$. 
Note the weak and {\em positive} dependence of the collapse rate on $\mathcal{M}_{h}$ -- this 
comes from our assumption of marginal stability for the disk 
as a whole -- at a fixed stability level, larger $\mathcal{M}$ means a broader 
density PDF, and so increases the collapse rate. 
We compare the resulting collapse rate as a function of $\mathcal{M}$ to 
full numerical hydrodynamic simulations: both simulations of small-scale, 
idealized turbulent boxes (in which self-gravitating regions at 
the resolution limit are identified as sink particles), and large-scale simulations of 
galaxy disks (without stellar feedback) in which self-gravitating regions become 
``star particles.'' In all cases we compare models with marginal stability on the 
largest scales. Our analytic calculation is in good agreement with the full 
numerical calculation.

\section{Discussion}
\label{sec:discussion}

We have used the fact that the ISM is super-sonically turbulent on a wide range 
of scales to develop a rigorous excursion-set model for the formation, structure, 
and time evolution of gas structures (e.g.\ GMCs, massive clumps/cores, 
and voids) in the ISM. We derive the conditions for self-gravitating collapse 
in a turbulent medium applicable on both small scales (the Jeans condition) and 
large scales (the Toomre criterion); together with the assumption that the 
density distribution in super-sonic turbulence is approximately lognormal, we 
use this to derive the statistical properties of the smoothed density field on 
all scales as a function of smoothing scale $R$. 
We show, then, that (with some 
appropriate modifications from the standard 
cosmological case) this becomes a well-defined barrier crossing problem 
(albeit one with a complicated barrier structure), 
for which the full methodology of excursion set theory can be applied. 

We use this model to calculate the mass function 
of bound gas structures (over the entire dynamic range from near the sonic length 
to masses well above the Jeans mass). We do so in a rigorous manner that 
explicitly resolves the ``cloud in cloud'' problem. 
We show that this agrees extremely well with observed GMC mass functions 
in the MW and other nearby galaxies. This prediction is nearly independent of any 
free parameters, with the only input being the mass and size of the galaxies. 
Even galaxies such as M33, which has been extensively discussed as apparently 
exhibiting a deviant GMC mass function slope, are accurately predicted. 
The generic properties of the mass function are rigorously derived: 
an exponential cutoff above the Jeans mass (because large-scale density 
fluctuations are suppressed by disk shear) and a faint-end power-law slope close to, but 
slightly shallower than, $-2$ (which deviates logarithmically with mass).
It is near $-2$ (equal mass on all scales) 
generically because the collapse threshold (being relative to $\ln{\rho}$) 
is only a logarithmic function of scale and gravity is scale free; but slightly shallower because 
collapse is more difficult on small scales for any realistic 
turbulent power spectrum. We show this is robust to large variations in mach number 
and velocity power spectrum shape, and even to large deviations from 
exact log-normality in the density PDF. 

The same model also predicts the linewidth-size and size-mass relations of 
these clouds, in good agreement with observations. The linewidth-size relation slope is a generic 
result of the assumed turbulent power spectrum, but its normalization is predicted 
by the assumption that the disk must be globally stable; the size-mass relation follows 
from the collapse criteria. Second-order corrections (from e.g.\ disk shear) make both 
less sensitive to the turbulent index in the range $p\sim5/3-2$.
Residuals from these relations naturally emerge as a function of the galaxy 
surface density, in good agreement with recent comparisons of GMC properties 
in the MW outskirts and center and in high-redshift galaxies. 

The excursion set theory also allows us to rigorously predict the spatial correlation function 
and clustering properties of these clouds; we predict that most of the mass in clouds 
should be weakly biased (i.e.\ trace the overall gas density), but the most massive 
clouds will preferentially be biased towards large-scale overdensities (e.g.\ spiral arms). 
We construct the auto-correlation function of GMCs from the catalogue of clouds in M33 
and show this agrees extremely well with that predicted for clouds of the same mass. 
If we assume that young star clusters should more or less trace the positions of their 
``parent'' clouds, then we can also compare their clustering. We show that both the 
star cluster-gas mass cross-correlation function, and the star cluster auto-correlation function 
observed in the youngest clusters in the Antennae and M51 agrees well with 
that predicted, over the observed range of scales $R <0.1\,h$ to $R>10\,h$.

Using similar logic as applied to the GMC mass function, we can predict the size and 
mass distributions of under-dense ``bubbles'' in the ISM. We show that a large fraction of 
the ISM should be in highly under-dense ``bubbles'' with $\rho\lesssim 0.01-0.1\,\rho_{0}$, 
and that the characteristic size should scale with scale height $h$, as a natural 
consequence of turbulent fluctuations. These require no additional ``power source'' 
other than whatever maintains the turbulence. If we consider the distribution of bubble/hole 
sizes below a density threshold such that they should be easily ionized by the galactic background, 
we show that this agrees very well with the HI ``hole'' size distribution observed in 
nearby galaxies such as the SMC, M31, and Holmberg II. 
The energetic cost of ``creating'' these bubbles is negligible, as compared to the nominally large $PdV$ work required if they were excavated by e.g.\ SNe explosions, and they do not require any internal stars/star clusters to power their expansion. Even if some are powered in this manner, it is clear that many are not. 
This resolves a long-standing problem, as follow-up observations of these ``holes'' have consistently failed to observe SNe remnants or other evidence of young stellar populations ``powering'' the hole expansion \citep[see e.g.][]{rhode:1999.sne.origin.HI.hole.test,weisz:2009.stellar.fb.HI.holes}. We stress that turbulence alone will not explain the gas in bubbles being {\em hot}: the fraction of holes predicted to have a cooling time much longer than a dynamical time from turbulent shocks alone is small. But it will explain their sizes, expansion, and densities. Where they are heated, it requires a much smaller amount of energy, at that point, to simply ``leak into'' the bubble. 

We generalize the excursion-set model of the ISM to allow the construction of {\em time-dependent} 
``merger/fragmentation trees'' which can be used to follow the evolution of 
clouds and construct semi-analytic models for GMC and star-forming populations. 
We provide explicit recipes to construct these trees.
We use a simple example 
to show that, if clouds were not destroyed by some feedback process 
in a timescale $\sim1-5$ crossing times, then all the ISM mass would 
be ``consumed'' (collapsing to arbitrarily high densities in bound objects) 
even if the large scale turbulent cascade were maintained. Absent 
such feedback, we show that our analytic 
calculation can predict
with reasonable accuracy the collapse rates seen in full nonlinear hydrodynamic 
(and MHD) simulations of both turbulent boxes and galaxies over a wide range of characteristic 
Mach numbers. 

It is striking that we can predict so many properties of a highly complex, chaotic, and -- unlike the cosmological case -- fully nonlinear system with a single model. This suggests that a wealth of properties of the ISM and GMC populations are generic consequences of collapse in a supersonically turbulent medium with a characteristic ``scale'' set by gravitational instability in a gaseous disk. This explains why different simulations \citep{ostriker:2001.gmc.column.dist,
dobbs:2008.gmc.collapse.bygrav.angmom,
bournaud:disk.clumps.to.bulge,
tasker:2009.gmc.form.evol.gravalone,
tasker:2011.photoion.heating.gmc.evol} have been able to reproduce various aspects of these observations, despite including very different physics for cooling/feedback/star formation/magnetic fields, and in some cases clearly failing to reproduce other (probably feedback-dependent) properties such as the observed star formation rates and galactic winds 
\citep[see e.g.\ the discussion in][and references therein]{hopkins:fb.ism.prop}. 
What is remarkable is that our theory allows us to calculate these nonlinear properties analytically, over a large dynamic range, and in {\em quantitative} agreement with the observations. 

We should also stress that this model does not necessarily imply or require that the ISM structure be rigorously self-similar or fractal: that may be true, but it is a much stronger statement about the structure of $S(R)$ and $B(R)$ (compared to our assumptions). In our default model those happen to be {\em approximately} scale free over some range, but there are at least two scales -- the sonic length and disk scale height -- above/below which behavior changes. The application of this model also does not necessarily imply the ISM is ``hierarchical'' either in the cosmological sense or the sense of \citet{vazquez-semadeni:1994.turb.density.pdf} (see \S~\ref{sec:intro}). In fact in the cosmological sense of the term, the predicted structure is more appropriately ``anti-hierarchical,'' in that collapse tends to proceed ``top-down.'' Large scales $\sim h$ are most unstable and contain most of the turbulent power, and we have shown that most self-gravitating objects on small scales are formed by ``fragmenting out'' of larger structures (i.e.\ form within parent GMCs; these are the low-mass GMCs predicted if we ignore the cloud-in-cloud problem, which are much more abundant than isolated counterparts). In contrast, in the cosmological case, small structures form first, and ``subhalos'' are only a small fraction of the population at most masses.

There are many interesting potential future directions for these models. 
Many of our assumptions can be made more general, and the model made more accurate. 
For example, if the gas is not isothermal, or when magnetic fields and global gravitational forces are strong, some deviations from lognormality are expected. We have argued that should not qualitatively change our conclusions, but it is possible to rigorously treat this regime, by extending the 
excursion set formalism to non-Gaussian fields 
\citep[as developed in e.g.][]{matarrese:2000.nongaussian.numdens,
afshordi:2008.nongaussian.collapsestatistics,
maggiore:2010.nongaussian.eps}. The Monte Carlo excursion set approach is also 
completely generalizeable to treat correlated random walks, so that arbitrary higher-order structure 
functions (i.e.\ correlations between fluctuations on different scales) can be incorporated -- it is only a convenient simplifying assumption to assume strict locality \citep[see the review in][]{zentner:eps.methodology.review}. 
Near and below the sonic length (or in the warm/diffuse ISM), when 
the turbulence is sub/transsonic, additional corrections to the 
power spectrum could be included. Magnetic fields can also be included as more than just a correction to the effective sound speed, if their power spectrum is well determined 
\citep[see e.g.][]{kim:2002.mhd.disk.instabilities}.
We have also assumed that the density and velocity field are not directly coupled, but it is 
in principle straightforward to allow for a direct correlation between the local density and 
velocity fluctuations, to follow both with a higher-dimensional random 
walk, and to incorporate this in the collapse criterion \citep[see e.g.][]{sheth:2002.linear.barrier}. 
We have neglected large-scale gradients in e.g.\ the disk surface density profile; 
this should not be important for most GMCs since their sizes are $\lesssim h$, 
but a more rigorous calculation of global properties (e.g.\ the large-scale spatial 
distribution of clouds) could consider each radial annulus in turn with some 
global mass profile. Our derivation of the collapse barrier also assumes spherical collapse, 
when in fact most GMCs are ellipsoidal or triaxial. In the cosmological 
context, ellipsoidal collapse is a well-studied problem \citep{shethtormen}, and can be incorporated 
via an appropriate change of the barrier shape (although the appropriate parameters 
are usually determined by reference to numerical simulations); however, if the 
cosmological case is any guide, the differences should be small (tens of percent level).

The lack of dependence of many of the predicted observables on the detailed properties of turbulence is, in one sense, reassuring (and explains agreement between previous models with different physics). On the other hand, unfortunately, it means that observations have a limited ability to discriminate between these different physical scenarios. 
Of course, the ISM is not all highly supersonic, and there will be some regimes in which the model here is not appropriate. Implicitly, our model assumes that the ISM can cool efficiently (cooling time short relative to the dynamical time), so that the turbulence remains super-sonic. This is easily satisfied inside the radii that include most star formation in galactic disks. However, at sufficiently large radii and very low gas densities, the 
galactic and cosmological ionizing backgrounds maintain gas disks as fully ionized with $Q>1$. 
Even in the star-forming disk, there are bubbles of hot gas that may escape before cooling. And a significant fraction of the mass in the ``warm'' ISM medium is turbulent and bound, but has comparable thermal sound speeds and turbulent velocities (i.e.\ is transsonic rather than super-sonic).
In this regime, it is necessary to account explicitly for the effects of heating and cooling, 
for example as in the model of \citet{ostriker:2010.molecular.reg.sf}, since only regions pushed to a critical density where cooling becomes efficient will behave as we assume. Even in this regime, however, the internal structure of those cooling regions/GMCs should be treatable with the model here, but it should be emphasized that this model is most applicable to the cold/rapidly cooling gas rather than the extended low-density gas.

There is also a huge space of physical predictions which can be explored with this model 
that we have not yet addressed, some of which may be more sensitive tests of the properties of turbulence and ISM structure formation. Using the time-dependent formulation, the growth of 
of GMCs via turbulent density fluctuations, mergers, and accretion can be rigorously 
analytically calculated. By allowing for global evolution of self-gravitating regions, it is possible to self-consistently follow features that nominally appear to contradict the model -- for example, following a simple 
model whereby, once ``detached'' from the background, a cloud collapses spherically 
will naturally predict a power-law tail to high densities (in collapsing regions), even though we can continue to use the same treatment for each cloud internally.  Together with a semi-analytic model for star formation in such units, their destruction via feedback can also be followed analytically in a self-consistent statistical model for the population. Global feedback effects can also be predicted -- for example, many authors have used the cosmological formulation of the problem to study the reionization history of the Universe and evolving size distribution of HII regions \citep[e.g.][]{haiman:2000.reion.bubble.eps,
furlanetto:2004.reion.bubble.eps}. It is straightforward to adapt their approach to the problem here to predict the properties of galactic HII regions, SNe blastwaves, and ionizing photon escape fractions. The model can be extended iteratively (downwards in scale) within GMCs, to calculate the properties of dense collapsing subregions (cores). Extended sufficiently in scale, the model can even be used to predict the stellar IMF in each subregion, following \citet{hennebelle:2008.imf.presschechter} -- with a model determined on {\em galactic} scales. These scales, being closer to the sonic length, should exhibit much stronger dependence on the actual turbulent structure than the galactic-scale quantities we calculate here \citep[as seen in other analytic calculations and simulations; see][]{ballesteros-paredes:2006.imf.turb.sims,hennebelle:2009.imf.variation}, and might be used to break degeneracies between different models for the ISM microphysics.

\acknowledgments 
We thank Chris McKee and Eliot Quataert for helpful discussions 
during the development of this work. We also thank the anonymous referee as well as Patrick Hennebelle, Gilles Chabrier, and Alyssa Goodman for a number of suggestions and thoughtful comments. Support for PFH was provided by the Miller Institute for Basic Research in Science, University of California Berkeley.
\\

\bibliography{/Users/phopkins/Documents/lars_galaxies/papers/ms}

\begin{thebibliography}{114}
\expandafter\ifx\csname natexlab\endcsname\relax\def\natexlab#1{#1}\fi

\bibitem[{{Afshordi} \&
  {Tolley}(2008)}]{afshordi:2008.nongaussian.collapsestatistics}
{Afshordi}, N., \& {Tolley}, A.~J. 2008, \prd, 78, 123507

\bibitem[{{Anathpindika}(2011)}]{anathpindika:2011.shell.frag}
{Anathpindika}, S. 2011, New Astronomy, 16, 323

\bibitem[{{Ballesteros-Paredes} {et~al.}(2006){Ballesteros-Paredes}, {Gazol},
  {Kim}, {Klessen}, {Jappsen}, \&
  {Tejero}}]{ballesteros-paredes:2006.imf.turb.sims}
{Ballesteros-Paredes}, J., {Gazol}, A., {Kim}, J., {Klessen}, R.~S., {Jappsen},
  A.-K., \& {Tejero}, E. 2006, \apj, 637, 384

\bibitem[{{Ballesteros-Paredes}
  {et~al.}(2011{\natexlab{a}}){Ballesteros-Paredes}, {Hartmann},
  {V{\'a}zquez-Semadeni}, {Heitsch}, \&
  {Zamora-Avil{\'e}s}}]{ballesteros-paredes:2011.core.collapse.sizemass}
{Ballesteros-Paredes}, J., {Hartmann}, L.~W., {V{\'a}zquez-Semadeni}, E.,
  {Heitsch}, F., \& {Zamora-Avil{\'e}s}, M.~A. 2011{\natexlab{a}}, \mnras, 411,
  65

\bibitem[{{Ballesteros-Paredes}
  {et~al.}(2011{\natexlab{b}}){Ballesteros-Paredes}, {Vazquez-Semadeni},
  {Gazol}, {Hartmann}, {Heitsch}, \&
  {Colin}}]{ballesteros-paredes:2011.dens.pdf.vs.selfgrav}
{Ballesteros-Paredes}, J., {Vazquez-Semadeni}, E., {Gazol}, A., {Hartmann},
  L.~W., {Heitsch}, F., \& {Colin}, P. 2011{\natexlab{b}}, \mnras, 416, 1436

\bibitem[{{Begelman} \&
  {Shlosman}(2009)}]{begelman:direct.bh.collapse.w.turbulence}
{Begelman}, M.~C., \& {Shlosman}, I. 2009, \apjl, 702, L5

\bibitem[{{Blitz} \& {Rosolowsky}(2006)}]{blitz:h2.pressure.corr}
{Blitz}, L., \& {Rosolowsky}, E. 2006, \apj, 650, 933

\bibitem[{{Block} {et~al.}(2010){Block}, {Puerari}, {Elmegreen}, \&
  {Bournaud}}]{block:2010.lmc.vel.powerspectrum}
{Block}, D.~L., {Puerari}, I., {Elmegreen}, B.~G., \& {Bournaud}, F. 2010,
  \apjl, 718, L1

\bibitem[{{Bolatto} {et~al.}(2008){Bolatto}, {Leroy}, {Rosolowsky}, {Walter},
  \& {Blitz}}]{bolatto:2008.gmc.properties}
{Bolatto}, A.~D., {Leroy}, A.~K., {Rosolowsky}, E., {Walter}, F., \& {Blitz},
  L. 2008, \apj, 686, 948

\bibitem[{{Boldyrev} {et~al.}(2002){Boldyrev}, {Nordlund}, \&
  {Padoan}}]{boldyrev:2002.sf.cloud.turb}
{Boldyrev}, S., {Nordlund}, {\AA}., \& {Padoan}, P. 2002, \apj, 573, 678

\bibitem[{{Bonazzola} {et~al.}(1987){Bonazzola}, {Heyvaerts}, {Falgarone},
  {Perault}, \& {Puget}}]{bonazzola:1987.turb.jeans.instab}
{Bonazzola}, S., {Heyvaerts}, J., {Falgarone}, E., {Perault}, M., \& {Puget},
  J.~L. 1987, \aap, 172, 293

\bibitem[{{Bond} {et~al.}(1991){Bond}, {Cole}, {Efstathiou}, \&
  {Kaiser}}]{bond:1991.eps}
{Bond}, J.~R., {Cole}, S., {Efstathiou}, G., \& {Kaiser}, N. 1991, \apj, 379,
  440

\bibitem[{Bournaud {et~al.}(2007)Bournaud, Elmegreen, \&
  Elmegreen}]{bournaud:disk.clumps.to.bulge}
Bournaud, F., Elmegreen, B.~G., \& Elmegreen, D.~M. 2007, The Astrophysical
  Journal, 670, 237

\bibitem[{{Bournaud} {et~al.}(2010){Bournaud}, {Elmegreen}, {Teyssier},
  {Block}, \& {Puerari}}]{bournaud:2010.grav.turbulence.lmc}
{Bournaud}, F., {Elmegreen}, B.~G., {Teyssier}, R., {Block}, D.~L., \&
  {Puerari}, I. 2010, \mnras, 409, 1088

\bibitem[{{Bower}(1991)}]{bower:1991.eps.groups}
{Bower}, R.~G. 1991, \mnras, 248, 332

\bibitem[{Bowman(1996)}]{bowman:inertial.range.turbulent.spectra}
Bowman, J.~C. 1996, Journal of Fluid Mechanics, 306, 167

\bibitem[{{Brinks} \& {Bajaja}(1986)}]{brinks:1986.m31.HI.holes}
{Brinks}, E., \& {Bajaja}, E. 1986, \aap, 169, 14

\bibitem[{Burgers(1973)}]{burgers1973turbulence}
Burgers, J. 1973, The nonlinear diffusion equation: asymptotic solutions and
  statistical problems (D. Reidel Pub. Co.)

\bibitem[{{Bussmann} {et~al.}(2008)}]{bussmann:2008.hcn32.sfr}
{Bussmann}, R.~S., {et~al.} 2008, \apjl, 681, L73

\bibitem[{{Chandrasekhar}(1951)}]{chandrasekhar:1951.turb.jeans.condition}
{Chandrasekhar}, S. 1951, Royal Society of London Proceedings Series A, 210, 26

\bibitem[{Cole {et~al.}(2000)Cole, Lacey, Baugh, \&
  Frenk}]{cole:durham.sam.initial}
Cole, S., Lacey, C.~G., Baugh, C.~M., \& Frenk, C.~S. 2000, \mnras, 319, 168

\bibitem[{{Deul} \& {den Hartog}(1990)}]{deul:1990.m33.HI.holes}
{Deul}, E.~R., \& {den Hartog}, R.~H. 1990, \aap, 229, 362

\bibitem[{{Dobbs}(2008)}]{dobbs:2008.gmc.collapse.bygrav.angmom}
{Dobbs}, C.~L. 2008, \mnras, 391, 844

\bibitem[{{Elmegreen}(1987)}]{elmegreen:1987.cloud.instabilities}
{Elmegreen}, B.~G. 1987, \apj, 312, 626

\bibitem[{{Elmegreen}(2002)}]{elmegreen:2002.fractal.cloud.mf}
---. 2002, \apj, 564, 773

\bibitem[{{Engargiola} {et~al.}(2003){Engargiola}, {Plambeck}, {Rosolowsky}, \&
  {Blitz}}]{engargiola:2003.m33.gmc.catalogue}
{Engargiola}, G., {Plambeck}, R.~L., {Rosolowsky}, E., \& {Blitz}, L. 2003,
  \apjs, 149, 343

\bibitem[{{Evans} {et~al.}(2009)}]{evans:2009.sf.efficiencies.lifetimes}
{Evans}, N.~J., {et~al.} 2009, \apjs, 181, 321

\bibitem[{{Evans}(1999)}]{evans:1999.sf.gmc.review}
{Evans}, II, N.~J. 1999, \araa, 37, 311

\bibitem[{{Federrath} {et~al.}(2010){Federrath}, {Roman-Duval}, {Klessen},
  {Schmidt}, \& {Mac Low}}]{federrath:2010.obs.vs.sim.turb.compare}
{Federrath}, C., {Roman-Duval}, J., {Klessen}, R.~S., {Schmidt}, W., \& {Mac
  Low}, M.-M. 2010, \aap, 512, A81+

\bibitem[{{Fukui} {et~al.}(2008){Fukui}, {Kawamura}, {Minamidani}, {Mizuno},
  {Kanai}, {Mizuno}, {Onishi}, {Yonekura}, {Mizuno}, {Ogawa}, \&
  {Rubio}}]{fukui:2008.lmc.gmc.catalogue}
{Fukui}, Y., {Kawamura}, A., {Minamidani}, T., {Mizuno}, Y., {Kanai}, Y.,
  {Mizuno}, N., {Onishi}, T., {Yonekura}, Y., {Mizuno}, A., {Ogawa}, H., \&
  {Rubio}, M. 2008, \apjs, 178, 56

\bibitem[{{Furlanetto} {et~al.}(2004){Furlanetto}, {Zaldarriaga}, \&
  {Hernquist}}]{furlanetto:2004.reion.bubble.eps}
{Furlanetto}, S.~R., {Zaldarriaga}, M., \& {Hernquist}, L. 2004, \apj, 613, 1

\bibitem[{{Gao} \& {Solomon}(2004)}]{gao:2004.hcn.sfr.relation}
{Gao}, Y., \& {Solomon}, P.~M. 2004, \apj, 606, 271

\bibitem[{{Goldsmith} {et~al.}(2008){Goldsmith}, {Heyer}, {Narayanan}, {Snell},
  {Li}, \& {Brunt}}]{goldsmith:2008.taurus.gmc.mapping}
{Goldsmith}, P.~F., {Heyer}, M., {Narayanan}, G., {Snell}, R., {Li}, D., \&
  {Brunt}, C. 2008, \apj, 680, 428

\bibitem[{{Goodman} {et~al.}(1998){Goodman}, {Barranco}, {Wilner}, \&
  {Heyer}}]{goodman:1998.lws.dependence.on.tracers}
{Goodman}, A.~A., {Barranco}, J.~A., {Wilner}, D.~J., \& {Heyer}, M.~H. 1998,
  \apj, 504, 223

\bibitem[{{Goodman} {et~al.}(2009{\natexlab{a}}){Goodman}, {Pineda}, \&
  {Schnee}}]{goodman:2009.gmc.column.dist}
{Goodman}, A.~A., {Pineda}, J.~E., \& {Schnee}, S.~L. 2009{\natexlab{a}}, \apj,
  692, 91

\bibitem[{{Goodman} {et~al.}(2009{\natexlab{b}}){Goodman}, {Rosolowsky},
  {Borkin}, {Foster}, {Halle}, {Kauffmann}, \&
  {Pineda}}]{goodman:2009.dendrogram.sims}
{Goodman}, A.~A., {Rosolowsky}, E.~W., {Borkin}, M.~A., {Foster}, J.~B.,
  {Halle}, M., {Kauffmann}, J., \& {Pineda}, J.~E. 2009{\natexlab{b}}, \nat,
  457, 63

\bibitem[{{Haiman} {et~al.}(2000){Haiman}, {Abel}, \&
  {Rees}}]{haiman:2000.reion.bubble.eps}
{Haiman}, Z., {Abel}, T., \& {Rees}, M.~J. 2000, \apj, 534, 11

\bibitem[{{Hennebelle} \& {Chabrier}(2008)}]{hennebelle:2008.imf.presschechter}
{Hennebelle}, P., \& {Chabrier}, G. 2008, \apj, 684, 395

\bibitem[{{Hennebelle} \& {Chabrier}(2009)}]{hennebelle:2009.imf.variation}
---. 2009, \apj, 702, 1428

\bibitem[{{Heyer} {et~al.}(2009){Heyer}, {Krawczyk}, {Duval}, \&
  {Jackson}}]{heyer:2009.gmc.trends.w.surface.density}
{Heyer}, M., {Krawczyk}, C., {Duval}, J., \& {Jackson}, J.~M. 2009, \apj, 699,
  1092

\bibitem[{{Hopkins}(2012)}]{hopkins:excursion.imf}
{Hopkins}, P.~F. 2012, \mnras, in press, arXiv:1201.4387

\bibitem[{{Hopkins} {et~al.}(2011){Hopkins}, {Quataert}, \&
  {Murray}}]{hopkins:rad.pressure.sf.fb}
{Hopkins}, P.~F., {Quataert}, E., \& {Murray}, N. 2011, \mnras, 417, 950

\bibitem[{{Hopkins} {et~al.}(2012){Hopkins}, {Quataert}, \&
  {Murray}}]{hopkins:fb.ism.prop}
---. 2012, \mnras, 421, 3488

\bibitem[{{Kim} \& {Ryu}(2005)}]{kim:2005.density.pwrspec.turb}
{Kim}, J., \& {Ryu}, D. 2005, \apjl, 630, L45

\bibitem[{{Kim} {et~al.}(1999){Kim}, {Dopita}, {Staveley-Smith}, \&
  {Bessell}}]{kim:1999.lmc.hi.holes}
{Kim}, S., {Dopita}, M.~A., {Staveley-Smith}, L., \& {Bessell}, M.~S. 1999,
  \aj, 118, 2797

\bibitem[{{Kim} {et~al.}(2002){Kim}, {Ostriker}, \&
  {Stone}}]{kim:2002.mhd.disk.instabilities}
{Kim}, W.-T., {Ostriker}, E.~C., \& {Stone}, J.~M. 2002, \apj, 581, 1080

\bibitem[{{Klessen}(2000)}]{klessen:2000.pdf.supersonic.turb}
{Klessen}, R.~S. 2000, \apj, 535, 869

\bibitem[{{Kowal} {et~al.}(2007){Kowal}, {Lazarian}, \&
  {Beresnyak}}]{kowal:2007.log.density.turb.spectra}
{Kowal}, G., {Lazarian}, A., \& {Beresnyak}, A. 2007, \apj, 658, 423

\bibitem[{{Kritsuk} {et~al.}(2007){Kritsuk}, {Norman}, {Padoan}, \&
  {Wagner}}]{kritsuk:2007.isothermal.turb.stats}
{Kritsuk}, A.~G., {Norman}, M.~L., {Padoan}, P., \& {Wagner}, R. 2007, \apj,
  665, 416

\bibitem[{{Krumholz} \& {McKee}(2005)}]{krumholz.schmidt}
{Krumholz}, M.~R., \& {McKee}, C.~F. 2005, \apj, 630, 250

\bibitem[{{Lacey} \& {Cole}(1993)}]{laceycole:halo.merger.rates.93}
{Lacey}, C., \& {Cole}, S. 1993, \mnras, 262, 627

\bibitem[{{Larson}(1981)}]{larson:gmc.scalings}
{Larson}, R.~B. 1981, \mnras, 194, 809

\bibitem[{{Lemaster} \& {Stone}(2009)}]{lemaster:2009.density.pdf.turb.review}
{Lemaster}, M.~N., \& {Stone}, J.~M. 2009, in Revista Mexicana de Astronomia y
  Astrofisica, vol. 27, Vol.~36, Revista Mexicana de Astronomia y Astrofisica
  Conference Series, 243--+

\bibitem[{{Li} {et~al.}(2004){Li}, {Norman}, {Mac Low}, \&
  {Heitsch}}]{li:2004.turb.reg.sfr}
{Li}, P.~S., {Norman}, M.~L., {Mac Low}, M.-M., \& {Heitsch}, F. 2004, \apj,
  605, 800

\bibitem[{{Li} \& {Nakamura}(2006)}]{li:2006.turb.reg.sfr}
{Li}, Z.-Y., \& {Nakamura}, F. 2006, \apjl, 640, L187

\bibitem[{{Lombardi} {et~al.}(2010){Lombardi}, {Alves}, \&
  {Lada}}]{lombardi:2010.larsonlaws.extinction.lognormal}
{Lombardi}, M., {Alves}, J., \& {Lada}, C.~J. 2010, \aap, 519, L7

\bibitem[{{Mac Low} \& {Klessen}(2004)}]{mac-low:2004.turb.sf.review}
{Mac Low}, M.-M., \& {Klessen}, R.~S. 2004, Reviews of Modern Physics, 76, 125

\bibitem[{{Maggiore} \&
  {Riotto}(2010{\natexlab{a}})}]{maggiore:2010.path.integral.non.tophat.eps.filter}
{Maggiore}, M., \& {Riotto}, A. 2010{\natexlab{a}}, \apj, 711, 907

\bibitem[{{Maggiore} \&
  {Riotto}(2010{\natexlab{b}})}]{maggiore:2010.nongaussian.eps}
---. 2010{\natexlab{b}}, \apj, 717, 526

\bibitem[{{Matarrese} {et~al.}(2000){Matarrese}, {Verde}, \&
  {Jimenez}}]{matarrese:2000.nongaussian.numdens}
{Matarrese}, S., {Verde}, L., \& {Jimenez}, R. 2000, \apj, 541, 10

\bibitem[{{McKee} \& {Ostriker}(2007)}]{mckee:2007.sf.theory.review}
{McKee}, C.~F., \& {Ostriker}, E.~C. 2007, \araa, 45, 565

\bibitem[{{Mo} \& {White}(1996)}]{mowhite:bias}
{Mo}, H.~J., \& {White}, S.~D.~M. 1996, \mnras, 282, 347

\bibitem[{{Nordlund} \& {Padoan}(1999)}]{nordlund:1999.density.pdf.supersonic}
{Nordlund}, {\AA}.~K., \& {Padoan}, P. 1999, in Interstellar Turbulence;
  Cambridge University Press, ed. {J.~Franco \& A.~Carraminana} (Cambridge, UK:
  Cambridge University Press), 218--+

\bibitem[{{Oey} \& {Clarke}(1997)}]{oey:1997.HI.hole.models}
{Oey}, M.~S., \& {Clarke}, C.~J. 1997, \mnras, 289, 570

\bibitem[{{Oka} {et~al.}(2001){Oka}, {Hasegawa}, {Sato}, {Tsuboi}, {Miyazaki},
  \& {Sugimoto}}]{oka:2001.mw.center.gmcs}
{Oka}, T., {Hasegawa}, T., {Sato}, F., {Tsuboi}, M., {Miyazaki}, A., \&
  {Sugimoto}, M. 2001, \apj, 562, 348

\bibitem[{{Ostriker} {et~al.}(1999){Ostriker}, {Gammie}, \&
  {Stone}}]{ostriker:1999.density.pdf}
{Ostriker}, E.~C., {Gammie}, C.~F., \& {Stone}, J.~M. 1999, \apj, 513, 259

\bibitem[{{Ostriker} {et~al.}(2010){Ostriker}, {McKee}, \&
  {Leroy}}]{ostriker:2010.molecular.reg.sf}
{Ostriker}, E.~C., {McKee}, C.~F., \& {Leroy}, A.~K. 2010, \apj, 721, 975

\bibitem[{{Ostriker} {et~al.}(2001){Ostriker}, {Stone}, \&
  {Gammie}}]{ostriker:2001.gmc.column.dist}
{Ostriker}, E.~C., {Stone}, J.~M., \& {Gammie}, C.~F. 2001, \apj, 546, 980

\bibitem[{{Padoan} {et~al.}(2004){Padoan}, {Jimenez}, {Nordlund}, \&
  {Boldyrev}}]{padoan:2004.turb.structure.functions}
{Padoan}, P., {Jimenez}, R., {Nordlund}, {\AA}., \& {Boldyrev}, S. 2004,
  Physical Review Letters, 92, 191102

\bibitem[{{Padoan} {et~al.}(2006){Padoan}, {Juvela}, {Kritsuk}, \&
  {Norman}}]{padoan:2006.perseus.turb.spectrum}
{Padoan}, P., {Juvela}, M., {Kritsuk}, A., \& {Norman}, M.~L. 2006, \apjl, 653,
  L125

\bibitem[{{Padoan} \& {Nordlund}(2002)}]{padoan:2002.density.pdf}
{Padoan}, P., \& {Nordlund}, {\AA}. 2002, \apj, 576, 870

\bibitem[{{Padoan} \& {Nordlund}(2011)}]{padoan:2011.new.turb.collapse.sims}
---. 2011, \apj, 730, 40

\bibitem[{{Padoan} {et~al.}(1997){Padoan}, {Nordlund}, \&
  {Jones}}]{padoan:1997.density.pdf}
{Padoan}, P., {Nordlund}, A., \& {Jones}, B.~J.~T. 1997, \mnras, 288, 145

\bibitem[{{Pan} \& {Scannapieco}(2010)}]{pan:2010.turbulent.mixing.times}
{Pan}, L., \& {Scannapieco}, E. 2010, \apj, 721, 1765

\bibitem[{{Passot} \& {Vazquez-Semadeni}(1998)}]{passot:1998.density.pdf}
{Passot}, T., \& {Vazquez-Semadeni}, E. 1998, PhRvE, 58, 4501

\bibitem[{{Pineda} {et~al.}(2009){Pineda}, {Rosolowsky}, \&
  {Goodman}}]{pineda:2009.clumpfind.issues}
{Pineda}, J.~E., {Rosolowsky}, E.~W., \& {Goodman}, A.~A. 2009, \apjl, 699,
  L134

\bibitem[{{Press} \& {Schechter}(1974)}]{pressschechter}
{Press}, W.~H., \& {Schechter}, P. 1974, \apj, 187, 425

\bibitem[{{Price} {et~al.}(2011){Price}, {Federrath}, \&
  {Brunt}}]{price:2011.density.mach.vs.forcing}
{Price}, D.~J., {Federrath}, C., \& {Brunt}, C.~M. 2011, \apjl, 727, L21

\bibitem[{{Puche} {et~al.}(1992){Puche}, {Westpfahl}, {Brinks}, \&
  {Roy}}]{puche:1992.holmbergII.hI.holes}
{Puche}, D., {Westpfahl}, D., {Brinks}, E., \& {Roy}, J.-R. 1992, \aj, 103,
  1841

\bibitem[{{Rhode} {et~al.}(1999){Rhode}, {Salzer}, {Westpfahl}, \&
  {Radice}}]{rhode:1999.sne.origin.HI.hole.test}
{Rhode}, K.~L., {Salzer}, J.~J., {Westpfahl}, D.~J., \& {Radice}, L.~A. 1999,
  \aj, 118, 323

\bibitem[{{Romeo}(1992)}]{romeo:1992.two.component.dispersion}
{Romeo}, A.~B. 1992, \mnras, 256, 307

\bibitem[{{Rosolowsky}(2005)}]{rosolowsky:gmc.mass.spectrum}
{Rosolowsky}, E. 2005, \pasp, 117, 1403

\bibitem[{{Rosolowsky} \& {Blitz}(2005)}]{rosolowsky:2005.gmcs.m64}
{Rosolowsky}, E., \& {Blitz}, L. 2005, \apj, 623, 826

\bibitem[{{Rosolowsky} {et~al.}(2008){Rosolowsky}, {Pineda}, {Kauffmann}, \&
  {Goodman}}]{rosolowsky:2008.dendrograms}
{Rosolowsky}, E.~W., {Pineda}, J.~E., {Kauffmann}, J., \& {Goodman}, A.~A.
  2008, \apj, 679, 1338

\bibitem[{{Scalo} {et~al.}(1998){Scalo}, {Vazquez-Semadeni}, {Chappell}, \&
  {Passot}}]{scalo:1998.turb.density.pdf}
{Scalo}, J., {Vazquez-Semadeni}, E., {Chappell}, D., \& {Passot}, T. 1998,
  \apj, 504, 835

\bibitem[{{Scheepmaker} {et~al.}(2009){Scheepmaker}, {Lamers}, {Anders}, \&
  {Larsen}}]{scheepmaker:2009.m51.cluster.clustering}
{Scheepmaker}, R.~A., {Lamers}, H.~J.~G.~L.~M., {Anders}, P., \& {Larsen},
  S.~S. 2009, \aap, 494, 81

\bibitem[{{Schmidt} {et~al.}(2009){Schmidt}, {Federrath}, {Hupp}, {Kern}, \&
  {Niemeyer}}]{schmidt:2009.isothermal.turb}
{Schmidt}, W., {Federrath}, C., {Hupp}, M., {Kern}, S., \& {Niemeyer}, J.~C.
  2009, \aap, 494, 127

\bibitem[{{Schmidt} {et~al.}(2008){Schmidt}, {Federrath}, \&
  {Klessen}}]{schmidt:2008.turb.structure.fns}
{Schmidt}, W., {Federrath}, C., \& {Klessen}, R. 2008, Physical Review Letters,
  101, 194505

\bibitem[{{Scoville} {et~al.}(1987){Scoville}, {Yun}, {Sanders}, {Clemens}, \&
  {Waller}}]{scoville:gmc.properties}
{Scoville}, N.~Z., {Yun}, M.~S., {Sanders}, D.~B., {Clemens}, D.~P., \&
  {Waller}, W.~H. 1987, \apjs, 63, 821

\bibitem[{{Sheth} {et~al.}(2001){Sheth}, {Mo}, \& {Tormen}}]{shethtormen}
{Sheth}, R.~K., {Mo}, H.~J., \& {Tormen}, G. 2001, \mnras, 323, 1

\bibitem[{{Sheth} \& {Tormen}(2002)}]{sheth:2002.linear.barrier}
{Sheth}, R.~K., \& {Tormen}, G. 2002, \mnras, 329, 61

\bibitem[{{Sheth} \& {van de Weygaert}(2004)}]{sheth:2004.void.eps}
{Sheth}, R.~K., \& {van de Weygaert}, R. 2004, \mnras, 350, 517

\bibitem[{{Slyz} {et~al.}(2005){Slyz}, {Devriendt}, {Bryan}, \&
  {Silk}}]{slyz:2005.kpc.box.sf.feedback}
{Slyz}, A.~D., {Devriendt}, J.~E.~G., {Bryan}, G., \& {Silk}, J. 2005, \mnras,
  356, 737

\bibitem[{{Somerville} \& {Kolatt}(1999)}]{somerville:merger.trees}
{Somerville}, R.~S., \& {Kolatt}, T.~S. 1999, \mnras, 305, 1

\bibitem[{{Stanimirovic} {et~al.}(1999){Stanimirovic}, {Staveley-Smith},
  {Dickey}, {Sault}, \& {Snowden}}]{stanimirovic:1999.smc.hi.pwrspectrum}
{Stanimirovic}, S., {Staveley-Smith}, L., {Dickey}, J.~M., {Sault}, R.~J., \&
  {Snowden}, S.~L. 1999, \mnras, 302, 417

\bibitem[{{Staveley-Smith} {et~al.}(1997){Staveley-Smith}, {Sault},
  {Hatzidimitriou}, {Kesteven}, \&
  {McConnell}}]{staveley-smith:1997.smc.HI.holes}
{Staveley-Smith}, L., {Sault}, R.~J., {Hatzidimitriou}, D., {Kesteven}, M.~J.,
  \& {McConnell}, D. 1997, \mnras, 289, 225

\bibitem[{{Swinbank} {et~al.}(2011){Swinbank}, {Papadopoulos}, {Cox}, {Krips},
  {Ivison}, {Smail}, {Thomson}, {Neri}, {Richard}, \&
  {Ebeling}}]{swinbank:clumps}
{Swinbank}, A.~M., {Papadopoulos}, P.~P., {Cox}, P., {Krips}, M., {Ivison},
  R.~J., {Smail}, I., {Thomson}, A.~P., {Neri}, R., {Richard}, J., \&
  {Ebeling}, H. 2011, \apj, 742, 11

\bibitem[{{Tasker}(2011)}]{tasker:2011.photoion.heating.gmc.evol}
{Tasker}, E.~J. 2011, \apj, 730, 11

\bibitem[{{Tasker} \& {Tan}(2009)}]{tasker:2009.gmc.form.evol.gravalone}
{Tasker}, E.~J., \& {Tan}, J.~C. 2009, \apj, 700, 358

\bibitem[{{Toomre}(1977)}]{toomre:spiral.structure.review}
{Toomre}, A. 1977, \araa, 15, 437

\bibitem[{{Vandervoort}(1970)}]{vandervoort:1970.dispersion.relation}
{Vandervoort}, P.~O. 1970, \apj, 161, 87

\bibitem[{{Vazquez-Semadeni}(1994)}]{vazquez-semadeni:1994.turb.density.pdf}
{Vazquez-Semadeni}, E. 1994, \apj, 423, 681

\bibitem[{{V{\'a}zquez-Semadeni} {et~al.}(2003){V{\'a}zquez-Semadeni},
  {Ballesteros-Paredes}, \& {Klessen}}]{vazquez-semadeni:2003.turb.reg.sfr}
{V{\'a}zquez-Semadeni}, E., {Ballesteros-Paredes}, J., \& {Klessen}, R.~S.
  2003, \apjl, 585, L131

\bibitem[{{V{\'a}zquez-Semadeni} \&
  {Garc{\'{\i}}a}(2001)}]{vazquez-semadeni:2001.nh.pdf.gmc}
{V{\'a}zquez-Semadeni}, E., \& {Garc{\'{\i}}a}, N. 2001, \apj, 557, 727

\bibitem[{{Vazquez-Semadeni} \&
  {Gazol}(1995)}]{vazquez-semadeni:1995.turb.jeans.instab}
{Vazquez-Semadeni}, E., \& {Gazol}, A. 1995, \aap, 303, 204

\bibitem[{{Walter} \& {Brinks}(1999)}]{walter:1999.ic2574.hI.holes}
{Walter}, F., \& {Brinks}, E. 1999, \aj, 118, 273

\bibitem[{{Weisz} {et~al.}(2009){Weisz}, {Skillman}, {Cannon}, {Dolphin},
  {Kennicutt}, {Lee}, \& {Walter}}]{weisz:2009.stellar.fb.HI.holes}
{Weisz}, D.~R., {Skillman}, E.~D., {Cannon}, J.~M., {Dolphin}, A.~E.,
  {Kennicutt}, Jr., R.~C., {Lee}, J., \& {Walter}, F. 2009, \apj, 704, 1538

\bibitem[{{Williams} \& {McKee}(1997)}]{williams:1997.gmc.prop}
{Williams}, J.~P., \& {McKee}, C.~F. 1997, \apj, 476, 166

\bibitem[{{Wilson} {et~al.}(2003){Wilson}, {Scoville}, {Madden}, \&
  {Charmandaris}}]{wilson:2003.supergiant.clumps}
{Wilson}, C.~D., {Scoville}, N., {Madden}, S.~C., \& {Charmandaris}, V. 2003,
  \apj, 599, 1049

\bibitem[{{Wong} {et~al.}(2009){Wong}, {Hughes}, {Fukui}, {Kawamura}, {Mizuno},
  {Ott}, {Muller}, {Pineda}, {Welty}, {Kim}, {Mizuno}, {Murai}, \&
  {Onishi}}]{wong:2009.lmc.co.detection}
{Wong}, T., {Hughes}, A., {Fukui}, Y., {Kawamura}, A., {Mizuno}, N., {Ott}, J.,
  {Muller}, E., {Pineda}, J.~L., {Welty}, D.~E., {Kim}, S., {Mizuno}, Y.,
  {Murai}, M., \& {Onishi}, T. 2009, \apj, 696, 370

\bibitem[{{Wong} {et~al.}(2008)}]{wong:2008.gmc.column.dist}
{Wong}, T., {et~al.} 2008, \mnras, 386, 1069

\bibitem[{Zentner(2007)}]{zentner:eps.methodology.review}
Zentner, A.~R. 2007, International Journal of Modern Physics D, 16, 763

\bibitem[{{Zhang} {et~al.}(2001){Zhang}, {Fall}, \&
  {Whitmore}}]{zhang:2001.antennae.starcluster.clustering}
{Zhang}, Q., {Fall}, S.~M., \& {Whitmore}, B.~C. 2001, \apj, 561, 727

\bibitem[{{Zuckerman} \& {Evans}(1974)}]{zuckerman:1974.gmc.constraints}
{Zuckerman}, B., \& {Evans}, II, N.~J. 1974, \apjl, 192, L149

\end{thebibliography}

\end{document}